\documentclass{fairmeta}
\usepackage{float}      %
\usepackage{multirow}
\usepackage{tabularx, booktabs} %
\usepackage{amsmath}
\usepackage[normalem]{ulem}
\usepackage{xspace}
\usepackage{pifont}%
\usepackage{diagbox} 
\usepackage{array}
\usepackage{xcolor} %
\usepackage{graphicx}
\usepackage{subcaption} %

\newcommand{\flexinput}[2][0.85\linewidth]{%
  \let\oldincludegraphics\includegraphics
  \renewcommand{\includegraphics}[2][]{%
    \oldincludegraphics[width=#1]{##2}%
  }%
  \input{#2}%
  \let\includegraphics\oldincludegraphics
}

\def\bf{\bfseries\sffamily}

\newcommand{\kindatiny}{\fontsize{6pt}{7.2pt}\selectfont}

\newlength\savewidth\newcommand\shline{\noalign{\global\savewidth\arrayrulewidth
  \global\arrayrulewidth 0.5pt}\hline\noalign{\global\arrayrulewidth\savewidth}}

\newcommand{\tablestyle}[2]{%
    \fontfamily{ptm}\selectfont%
    \let\itold\it%
    \def\it{\itold \fontfamily{ptm}\selectfont}%
    \setlength{\tabcolsep}{#1}\renewcommand{\arraystretch}{#2}\centering\kindatiny%
    \let\citeold\cite%
    \renewcommand{\cite}[1]{\normalfont\fontfamily{ptm}\selectfont\tiny\citeold{##1}}%
}
\newcommand{\bigtablestyle}[2]{%
    \fontfamily{ptm}\selectfont%
    \let\itold\it%
    \def\it{\itold \fontfamily{ptm}\selectfont}%
    \setlength{\tabcolsep}{#1}\renewcommand{\arraystretch}{#2}\centering\footnotesize%
    \let\citeold\cite%
    \renewcommand{\cite}[1]{\normalfont\fontfamily{ptm}\selectfont\footnotesize\citeold{##1}}%
}

\renewcommand{\paragraph}[1]{\vspace{1.25mm}\noindent\textbf{#1}}

\newcolumntype{x}[1]{>{\centering\arraybackslash}p{#1pt}}
\newcolumntype{y}[1]{>{\raggedright\arraybackslash}p{#1pt}}
\newcolumntype{z}[1]{>{\raggedleft\arraybackslash}p{#1pt}}
\newcolumntype{w}{>{\centering\arraybackslash}p{18pt}}
\newcolumntype{a}{>{\centering\arraybackslash}p{16pt}}
\newcolumntype{b}{>{\centering\arraybackslash}p{14pt}}
\newcommand{\flexwraptable}[5]{%
  \begin{wraptable}{#1}{#2}%
    \vspace{#3}%
    \centering
    {%
      #5%
      \vspace{#4}%
    }%
  \end{wraptable}%
}

\newcommand{\flexwrapfigure}[5]{%
  \begin{wrapfigure}{#1}{#2}%
    \vspace{#3}%
    \centering
    {%
      #5%
      \vspace{#4}%
    }%
  \end{wrapfigure}%
}

\definecolor{c0-title-bkg}{HTML}{ffffff}
\definecolor{c0-title-text}{HTML}{000000}
\definecolor{c0-item-bkg}{HTML}{ffffff}
\definecolor{c0-item-text}{HTML}{676767}

\definecolor{c1-title-bkg}{HTML}{d1e2dd}
\definecolor{c1-title-text}{HTML}{005953}
\definecolor{c1-item-bkg}{HTML}{e6efec}
\definecolor{c1-item-text}{HTML}{2d7b6d}

\definecolor{c2-title-bkg}{HTML}{cfe1e1}
\definecolor{c2-title-text}{HTML}{005760}
\definecolor{c2-item-bkg}{HTML}{e4eeed}
\definecolor{c2-item-text}{HTML}{24797b}

\definecolor{c3-title-bkg}{HTML}{cddfe5}
\definecolor{c3-title-text}{HTML}{00536b}
\definecolor{c3-item-bkg}{HTML}{e2ecef}
\definecolor{c3-item-text}{HTML}{287687}

\definecolor{c4-title-bkg}{HTML}{cedce8}
\definecolor{c4-title-text}{HTML}{124e74}
\definecolor{c4-item-bkg}{HTML}{e1eaf1}
\definecolor{c4-item-text}{HTML}{3a7190}

\definecolor{c5-title-bkg}{HTML}{d0d9eb}
\definecolor{c5-title-text}{HTML}{324779}
\definecolor{c5-item-bkg}{HTML}{e1e8f3}
\definecolor{c5-item-text}{HTML}{4d6b97}

\definecolor{c6-title-bkg}{HTML}{d3d5ed}
\definecolor{c6-title-text}{HTML}{493e7b}
\definecolor{c6-item-bkg}{HTML}{e3e5f5}
\definecolor{c6-item-text}{HTML}{61639b}

\definecolor{c7-title-bkg}{HTML}{dad1ed}
\definecolor{c7-title-text}{HTML}{5a3477}
\definecolor{c7-item-bkg}{HTML}{e5e1f5}
\definecolor{c7-item-text}{HTML}{725b99}

\definecolor{c8-title-bkg}{HTML}{ded1ec}
\definecolor{c8-title-text}{HTML}{633273}
\definecolor{c8-item-bkg}{HTML}{ebe2f6}
\definecolor{c8-item-text}{HTML}{7c5997}

\definecolor{c9-title-bkg}{HTML}{e5d1eb}
\definecolor{c9-title-text}{HTML}{6c2f6b}
\definecolor{c9-item-bkg}{HTML}{f0e0f6}
\definecolor{c9-item-text}{HTML}{885591}

\definecolor{c10-title-bkg}{HTML}{ebd1e7}
\definecolor{c10-title-text}{HTML}{722e5f}
\definecolor{c10-item-bkg}{HTML}{f5e2f3}
\definecolor{c10-item-text}{HTML}{915487}

\definecolor{avg-title-bkg}{HTML}{f3f3f3}
\definecolor{avg-title-text}{HTML}{000000}
\definecolor{avg-item-bkg}{HTML}{f3f3f3}
\definecolor{avg-item-text}{HTML}{000000}

\newcommand{\addpadding}{%
  \rule{0pt}{\dimexpr\normalbaselineskip-1pt\relax}%
}

\newcommand{\ct}[2][c0]{\addpadding{\cellcolor{#1-item-bkg}\textcolor{#1-title-text}{#2}}}

\definecolor{promptcolor}{HTML}{D1D0F2}
\definecolor{promptcolorheader}{HTML}{bdbcec}
\newcommand{\promptbox}[2]{
    \begin{tcolorbox}[
        top=0.3em,bottom=0.3em,left=0.5em,right=0.5em,
        toptitle=0.3em,bottomtitle=0.2em,boxsep=0pt,
        colframe=promptcolorheader,colback=promptcolor!50,boxrule=0.5pt,
        title={\footnotesize \fontfamily{zi4}\selectfont #1}
    ]
        \footnotesize
        {\fontfamily{zi4}\selectfont #2}
    \end{tcolorbox}
}

\definecolor{democolor}{HTML}{e6eeeb}
\definecolor{democolorheader}{HTML}{a0cdc9}

\ExplSyntaxOn

\cs_generate_variant:Nn \coffin_typeset:Nnnnn {Nffff}

\NewDocumentCommand\rotbox{ O{l,H} D<>{0pt,0pt} m m}{
    \hcoffin_set:Nn \l_tmpa_coffin {#4}
    \coffin_rotate:Nn \l_tmpa_coffin {#3}
    \coffin_typeset:Nffff \l_tmpa_coffin 
        {\clist_item:nn{#1}{1}}
        {\clist_item:nn{#1}{2}}
        {\clist_item:nn{#2}{1}}
        {\clist_item:nn{#2}{2}}
}
\ExplSyntaxOff

\DeclareRobustCommand{\PEav}{PE$_\text{AV}$}
\DeclareRobustCommand{\PEcore}{PE$_\text{core}$}
\DeclareRobustCommand{\PEcore}{PE-}
\DeclareRobustCommand{\PE}{PE}

\DeclareRobustCommand{\PEaframe}{PE$_\text{A-Frame}$}

\newlength{\ccustomlen}
\newcommand{\ccustom}[3][c0]{%
    \cellcolor{#1-item-bkg}{%
        \rotbox[l,t]{90}{%
            \parbox[t]{\ccustomlen}{%
                \ifthenelse{\isempty{#3}}{%
                    \mbox{%
                        \kindatiny\textcolor{#1-title-text}{#2}%
                    }%
                }{%
                    \kindatiny\textcolor{#1-title-text}{#2} \\%
                    \tiny{\textcolor{#1-item-text}{\it #3}}%
                }%
            }%
        }%
    }%
}

\newcommand{\cz}[3][c0]{%
    \setlength{\ccustomlen}{1.2cm}%
    \ccustom[#1]{#2}{#3}%
}

\newcommand{\ca}[1]{\cellcolor{avg-item-bkg}{#1}}

\newcommand{\bigO}[1]{$\mathcal{O}\!\left(#1\right)$}
\newcommand{\cmark}{\ding{51}}%
\newcommand{\xmark}{\ding{55}}%

\def\eg{\emph{e.g.}\xspace}
 
\def\vs{\emph{vs.}\xspace}

\makeatother

\usepackage{amssymb}
\usepackage{xifthen}
\usepackage{wrapfig}
\usepackage{makecell}
\usepackage[percent]{overpic}
\usepackage{multicol}
\usepackage{hyperref}

\usepackage{pgfplots}
\pgfplotsset{compat=1.18}
\usetikzlibrary{calc}

\usepackage[nolist]{acronym}

\begin{acronym}
\acro{ASR}{automatic speech recognition}
\acro{DNN}{deep neural network}
\acro{SED}{sound event detection}
\acro{FL-CLAP}{frame-level contrastive language-audio pre-training}
\acro{PSDS}{polyphonic sound detection score}
\acro{AUROC}{area under the receiver operating characteristic}
\acro{LID}{language identification}
\acro{SSL}{self-supervised learning}
\acro{PE}{Perception Encoder}
\end{acronym}

\title{Pushing the Frontier of Audiovisual Perception \\with Large-Scale Multimodal Correspondence Learning}

\author[*]{Apoorv Vyas}
\author[*]{Heng-Jui Chang}
\author[*]{Cheng-Fu Yang}
\author[*]{Po-Yao Huang}
\author[*]{Luya Gao}
\author[*]{Julius Richter}
\author[\dagger]{Sanyuan Chen}
\author[\dagger]{Matt Le}
\author[\ddagger]{Piotr Dollár}
\author[\ddagger]{Christoph Feichtenhofer}
\author[\ddagger]{Ann Lee}
\author[\ddagger]{Wei-Ning Hsu}

\affiliation[]{Meta Superintelligence Labs}

\contribution[*]{Core contributors (random order from second author onward)}
\contribution[\dagger]{Contributors (random order)}
\contribution[\ddagger]{Project leads (random order)}

\abstract{We introduce {Perception Encoder Audiovisual}, \PEav{}, a new family of encoders for audio and video understanding trained with scaled contrastive learning. 
Built on PE~\citep{pe}, \PEav{} makes several key contributions to extend representations to audio, and natively support joint embeddings across audio–video, audio–text, and video–text modalities. 
\PEav{}'s unified cross-modal embeddings enable novel tasks such as speech retrieval, and set a new state of the art across standard audio and video benchmarks. 
We unlock this by building a strong audiovisual data engine that synthesizes high-quality captions for $\mathcal{O}(100\text{M})$ audio–video pairs, enabling large-scale supervision consistent across modalities.
Our audio data includes speech, music, and general sound effects—avoiding single-domain limitations common in prior work. 
We exploit ten pairwise contrastive objectives, showing that scaling cross-modality and caption-type pairs strengthens alignment and improves zero-shot performance. 
We further develop \PEaframe{} by fine-tuning \PEav{} with frame-level contrastive objectives, enabling fine-grained audio-frame-to-text alignment for tasks such as sound event detection.}

\metadata[Code]{\\
\url{https://github.com/facebookresearch/perception_models}\\ \url{ https://huggingface.co/collections/facebook/perception-encoder-audio-visual}}

\begin{document}

\maketitle

\nocite{sam-audio}
\vspace{-4mm}
\section{Introduction} \label{sec:intro}
Vision, audio, and language are fundamental modalities in human perception. Audio often provides complementary or disambiguating cues for visually subtle or ambiguous actions (\eg, distinguishing ``speaking'' vs. ``whispering''). Audio can also carry distinct visual concepts, such as an ambulance siren warning us of an emergency vehicle before it comes into sight while driving. Perceptual and neuroscience studies further suggest that audio and visual signals are tightly coupled in the brain—as \eg illustrated by the McGurk effect~\cite{mcgurk1976hearing}, where listening to an audio clip (\eg, sounding ``ba-ba'') while watching fabricated lip movements (indicating ``va-va'') changes the perceived sound (from ``ba-ba'' to ``va-va'').

Multimodal correspondences can therefore be used to learn \textit{semantic} representations that align text, audio, and vision. Consider, \eg, that an ambulance vehicle representation could be learned from any combination of video of an ambulance, its distinctive sound, or a text  description.
Such representations can, in turn, be directly used for downstream applications such as classification and retrieval.

Contrastive models demonstrate strong performance in aligning individual modalities (\eg, images or audio) to text~\cite{clip,metaclip,laion_clap,speech_clap}.  Recent works like ImageBind~\cite{girdhar2023imagebind}, LanguageBind~\cite{langbind}, and InternVideo2~\cite{internvideo2} connect more modalities via an ``anchor'' modality. However, imbalanced scale and diversity across modality pairs continue to limit performance. While there has been large progress in vision-language learning with CLIP~\cite{clip} and its follow-up works, the audio-video domain remains underrepresented and lags behind in performance.   

\begin{figure}[t!]
    \centering
    \includegraphics[width=0.75\linewidth, trim=1mm 1mm 2mm 1mm, clip]{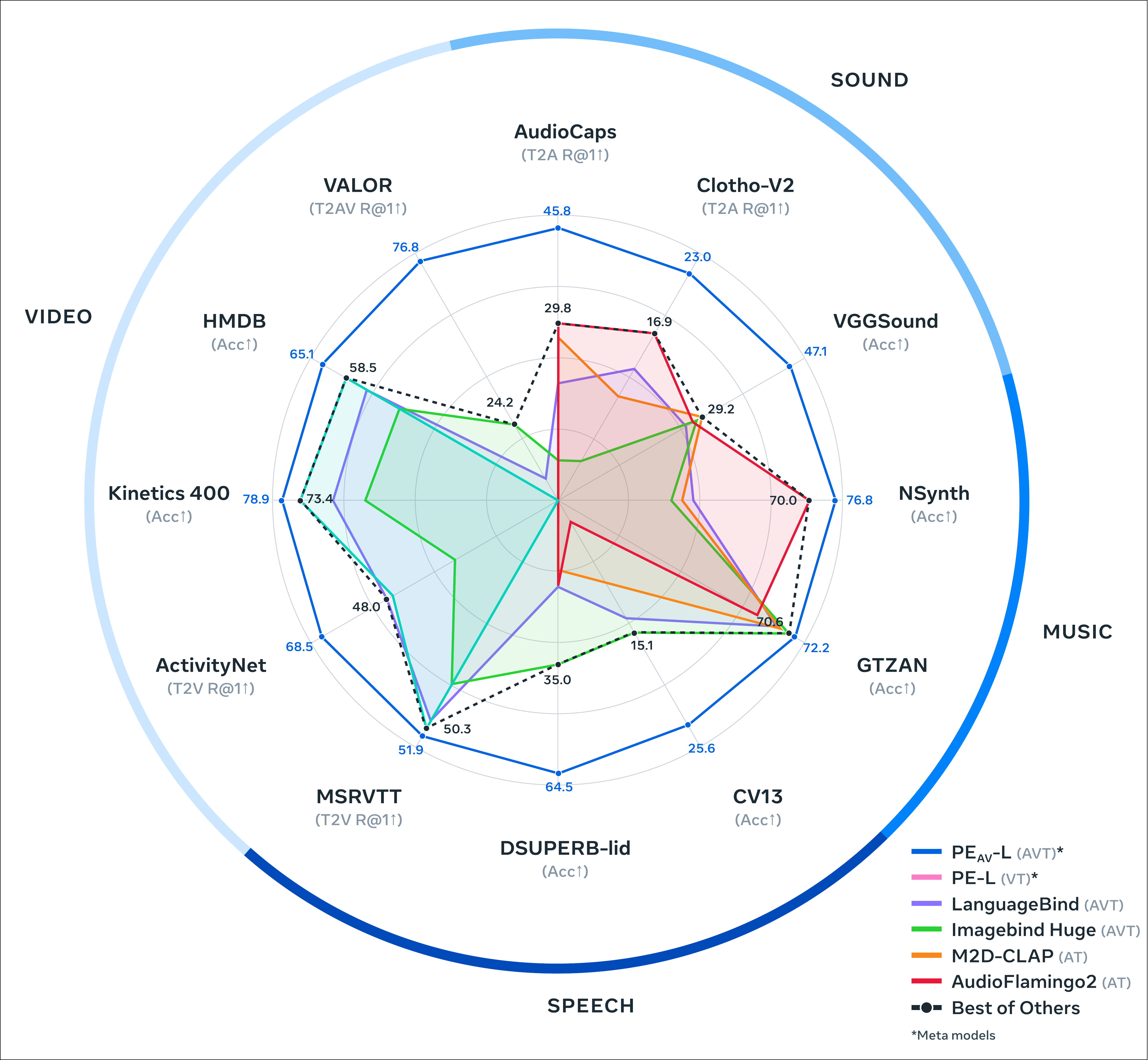}
    \vspace{-5pt}
    \caption{{\bf Perception Encoder Audiovisual (\PEav{})} is a family of audio-video-text (AVT) encoders. 
    By scaling coverage of contrastive learning, model size, and training with synthetically aligned video-audio-text pairs across diverse domains, \PEav{} achieves state-of-the-art performance on a wide range of zero-shot sound, music, speech, and video classification and retrieval tasks.
    \label{fig:teaser}}
    \vspace{-10pt}
\end{figure}

In this work, we build Perception Encoder Audiovisual (\PEav), a family of audio-visual-text aligned encoders trained with a simple contrastive objective across modalities.
Our approach is centered on the \textit{scale} and \textit{quality} of synthetic data, data types, loss pairs, and model size.

We build a robust audiovisual data engine that generates high‑quality synthetic captions at scale.
We start by using an LLM to combine information from multiple weak audio captioning models along with their confidence scores, and video captions to produce captions for audio, visual, and audio‑visual information in \textit{unlabeled} video clips. 
We train an initial version of \PEav~on these synthetic labels. 
This \PEav~is then used with an LLM decoder to generate refined audiovisual captions.
Our two‑stage data engine yields reliable captions for \(\mathcal{O}(100\text{M})\) audio–video pairs, dramatically
expanding beyond existing data while producing a corpus well balanced across modalities.

In the modeling part, we scale the contrastive objective to encompass up to \textit{ten} cross-modal pairs among video, audio, and diverse text captions. Expanding the coverage of modality pairs consistently enhances alignment and enables joint embeddings for audio–text, video–text, and audio–video.
For the model architecture, we employ separate audio and vision towers, followed by a joint audio-visual encoder. We use PE~\cite{pe} as a video frame encoder, followed by a stack of temporal transformer blocks to capture temporal dynamics, as an efficient vision architecture for encoding videos.

We train \PEav{}'s audio encoder at three scales (0.1-1.1 billion parameters) and see consistent gains in zero-shot retrieval and classification of video, sound, music, and speech. As shown in Fig.~\ref{fig:teaser}, \PEav{L} sets a new state-of-the-art (SoTA) performance on multiple audio–video benchmarks compared to recent audio-text~\cite{laion_clap,ms_clap,af2,m2d}, and audio-video-text models~\cite{girdhar2023imagebind,langbind}. 
Notably, on AudioCaps, text-to-audio improves from 35.4 R@1 to 45.8 R@1; and on VGGSound, classification accuracy improves from 36.0 to 47.1. \PEav{} is the only model to enable speech retrieval (85.6 while others are near 0). 
On video benchmarks, \PEav{L} improves ActivityNet text-to-video retrieval from 60.4 R@1 to 66.5 R@1, and video classification on Kinetics-400 from 76.9 to 78.9 accuracy, surpassing models $2$–$4\times$ larger~\cite{pe, internvideo2}.

In summary, \PEav{}'s main contributions are:
\begin{itemize}
  \item \textbf{A strong multimodal data engine.} 
  Our audiovisual data engine produces high-quality, diverse synthetic captions at scale that outperform real captions in synthetic \vs real comparisons. Their complementary nature further boosts accuracy when combined.
  \item \textbf{A broad learning paradigm.} Ten training objectives improve alignment across modalities and data types.
  \item \textbf{Unprecedented domain coverage.} 
  Our audio encoder supports speech, music, and general sound‑effects, unlike prior audio models that specialize in a single domain.
  \item \textbf{Unified cross-modal embeddings:} Beyond separate audio, video, and text embeddings, \PEav{} learns to jointly encode audio–video, audio–text, and video–text at scale, achieving SoTA zero-shot performance on video classification and retrieval, sound and music tasks, and speech benchmarks. We release our models and code to support reproducibility and future research and applications.
\end{itemize}

Finally, we introduce \emph{Perception Encoder Audio-Frame} (\PEaframe), which fine-tunes \PEav{} using a frame-level contrastive loss. Moving beyond utterance-level, \PEav{} enables fine-grained frame-level audio-to-text alignments.
\PEaframe{} performs strongly across open- and closed-vocabulary \ac{SED}, achieving top results in identifying target-sound events across benchmarks. It yields the best performance on all open-vocabulary tests and real-world benchmark such as the closed-set DESED~\cite{turpault2019sound}.

\flexinput[0.6\linewidth]{fig/architecture_arxiv}

\section{Perception Encoder for Audio and Video} \label{sec:core}

As shown in Fig.~\ref{fig:model_overview},  \PEav{} consists of a text encoder,
a video-frame encoder, a video encoder, an audio feature extractor, an audio
encoder, and an audio-video fusion encoder.  \PEav{} is trained using
contrastive objectives on videos with synthetic captions \bigO{100M} and a few
real datasets \bigO{5M}. The synthetic audio, visual, and audio-visual captions
are powered by an audiovisual data engine described next.

\subsection{\PEav{} Data Engine}\label{sec:aed}
For learning audio-video-text representations, conventional single-anchor
models struggle when the anchoring modality is absent. For example,
text-anchored LanguageBind underperforms in audio–video tasks, while
image-anchored ImageBind\cite{girdhar2023imagebind} performs poorly in
audio–text tasks (see Tab.~\ref{tab:exp:core:sound_results}). These limitations
stem from asymmetry due to mismatched cross-modal data scales and the inherent
brittleness of binding all modalities to a single hub.

To address this, we develop an audiovisual data engine that \emph{generates
missing audio, video, and audio-visual captions} at scale and strengthens
alignments between \emph{all} modality pairs.  Current visual captioners can
accurately describe details and enable efficient data scaling while audio
captioners remain weak. Therefore, we leverage multiple weak audio captions
along with video captions, and apply LLM-rewriting to improve coverage and
quality of captions.

This richer supervision enables scaling the contrastive objective to cover
\emph{more} cross-modal pairs; empirically, increasing the number of pairs
consistently improves cross-modal alignment
(Tab.~\ref{tab:ablation:clip_siglip}).  Practically, we designed a two-stage
data engine described below. 

\paragraph{Stage-1: Synthetic Captioning Pipeline.}\label{sec:stage1}
As shown in Fig.~\ref{fig:data_engine_synthetic_captions} stage-1 synthetic
captioning pipeline uses Llama 3.1 8B~\cite{llama3} to combine outputs from
two weak audio captioning models, EnCLAP~\cite{enclap} and
CoNeTTE~\cite{conette}, along with the corresponding confidence scores from
Joint-CLAP\cite{vyas2023audiobox} and video captions.  We use the pipeline to
generate captions for $O(100\mathrm{M})$ videos chunked at 30-second intervals.

\flexinput[0.48\linewidth]{fig/stage1_arxiv}

\begin{figure*}[ht!]
  \centering
  \resizebox{\textwidth}{!}{%
  \begin{minipage}{\textwidth}
  \begin{subfigure}[b]{0.95\textwidth}
    \centering
    \includegraphics[width=0.8\textwidth]{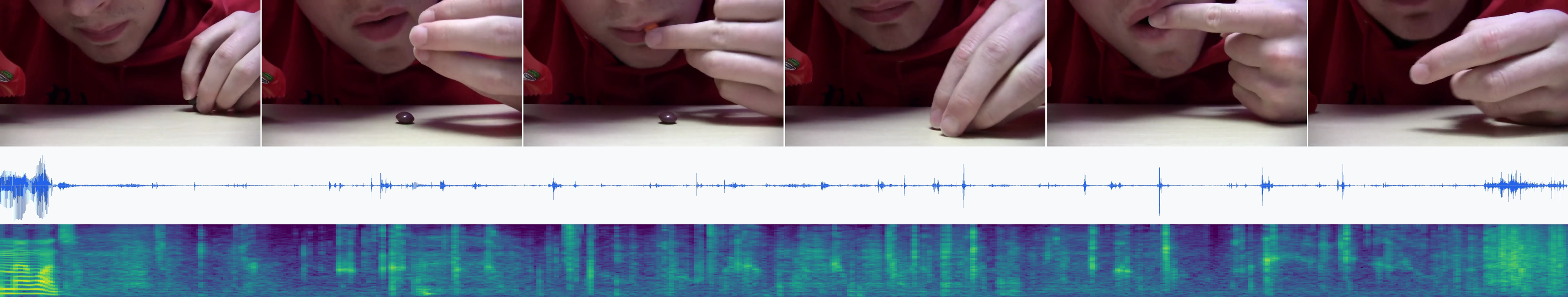}
    {\footnotesize
    \begin{tabular}{@{}p{0.07\textwidth}p{0.92\textwidth}@{}}
      {EnCLAP:} & a man speaks followed by some crinkling and crumpling (score: 0.48) \\
      {CoNeTTE:} & a man is speaking and chewing on something (score: 0.57) \\
      {Video:} & The video shows a close-up shot of a man eating a small, round candy. In the background, he sits at a white table. The man has fair skin, brown hair and a red shirt. He sits at the table with his hands resting on the edge. He holds an orange candy in his hand to the right. The candy has a blue wrapper with white writing on it. The man puts the candy in his mouth. He chews the candy and swallows it. He then picks up a brown candy from the table. He puts the brown candy in his mouth. He chews the candy and swallows it. He then picks up another orange candy from the table. He puts the orange candy in his mouth. He chews the candy and swallows it. He then picks up another brown candy from the table. He puts the brown candy in his mouth. He chews the candy and swallows it. He then picks up another orange candy from the table. He puts the orange candy in his mouth. He chews the candy and swallows it. The video ends with the man chewing the candy. \\
    \end{tabular}
    }
  \end{subfigure}
  \hfill
  \begin{subfigure}[b]{0.95\textwidth}
    \centering
    \includegraphics[width=0.8\textwidth]{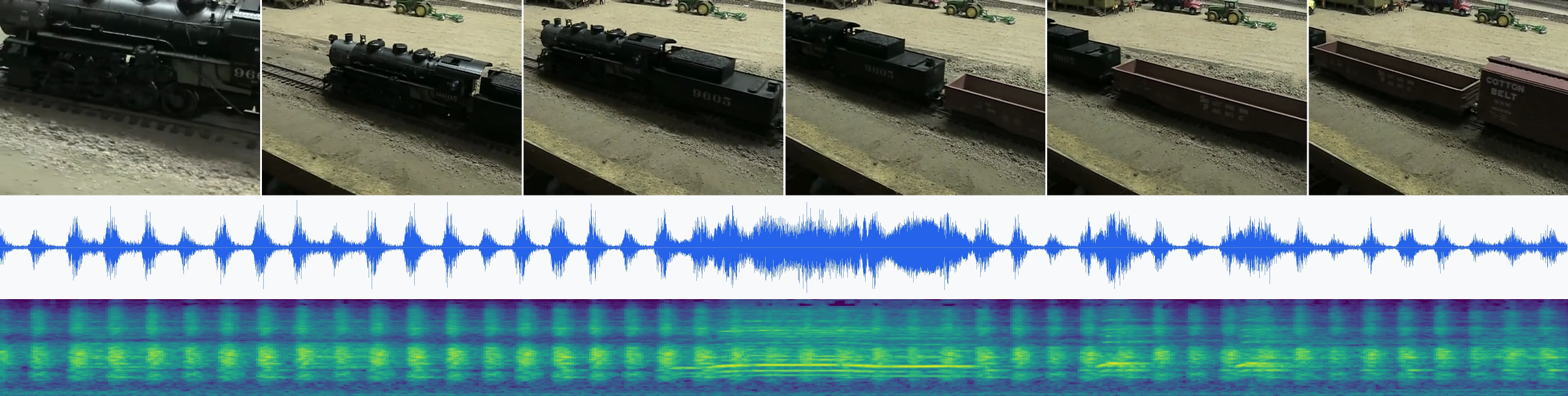}
    {\footnotesize
    \begin{tabular}{@{}p{0.07\textwidth}p{0.92\textwidth}@{}}
      {EnCLAP:} & a train is chugging along and a steam whistle goes off (score: 0.50) \\
      {CoNeTTE:} & a person is using a wrench on a typewriter (score: -0.08) \\
      {Video:} & The video shows a train traveling on a track. The train has a large, black engine with a black caboose behind it. The caboose has white lettering on the side that reads "COTTON BELT" and "9605." The train is traveling on a track that runs along the bottom of the frame. The background shows a sandy area with several pieces of farm equipment. The video begins with the train traveling toward the left side of the frame. As the video progresses, the train continues to travel along the track. The video ends with the train moving toward the left side of the frame. \\
      \\
    \end{tabular}
    }
  \end{subfigure}
  \hfill
  \begin{subfigure}[b]{0.95\textwidth}
    \centering
    \includegraphics[width=0.8\textwidth]{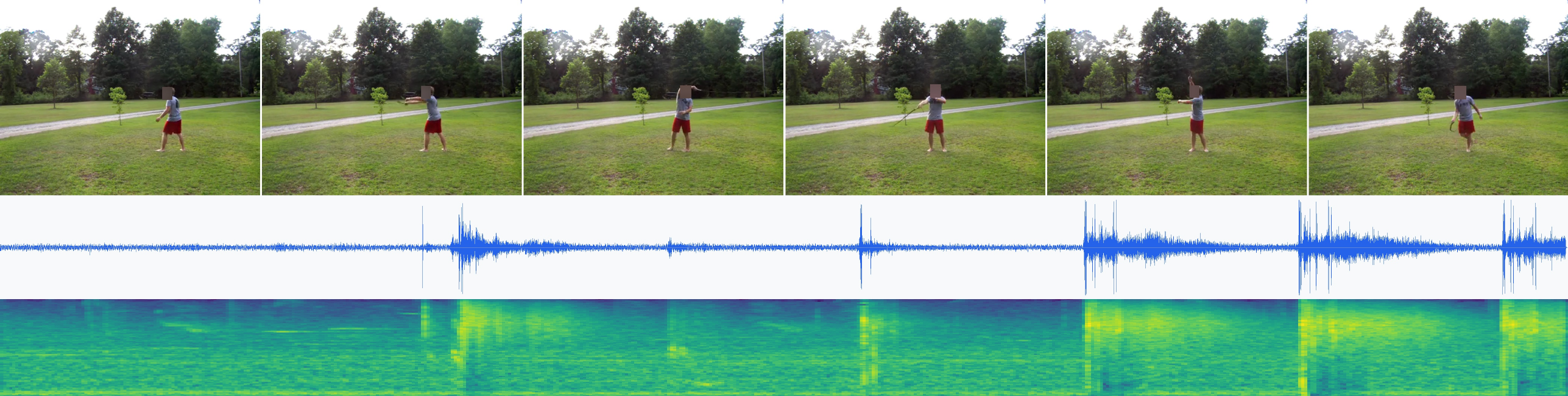}
    {\footnotesize
    \begin{tabular}{@{}p{0.07\textwidth}p{0.92\textwidth}@{}}
      {EnCLAP:} & a toy helicopter is flying around and making a lot of noise (score: -0.02) \\
      {CoNeTTE:} & a whip is being struck in the distance while birds are chirping in the background (score: 0.43) \\
      {Video:} & The video shows a young man playing with a whip in a grassy field. In the background, there is a gravel path and a line of trees with a house behind them. The man has fair skin and short brown hair. He is wearing a gray t-shirt and red shorts. The whip is long and thin. At the beginning, the man faces the right side of the video while holding the whip in his hand on the left side of the video. He swings the whip over his head and releases it. It whips around the man's body, then moves back toward the camera. The man moves toward the camera as the whip approaches. He catches the whip with his hand on the right side of the video and swings it over his head. The whip whips around the man's body, then moves back toward the camera. The man catches the whip with his hand on the left side of the video. The video is shot by a handheld camera. \\
    \end{tabular}
    }
  \end{subfigure}
  \end{minipage}
  }
  \caption{EnCLAP and CoNeTTE captions often provide complementary information and the confidence scores reflect the accuracy reasonably, making them favorable to combine with an LLM. Video captions provide strong context. Together this provides strong audio and visual cues for LLM rewriting.
  }
  \label{fig:video_context_combined}

\end{figure*}

Fig.~\ref{fig:video_context_combined} shows the captions for EnCLAP, CoNeTTE,
and our internal video captioning model. 
We find that (1) EnCLAP and CoNeTTE tend to make different errors,
which are reflected in the confidence scores of a CLAP-based model, and (2)
video captions  can provide useful text context, \eg TV-show, to help
disambiguate audio events and improve caption quality.
With these observations, we prompt an LLM to combine information from audio and
video captions along with discretized confidence scores (low/medium/high) to
rewrite audio, visual, and audio–visual caption.
Summarizing the visual captions also reduces some LLM decoding errors present
in raw video captions. In a small-scale blind subjective evaluation on $\sim$50
utterances, LLM audio captions are strictly better than EnCLAP on
\textbf{65.2\%} of utterances, similar on \textbf{28.3\%}, and worse only on
\textbf{6.5\%}.

\begin{figure}[h]
  \centering
  \begin{subfigure}[b]{0.48\linewidth}
    \centering
    \includegraphics[width=\linewidth]{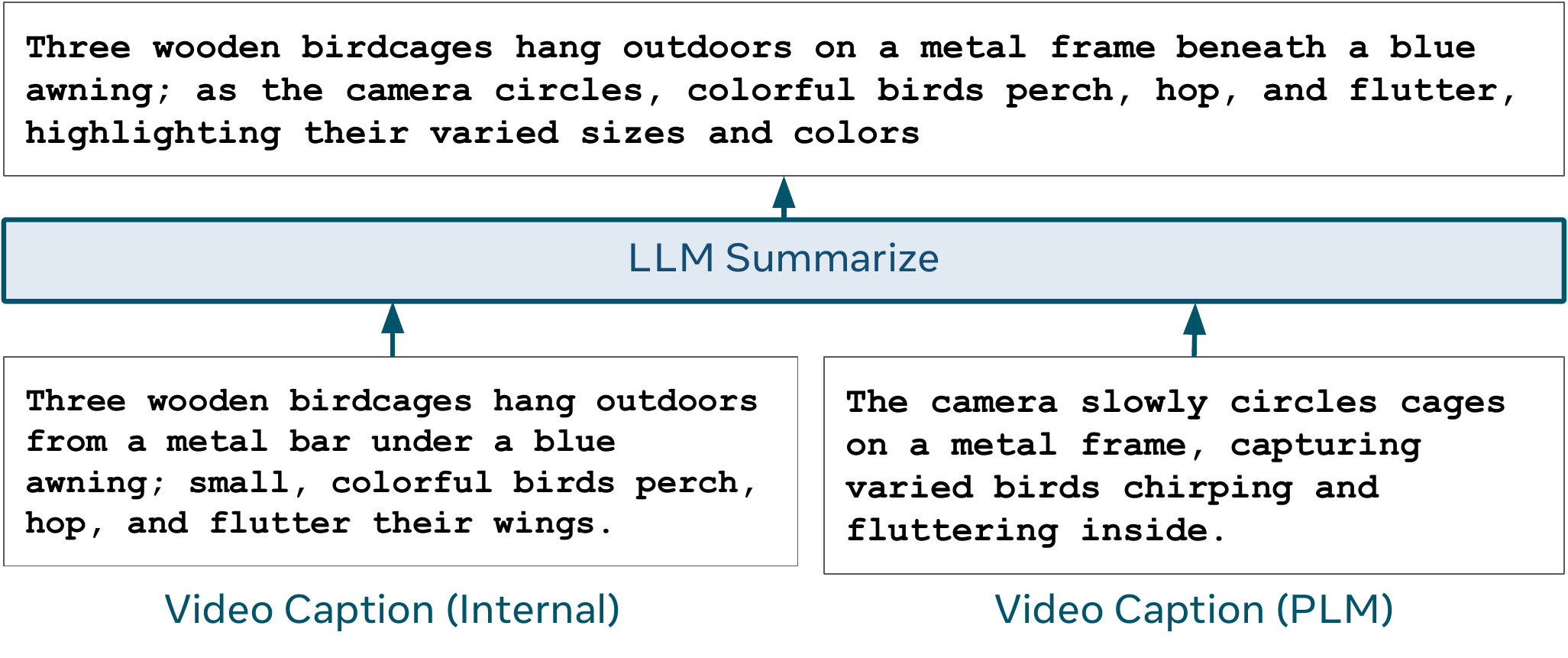}
    \caption{Improved Video Captions}
    \label{fig:data_engine_synthetic_captions_r2}
  \end{subfigure}\hfill
  \begin{subfigure}[b]{0.48\linewidth}
    \centering
    \includegraphics[width=\linewidth]{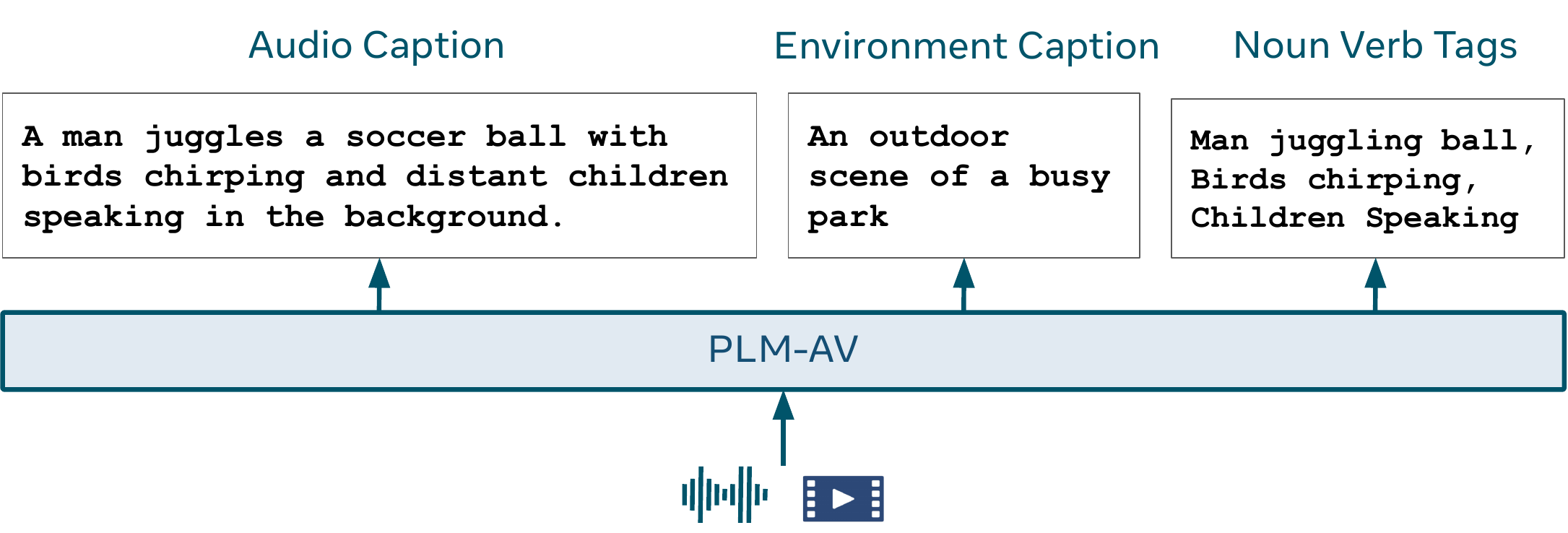}    
    \caption{Improved Audio Captions}
    \label{fig:data_engine_plm_audio}
  \end{subfigure}
\caption{{\bf Data Engine: Stage-2 Improved Captions:} We improve both the audio and visual captions in the second stage. For the visual captions, we use PLM~\cite{plm} to generate the video captions and an LLM to summarize the first stage and PLM captions. For audio captions, we follow the PLM recipe~\cite{plm} to train a PLM-AV model to generate three different audio caption variants focused on audio events, caption, and acoustic environment.}
  \label{fig:data_engine}
  \vspace{-6pt}
\end{figure}

\paragraph{Stage-2: Improved Audio and Visual Captions.}
Fig.~\ref{fig:data_engine} shows the second stage pipeline to fuse and improve
the audio and visual captions from the first stage.  To improve video captions,
we utilize the video data engine- a PLM-based model~\cite{plm} used in
PE~\cite{pe} that focused on fine-grained spatial-temporal visual events.  This
model processes video metadata along with 32 sampled frames to produce
fine-grained video captions.  Next, we employ Llama 3.1 (8B)~\cite{llama3} to
summarize both the stage-1 output and the fine-grained PLM captions, resulting
in the final improved video captions.  Similarly, for audio captions, we follow
the PLM recipe~\cite{plm} to train an early version of PLM-AV,  a multimodal
LLM with a \PEav{} trained on stage-1 synthetic captions as the audio-visual
encoder and Llama \cite{llama3} as the text decoder. We focus on improving
audio understanding with additional visual context.  

\begin{figure}[h]
  \centering
    \begin{subfigure}[b]{0.95\textwidth}
    \centering
\includegraphics[width=0.9\textwidth]{fig/data_engine_assl/sMPHFVRrJVc_500000_510000.pdf}
    \caption*{}
    \vspace{-1em}
    {\footnotesize
    \setlength{\tabcolsep}{5pt}%
    \renewcommand{\arraystretch}{1.0}%
    \begin{tabularx}{\linewidth}{@{}l X@{}}
      \toprule
      \textbf{Type} & \textbf{Caption} \\
      \midrule
      Audio &  A man is speaking and eating candies, accompanied by the sounds of crinkling and crumpling wrappers. \\
      Video & A man with fair skin, brown hair, and a red shirt eats various candies at a white table, alternating between orange and brown candies with blue and white wrappers. \\
      Audio--Visual & The video shows a man eating various candies at a table, with sounds of him speaking and chewing. He picks up and consumes multiple orange and brown candies with blue and white wrappers, accompanied by some sounds of crinkling and crumpling. \\
      \bottomrule
    \end{tabularx}}
  \end{subfigure}
  \hfill
  \begin{subfigure}[b]{0.95\linewidth}
    \centering
    \includegraphics[width=0.9\linewidth]{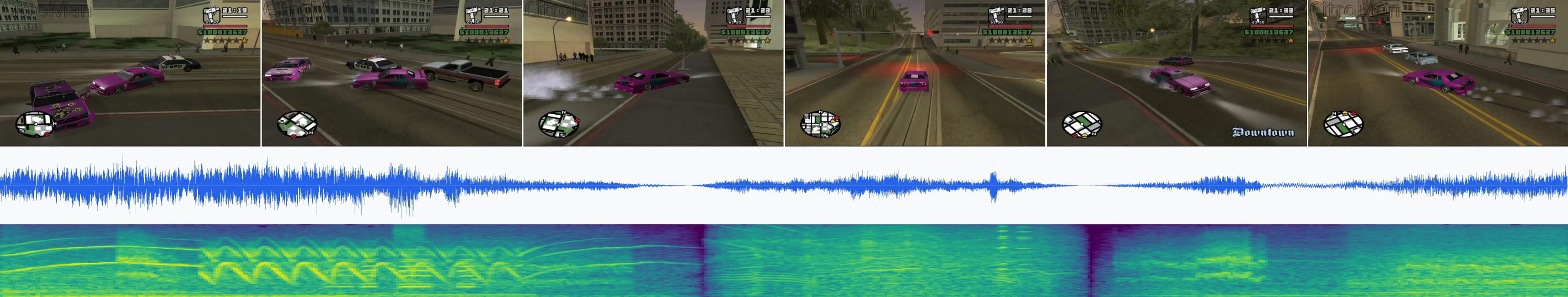}
    \caption*{}
    {\footnotesize
    \setlength{\tabcolsep}{5pt}%
    \renewcommand{\arraystretch}{1.0}%
    \begin{tabularx}{\linewidth}{@{}l X@{}}
      \toprule
      \textbf{Type} & \textbf{Caption} \\
      \midrule
      Audio & A police vehicle is moving, triggering a siren, as cars drive by on the city streets. \\
      Video & A purple car with blue and black design drives through the city, passing a black and white police car with flashing red and blue lights, a black and red car, and a white and black car, with the city's buildings, road, and pedestrians in the background. \\
      Audio--Visual & The video shows a purple car racing through the streets of a city, passing by various other cars, including a police car with flashing red and blue lights. The audio captures the sound of a police siren being triggered and a vehicle moving. \\
      \bottomrule
    \end{tabularx}}
  \end{subfigure}
  \caption{Examples of Audio, Video, and Audio--Visual captions generated using the data engine.}
  \label{fig:final_examples}

\end{figure}

Furthermore, we include speech-related
attributes–transcript, language ID~(LID), and accent–to enhance \PEav{}'s
capability in speech processing.  
We use Whisper Large-v3 and Medium ASR models~\cite{whisper} and keep only English transcripts where the two agree (low word error rate).
Similarly, we create LID labels with MMS LID models with 126 and 256
languages~\cite{mms}, and keep the labels where the two models share the top-1
prediction.  
For accent, we train an English accent classifier on Common Voice 13~\cite{commonvoice} and apply it to clips with English LID.
During training, we randomly inject transcript, LID, and accent into the audio
caption, and assign transcripts to a subset of data, always replacing the
original caption; this improves transcript retrieval without degrading other
tasks.
Fig.~\ref{fig:final_examples} presents examples of the raw and two-stage
processed captions. 

\subsection{\PEav{} Model and Training}

\paragraph{Architecture.}
\PEav{} builds on pre-trained feature extractors—Perception Encoder~\cite{pe}
for video frames, DAC-VAE~\cite{moviegen} for audio, and
ModernBERT~\cite{modernbert} for text—to obtain video, audio, and text tokens,
which are then encoded for contrastive learning.  For video encoding, we use
\PEcore{L} as the frame encoder to extract embeddings at 30 frames-per-second
(FPS).  For audio encoding, we extract audio tokens using DAC-VAE at 25~Hz.
For text encoding, we leverage ModernBERT with a context length of 512 and use
the 22\textsuperscript{nd} layer, which we found to adapt better to audio and
speech tasks than the original PE-L text encoder.

More formally, let $x^a$, $x^v$, $x^t$ denote raw audio, video, and text inputs, respectively. We first extract the sequential features as follows:
\begin{align*}
\mathbf{x}^{a} &= \text{DAC-VAE}(x^a) \in \mathbb{R}^{L_a \times C_a} \\
\mathbf{x}^{v} &= \text{PE-L}(x^v) \in \mathbb{R}^{L_v \times C_v}\\
\mathbf{x}^{t} &= \text{ModernBERT}(x^t) \in \mathbb{R}^{L_t \times C_t}
\end{align*}
where $C_a = 128$, $C_v = 1024$, and $C_t = 1024$ are the feature dimensions of DAC-VAE, \PEcore{L}, and ModernBERT, respectively, and $L_a$, $L_v$, and $L_t$ are the corresponding sequence lengths.
For a 30-second clip with 25\,Hz audio features and 30\,FPS video frames, we have $L_a = 750$ and $L_v = 900$.

For text, we use the \text{[CLS]} token from ModernBERT. For audio encoding, we first concatenate a learnable \text{[CLS]} token to the projected DAC-VAE tokens. We input this sequence into a $N$-layer audio Transformer $\text{T}_{a}(.)$ with rotary positional embeddings (RoPE)~\cite{rope}. The hidden dimension of the audio Transformer is set to $64N$.
To further capture temporal context in videos, we concatenate a \text{[CLS]} token to the projected frame embeddings, and use a shallow Transformer $\text{T}_{v}(.)$ as the video encoder.
The video Transformer shares the same configurations as the audio Transformer except for depth.
For brevity, we do not formalize the increasing sequence length of one by the \text{[CLS]} tokens.
The encoded audio and video are denoted as:
\begin{align*}
\mathbf{e}^a &= \text{T}_{a}(\text{Proj}(\mathbf{x}^a)) \in \mathbb{R}^{L_a \times C_e} \\
\mathbf{e}^v &= \text{T}_{v}(\text{Proj}(\mathbf{x}^v)) \in \mathbb{R}^{L_v \times C_e}
\end{align*}
where $C_e$ is the dimension of the transformer outputs.

Subsequently, we temporally align the video to the audio tokens using nearest-neighbor interpolation:
\begin{align*}
\tilde{\mathbf{e}}^v &= \text{NearestNeighbor}(\mathbf{e}^a, \mathbf{e}^v) \in \mathbb{R}^{L_a \times C_e} \\
\tilde{\mathbf{e}}^{av} &= \text{Proj}(\text{ChannelConcat}(\mathbf{e}^a, \tilde{\mathbf{e}}^v)) \in \mathbb{R}^{L_a \times C_e}
\end{align*}
where $\text{ChannelConcat}(\cdot,\cdot)$ concatenates along the channel dimension, and $\text{Proj}(\cdot)$ produces a audiovisual representation.

We then prepend a learnable \text{[CLS]} token and process the sequence with a shallow Transformer $\text{T}_{av}(.)$ that models joint audiovisual context and produces the fused audiovisual feature for contrastive learning:
\begin{align*}
\mathbf{e}^{av} = \text{T}_{av}(\tilde{\mathbf{e}}^{av}) \in \mathbb{R}^{L_a \times C_e}
\end{align*}

To compute the contrastive loss, we extract the [CLS] outputs from each
encoder—text, audio, video, and audio–video. For captions, we use three
separate text projection heads for audio, video, and audio–visual captions,
yielding $\mathbf{h}^{t_a},\mathbf{h}^{t_v},\mathbf{h}^{t_{av}}$, while the
audio, video, and audio–video encoders are projected to $\mathbf{h}^a,
\mathbf{h}^v, \mathbf{h}^{av}$ in the same shared embedding space. Here, all
$\mathbf{h}$ vectors are in $\mathbb{R}^{C_h}$ where $C_h$ is the dimension of
the shared embedding space set to $1024$ for all model variants.

In addition, to enable text-conditioned cross-modal retrieval
(V+T$\rightarrow$A, A+T$\rightarrow$V)—where the text supplies cues or intent
that may be missing in the raw query—we compose the query modality with text to
form jointly encoded features. Specifically, given class embeddings
$\mathbf{c}^{a}$, $\mathbf{c}^{v}$, and $\mathbf{c}^{t}$ for audio, video, and
text, we perform channel concatenation with text and project to the shared
space to obtain
\begin{align*}
\mathbf{h}^{vt} &= \text{Proj}\big(\text{ChannelConcat}(\mathbf{c}^{v}, \mathbf{c}^{t})\big) \in \mathbb{R}^{C_h}, \\
\mathbf{h}^{at} &= \text{Proj}\big(\text{ChannelConcat}(\mathbf{c}^{a}, \mathbf{c}^{t})\big) \in \mathbb{R}^{C_h}.
\end{align*}
We include these joint features in stage-2 fine-tuning (Tab.~\ref{tab:training_data:stats}) and contrast them with $\mathbf{h}^{a}$ and $\mathbf{h}^{v}$.
This enables text-conditioned retrieval such as V+T$\rightarrow$A (e.g., retrieving the correct music for a video using text describing the mood) and A+T$\rightarrow$V.

\paragraph{Training Objectives.}
For any modality pair, we pre-train using a sigmoid contrastive objective,
similar to~\cite{siglip}, to align the corresponding embeddings.
Formally, let $\mathbf{h}_b^{a}$ and $\mathbf{h}_b^{v}$ denote the embeddings
for the $b$-th sample for the modalities \emph{a} and \emph{v}, respectively,
and let $B$ denote the batch size. We then calculate the following sigmoid contrastive loss
for the alignment of the modalities:
\begin{equation}
\mathcal{L}\big(\mathbf{h}^a, \mathbf{h}^v\big)
= -\frac{1}{B}\sum_{b=1}^{B}\sum_{b'=1}^{B} \log \sigma\Big(z_{bb'}\big(-\alpha_{av}\,\mathbf{h}_b^a \cdot \mathbf{h}_{b'}^v + \beta_{av}\big)\Big),
\end{equation}
where ${\alpha}_{av}$ and ${\beta}_{av}$ denote the temperature and the bias
for the pair of modalities $a$ and $v$. $z_{bb'}$ is the indicator function;
$z_{bb'}=1$ when $b=b'$ and $z_{bb'}=-1$ elsewhere.

We compute the contrastive loss across the following pairs:
(1) Audio to audio caption;
(2) Audio to video;
(3) Audio to audio-video caption;
(4) Audio-video to audio caption;
(5) Audio-video to audio-video caption;
(6) Video to audio caption;
(7) Video to video caption;
(8) Video to audio-video caption.
As to be shown in Tab.~\ref{tab:ablation:clip_siglip}, we observe that maximum
performance is achieved when considering all the pairs of modality and caption
types.

In the fine-tuning stage, we additionally include two joint embeddings:
(9) Audio with video-caption to video; and
(10) Video with audio-caption to audio,
leading to ten pairs in total.
These additional pairs provide finer control over using only text, video, or
audio for retrieval and enable applications such as retrieving the correct
music for a video using text describing the mood.

\begin{wraptable}{r}{0.42\textwidth}
\vspace{-10pt}
\centering
{
\tablestyle{2pt}{1.3} 
\scriptsize
    \begin{tabular}{@{}lrrrr@{}}
    \toprule
    & \multicolumn{2}{c}{\textbf{Pre-training}} & \multicolumn{2}{c}{\textbf{Fine-tuning}} \\
    \cmidrule(lr){2-3}\cmidrule(lr){4-5}
    \textbf{Caption Type} & \textbf{Synthetic} & \textbf{Real} & \textbf{Synthetic} & \textbf{Real} \\
    \midrule
    a) Audio caption (speech)         &  2.0M & 1.5M & 2.0M &  5.5M \\
    b) Audio caption (sound-effects)  &  88.3M & 2.3M  &  13.9M & 3.0M \\
    c) Audio caption (music)          &  1.5M &  1.5M & 1.5M  & 4.3M \\
    d) Video caption                  &   2.9M &   -  & 1.5M & 8.8M \\
    e) Audio--Visual caption          &  87.8M & -  &  13.1M & 1.5M \\
    \bottomrule
  \end{tabular}
 
  \caption{{\bf Pre-training and fine-tuning data statistics}. In total, \PEav{} uses 92M unique audios/videos for stage-1 pre-training and additional 32M unique audios/videos for stage-2 fine-tuning.}
  \label{tab:training_data:stats}
  \vspace{-15pt}
}
\end{wraptable}

\paragraph{Training Data.}
We employ a 2-stage training recipe.  In the stage-1 pre-training, we focus on
scaling the training data with a diverse collection of audio and video samples
at scale.  We exploit an audiovisual data engine developed in \S\ref{sec:aed}
to generate different types of refined synthetic captions (audio, video, and
audiovisual) for audio and video data without annotations. 
In addition, we incorporate real captions from public datasets for training. 
In total, we use 92M unique audio and video samples for stage-1.

In the stage-2 fine-tuning, we utilize a smaller training set well-balanced
across domains and modalities. We adjust the data distribution with an emphasis
on 1) speech data and the corresponding transcripts, and 2) video data.  For
the speech data, we include annotated English speech corpora to improve
\PEav{}'s capability in speech transcript retrieval. We observe that a short
fine-tuning schedule is sufficient to achieve effective speech-to-transcript
association.  For the video data, inspired by~\cite{dino2, clip, metaclip, pe},
we include a subset of video-text data by curating videos that contain
important visual concepts from the stage-1 videos as well as from~\cite{pe,
plm}. We up-sample these videos for fine-tuning.  Overall, we use 
32M unique audio/video samples for fine-tuning.
Tab.~\ref{tab:training_data:stats} summarizes the composition of the training
data.  In addition, we fine-tune \PEav{} with contrastive objective at
audio-frame-level to enable fine-grained sound–to–text alignment, dubbed as
\PEaframe{}, for speech event detection (SED) 
described next.

\section{\PEaframe{}: Audio-Frame Level Language Alignment}
\label{sec:pe_a_frame}

Typical language–audio models produce a single utterance-level embedding (a global token) for each modality.
Then, they apply a contrastive loss to the global class tokens, which achieves coarse-grained cross-modal alignment but overlooks fine-grained interactions.
As a result, the correspondence between audio at the frame-level and language remains underexplored, leading to low performance on tasks requiring detailed temporal alignment. 
To address this limitation, we propose \emph{Perception Encoder Audio-Frame} (\PEaframe{}), a model fine-tuned from \PEav{} with a frame-level audio to language contrastive loss.

\paragraph{Training.} We train \PEaframe{} to predict the specific frames within an audio signal $x^a$ that contain the sound described by the free-form text description $x^t$.
Building on the pre-trained \PEav{} model, we fine-tune frame-level audio and instance-level text encoders 
by adopting a frame-level variant of the sigmoid contrastive loss~\cite{siglip}.
For each frame $l$, we compute the logit between the frame-level audio embedding $\mathbf{e}^a_l$ and the global text embedding $\mathbf{h}^{t_a}$ as
$\tilde{h}_{l} = \mathbf{e}^a_l \cdot \mathbf{h}^{t_a}$.

\paragraph{Input Data and Ground-Truth Label.}
Each element in a batch of size $B$ consists of an audio sample $x_b^a$ and a
single sound event described by its associated text description $x_b^t$.
Although an audio clip may contain multiple sound events, we sample only one
text description per element in each batch to simplify implementation.
Nevertheless, we provide all annotated sound events and their corresponding
frame-level activity masks for every audio sample. Accordingly, each batch
element contains
\begin{itemize}
    \item an audio clip $x_b^a$,
    \item a sampled text query $x_b^t$ describing one sound event in $x_b^a$, and
    \item an event-activity mask $m_b$ encoding all annotated events for $x_b^a$.
\end{itemize}
For the $b$-th audio with $K_b$ annotated events $\{x^t_{b,1}, \dots,
x^t_{b,K_b}\}$, we define $m_b \in \{0,1\}^{L_a \times K_b}$, where $m_{b,l,k}
= 1$ indicates that event $x^t_{b,k}$ is active at frame $l$.
Even though only one text query per audio is used for contrastive learning, we
leverage all annotated events and their \emph{ontology-aware}
\footnote{By “ontology” we mean a hierarchical taxonomy of sound event labels, such as the
AudioSet ontology.} 
semantic expansions to construct frame-level supervision.

Let $\text{Ont}(x^t)$ be the set of ontology-linked variants of event $x^t$ (e.g., “speech” includes “female speech”, and “dog” includes “barking”).
Then, for each audio frame $l$ in $x^a_b$ and each text query $x^t_{b'}$ in the batch, we assign:
\begin{equation}
\label{eq:ontology_label}
z_{b,l,b'} =
\begin{cases}
+1, &
\begin{aligned}[t]
&\exists\, k \in \{1,\dots,K_b\}:\ x^t_{b'} \in \text{Ont}(x^t_{b,k}) \quad \text{and}\ m_{b,l,k}=1,
\end{aligned}\\[4pt]
-1, & \text{otherwise.}
\end{cases}
\end{equation}
Thus, semantically equivalent sound expressions activate the same frames in supervision, improving robustness to linguistic variation and reinforcing hierarchical concept generalization.

\paragraph{Frame-Level Objective.}
For this task, the model must learn not only \emph{which} sound events are present in an audio clip but also \emph{when} they occur over time. 
To capture both aspects, we employ two complementary objectives: a \emph{local-activity loss} (computed per batch item) that emphasizes fine-grained temporal localization, identifying \emph{when} events occur within each audio sample, and a \emph{global-activity loss} (computed across the batch) that introduces contrastive context between samples to determine \emph{which} events are active.
The local-activity loss helps the model detect event boundaries, 
while the global-activity loss promotes global event understanding and cross-sample alignment.

The resulting local-activity loss (per-batch-item) yields logits of shape $(B,L)$, while the global-activity loss (across-batch) yields $(B,L,B)$:

\begin{equation}
\tilde h_{b,l(,b')} =
\left\{
\begin{array}{@{}l@{\;}l@{}}
\mathbf{e}^a_l(x_b)\cdot \mathbf{h}^{t_a}({x^t_b}) &
\text{ (local-activity)} \\[4pt]
\mathbf{e}^a_l(x_b)\cdot \mathbf{h}^{t_a}({x^t_{b'}}) &
\begin{aligned}[t]
\forall, b' \in {1,\dots,B}\quad \text{(global-activity)},
\end{aligned}
\end{array}
\right.
\end{equation}

A learnable logit scale $\alpha$ and bias $\beta$ are applied to obtain the final scaled logits $h = \alpha \tilde{h} + \beta$.
The frame-level SigLIP loss is computed over these logits.
For the local-activity case, the loss is defined as
\begin{equation}
\mathcal{L}
= - \frac{1}{B{L}} \sum_{b,l}
\log \sigma\big({z_{b,l}}  (\alpha \tilde{h}_{b,l} + \beta)\big),
\end{equation}
where ${z_{b,l}} \in {\pm1}$ is the binary label indicating whether frame $l$ of audio $x_b^a$
corresponds to the paired text ${x^t_b}$. This formulation naturally generalizes to the global-activity case by computing the loss over
$\tilde{h}_{b,l,b'}$ and averaging across all text queries ${x^t_{b'}}$ in the batch.
During training, we probabilistically sample between the two objectives at each iteration, with the probability $p_{\text{local}}$ for the local-activity loss.
This stochastic choice allows the model to balance precise event-boundary detection and global event activity modeling.

\begin{wraptable}{r}{0.42\textwidth}
\vspace{-10pt}
\centering
{
\tablestyle{8pt}{1.0} 
  \centering
  \tablestyle{10pt}{0.9}
  \scriptsize
  \renewcommand{\arraystretch}{1.0}
    \begin{tabular}{@{}lrrrr@{}}
   
    \toprule
    & \multicolumn{2}{c}{\textbf{Duration} [hours]} & \multicolumn{2}{c}{\textbf{Sound Events}} \\
    \cmidrule(lr){2-3}\cmidrule(lr){4-5}
    \textbf{Type}  & \textbf{Real} & \textbf{Synthetic}  & \textbf{Real} & \textbf{Synthetic} \\
    \midrule
    Speech         & 0.4$\,$k & 0.3$\,$k  & 70.3$\,$k & 108.7$\,$k \\
    Music          & 0.2$\,$k & 0.2$\,$k & 0.4$\,$M & 0.4$\,$M \\
    General        & 0.6$\,$k & 12.3$\,$k & 0.8$\,$M & 4.1$\,$M \\ \midrule
    Total          & 1.2$\,$k & 12.8$\,$k & 1.3$\,$M & 4.6$\,$M \\
    \bottomrule
  \end{tabular}
  \caption{{\bf \PEaframe Training Data}. 
  Durations and sound event counts for real and synthetic recordings across three sound-type categories. 
  }
  \label{tab:sed_data}

  \vspace{-15pt}
}
\end{wraptable}
\paragraph{Training Data.} For training \PEaframe{}, we use a combination of real-world audio mixtures annotated by humans and synthetic mixtures automatically generated from diverse isolated audio sources.
To enhance robustness to reverberation and spatial variability, the audio mixtures are convolved with room impulse responses collected from a variety of acoustic environments.
This process allows the model to better generalize across different recording conditions and scene types.

Table \ref{tab:sed_data} provides a summary of the training data, reporting the total durations and number of sound events for both real and synthetic recordings. The \emph{Speech} and \emph{Music} subsets correspond to data explicitly annotated with these categories, whereas the \emph{General} subset comprises a broader range of sounds, which may also include speech or music instances not specifically labeled as such.
We use dataset-specific sampling ratios to ensure a balanced mixture of sound events, avoid over-representation, and maintain comprehensive coverage.

\section{\PEav{} Experiments}

\subsection{Experimental Setups}
\paragraph{Datasets.} We evaluate \PEav{} under the zero-shot setting across sound, music, speech, and video benchmarks.
For sound and music, we use VGGSound~\cite{chen2020vggsound}, GTzan~\cite{gtzan}, US8K~\cite{us8k}, Nsynth~\cite{nsynth}, ESC50~\cite{esc50k} and CREMA-D~\cite{cremad} classification. AudioCaps~\cite{kim2019audiocaps}, Clotho-V2~\cite{drossos2020clotho}, and VALOR~\cite{chen2023valor} are used for sound-text retrieval. 
Additionally, we use an internal video dataset for video-to-music retrieval. 
Unlike public datasets such as VGG-Sound and AudioCaps, where video durations are around 10 seconds, the internal dataset exhibits durations ranging from 5 to 30 seconds. We observe that models trained with a fixed number of frames tend to underperform on audio-to-video retrieval when input durations vary widely.
For speech tasks, we use Dynamic-SUPERB~\cite{dsuperb} to evaluate speech classification on accent, language identification (LID), speech emotion (EMO), and vocal sound detection. VCTK~\cite{vctk} is used for speech-to-transcript retrieval.
For video tasks, we have Kinetics-400~\cite{kay2017kinetics}, Kinetics-700~\cite{carreira2019short}, and HMDB~\cite{kuehne2011hmdb} for classification, and MSR-VTT~\cite{vtt}, MSVD~\cite{msvd}, ActivityNet~\cite{anet} and DiDemo~\cite{didemo} for video-text retrieval.

\paragraph{Evaluation Protocols.} We integrate public baselines and evaluate them using the same evaluation pipeline for a fair comparison,. 
Following~\cite{pe,internvl}, we apply Dual Softmax Loss~\cite{dsl} to re-weight retrieval results. 
For pre-training, we employ the zero-shot protocol to deduplicate and exclude samples from downstream datasets. 
For fine-tuning, we consider two protocols: one using only out-of-domain (OOD) data, and another that additionally includes training splits from downstream datasets to match prior setups (\eg, CLAP~\cite{laion_clap}) for fair comparison on audio benchmarks.

\subsection{Main Results}
\label{sec:core_results}

\paragraph{Zero-Shot Sound, Music, and Speech Results.}
Tab.~\ref{tab:exp:core:sound_results} reports zero-shot classification and retrieval results on sound, music, and speech benchmarks. 
We compare \PEav{} with recent audio encoders including CLAP~\cite{laion_clap}, CLAP-Fusion~\cite{enclap}, MS-CLAP~\cite{ms_clap}, M2D-CLAP~\cite{m2d}, and AudioFlamingo2~\cite{af2}, and audio-visual encoders including ImageBind~\cite{girdhar2023imagebind} and LanguageBind\cite{langbind}.
As illustrated in the last two rows, the proposed two-stage training improves \PEav{}'s performance, yielding substantial gains in stage-2 fine-tuning when additional speech and video data are incorporated—especially for speech-to-transcript retrieval in VCTK (16.7$\rightarrow$85.6 R@1).
Notably, the small, base, and large versions of \PEav{} consistently and significantly outperform all baselines across all zero-shot sound, music, and speech benchmarks. 
Even under the out-of-domain (OOD) setup, \PEav{} outperforms other baselines (\eg CLAP and AudioFlamingo2) trained with in-domain data. 
This represents a notable achievement: to the best of our knowledge, \PEav{} is the first audio-video-text encoder to achieve state-of-the-art results on \textit{all} types of sound tasks, surpassing both audio-focused (\eg CLAP~\cite{laion_clap,speech_clap,m2d,ms_clap}) and audio-visual models~\cite{langbind,girdhar2023imagebind}. 

\begin{table*}[t!]
    \centering
        \scriptsize
    \makebox[\linewidth][c]{
    \tablestyle{-0.4pt}{1.02} 
    \begin{tabular}{y{48}wx{12} awww w ww w w w awww wwwwww  wwww}
        \shline
        \multirow{2}{*}{\vspace{-2.2cm} Model}  && %
        & \multicolumn{10}{c}{\ct[c4]{\it Zero-Shot Retrieval}}
        & \multicolumn{14}{c}{\ct[c5]{\it Zero-Shot Classification}} \\
        & \cz{A-Enc Params.}{}
        & \cz{Data (M)}{}
        & \cz[c4]{\textit{\textbf{Avg Retrieval}}}{}
        & \cz[c4]{AudioCaps}{T$\rightarrow$A~\cite{kim2019audiocaps}}
        & \cz[c4]{AudioCaps}{T$\rightarrow$V~\cite{kim2019audiocaps}}
        & \cz[c4]{AudioCaps}{V$\rightarrow$A~\cite{kim2019audiocaps}}
        & \cz[c4]{Clotho}{T$\rightarrow$A~\cite{drossos2020clotho}}
        & \cz[c4]{Valor}{T$\rightarrow$A~\cite{chen2023valor}}
        & \cz[c4]{Valor}{T$\rightarrow$V~\cite{chen2023valor}}
        & \cz[c4]{VCTK}{A$\rightarrow$T~\cite{vctk}}
        & \cz[c4]{VGGSound}{V$\rightarrow$A~\cite{chen2020vggsound}}
        & \cz[c4]{Internal}{V$\rightarrow$A}
        & \cz[c4]{\textit{\textbf{Avg Class.}}}{}
        & \cz[c5]{VGGSound}{A$\rightarrow$T~\cite{chen2020vggsound}}
        & \cz[c5]{VGGSound}{V$\rightarrow$T~\cite{chen2020vggsound}}
        & \cz[c5]{UrbanSound}{8K~\cite{us8k}}
        & \cz[c5]{NSynth}{1K~\cite{nsynth}}
        & \cz[c5]{GTzan}{10~\cite{gtzan}}
        & \cz[c5]{ESC}{50~\cite{esc50k}}
        & \cz[c5]{CREMA-D}{6~\cite{cremad}} 
        & \cz[c5]{Expresso}{emo~\cite{nguyen2023expresso}}
        & \cz[c5]{CV13}{accent~\cite{commonvoice}}
        & \cz[c5]{D-SUPERB}{lid~\cite{dsuperb}}
        & \cz[c5]{D-SUPERB}{mspkr~\cite{dsuperb}}
        & \cz[c5]{D-SUPERB}{emo~\cite{dsuperb}}
        & \cz[c5]{D-SUPERB}{vocal~\cite{dsuperb}} \\
        \hline
\textit{Baselines} & & & \ca{} & & & & & & & & & & \ca{} & \\ 
 AFlamingo2~\cite{af2} & 0.3B & 8 & \ca{-}& 29.8 & - & - & 16.9 & 7.3 & - & 0.3 & - & - & \ca{-} & 27.4 & - & 78.5 & 70.0 & 60.9 & 93.6 & 25.5 & 39.2 & 2.5 & 20.0 & 67.5 & 21.3 & 38.1 \\
ImageBind~\cite{girdhar2023imagebind} & .09B & 3 & \ca{13.9} & 6.6 & 7.6 & 51.3 & 3.9 & 5.4 & 36.1 & 0.4 & 10.8 & 2.8 & \ca{40.8} & 28.2 & 40.4 & 53.3 & 31.5 & 70.6 & 67.4 & 24.7 & 34.5 & 15.1 & 39.0 & 66.5 & 29.6 & 29.9 \\
CLAP-Fusion~\cite{laion_clap} & .03B & 3 & \ca{-} & 35.4 & - & - & 17.7 & 5.5 & - & 0.3 & - & - & \ca{-} & 23.5 & - & 81.4 & 37.3 & 35.1 & 92.2 & 27.8 & 36.9 & 6.7 & 17.5 & 62.5 & 22.5 & 76.9 \\
CLAP~\cite{laion_clap} & .03B & 3 & \ca{-} & 31.6 & - & - & 16.6 & 5.8 & - & 0.2 & - & - & \ca{-} & 24.8 & - & 80.3 & 39.9 & 40.8 & 93.3 & 28.4 & 33.8 & 13.5 & 23.0 & 72.5 & 22.5 & 77.4 \\
LangBind~\cite{langbind} & 0.3B & 10 & \ca{12.1} & 19.7 & 10.6 & 9.1 & 13.3 & 6.5 & 46.8 & 0.2 & 1.6 & 1.4 & \ca{44.1} & 26.0 & 45.4 & 71.9 & 37.6 & 66.5 & 88.3 & 26.6 & 23.1 & 13.5 & 20.5 & 89.0 & 14.6 & 50.8 \\
M2D-CLAP~\cite{m2d} & .09B & 2 & \ca{-} & 27.4 & - & - & 10.5 & 6.3 & - & 0.1 & - & - & \ca{-} & 29.2 & - & 73.4 & 34.5 & 68.1 & 77.8 & 15.8 & 41.7 & 9.2 & 16.5 & 57.5 & 12.9 & 61.7 \\
MS-CLAP\textsuperscript{23'}~\cite{ms_clap} & .08B & 0.1 & \ca{-} & 23.4 & - & - & 17.8 & 5.9 & - & 0.3 & - & - & \ca{-} & 36.0 & - & \textbf{86.2} & 61.2 & 43.1 & 95.1 & 29.1 & 26.4 & 11.3 & 36.0 & 74.5 & 25.0 & 82.4 \\
\hline
\textit{16 Frames} & & & \ca{} & & & & & & & & & & \ca{} & \\ 
\PEav{S} & .09B & 124 & \ca{45.2} & 41.2 & 18.6 & 75.4 & \bf{24.0} & 29.8 & 70.1 & \textbf{96.1} & 34.1 & 17.9 & \ca{60.9} & 43.0 & 46.5 & 81.3 & 69.9 & 73.6 & 95.1 & 36.7 & 63.5 & 17.2 & 62.0 & 69.5 & \textbf{52.1} & 81.7 \\
\PEav{B} & .2B & 124 & \ca{47.0} & 43.1 & 19.8 & 80.6 & 23.4 & 31.9 & 70.0 & 94.8 & 39.0 & 20.4 & \ca{61.7} & 45.2 & 47.3 & 82.5 & 68.2 & 73.9 & 95.6 & 35.5 & 54.4 & 24.0 & 64.0 & 75.5 & 51.3 & 84.2 \\
\PEav{L} & 1.1B & 124 & \ca{48.2} & 44.7 & 19.5 & 86.1 & 22.8 & 35.0 & 70.9 & 85.6 & 45.2 & 23.9 & \ca{\textbf{63.7}} & 46.7 & 47.3 & 83.2 & 72.8 & 72.3 & 95.0 & 42.0 & \textbf{70.1} & 23.1 & 62.0 & 80.0 & 47.9 & 85.4 \\
\hline
\textit{30 FPS} & & & \ca{} & & & & & & & & & & \ca{} & \\ 
\PEav{L}-OOD & 1.1B & 114 & \ca{45.9} & 43.4 & 18.2 & 86.1 & 23.7 & 34.2 & 70.2 & 50.7 & 36.5 & \textbf{50.3} & \ca{58.5} & 43.9 & 46.7 & 82.6 & 35.6 & 70.8 & 94.7 & 42.2 & 42.4 & 24.4 & 59.5 & \bf{89.5} & 46.3 & 82.5 \\
\PEav{S} & .09B & 124 & \ca{48.1} & 41.8 & 18.8 & 77.4 & 23.9 & 29.3 & 70.9 & 94.9 & 35.4 & 40.5 & \ca{61.6} & 43.0 & 47.3 & 81.0 & 73.5 & \textbf{74.1} & 95.2 & 31.5 & 62.3 & 19.3 & \textbf{67.5} & 72.5 & \textbf{52.1} & 81.4 \\
\PEav{B} & 0.2B & 124 & \ca{50.2} & 42.7 & 19.6 & 83.7 & 23.8 & 30.8 & \textbf{71.2} & 94.9 & 40.7 & 44.6 & \ca{62.1} & 44.5 & 47.8 & 83.3 & 73.1 & 70.1 & 95.1 & 36.9 & 53.9 & 24.0 & 66.5 & 82.0 & 44.2 & 85.6 \\
\PEav{L} (PT) & 1.1B & 92 & \ca{36.5} & 33.7 & 14.7 & 83.3 & 17.5 & 24.0 & 57.1 & 16.7 & 33.9 & 47.8 & \ca{55.7} & 42.4 & 46.2 & 82.2 & 39.3 & 72.0 & 94.4 & 40.0 & 37.7 & 21.0 & 47.0 & 88.0 & 35.8 & 78.5 \\
\PEav{L} & 1.1B & 124 & \ca{\textbf{51.6}} & \textbf{45.8} & \textbf{20.8} & \textbf{88.3} & 23.0 & \textbf{35.1} & 70.9 & 85.6 & \textbf{48.3} & 46.5 & \ca{\textbf{63.7}} & \textbf{47.1} & \textbf{48.0} & 83.6 & \textbf{76.8} & 72.2 & \textbf{96.0} & \textbf{43.3} & 69.4 & \bf{25.6} & 64.5 & 72.0 & 43.8 & \textbf{86.1} \\
        \shline
    \end{tabular}
    }

    \caption{{\bf Zero-Shot Audio Results}. 
    A: Audio, V: Video, T: Audio/Video Caption, PT: pre-training only.
    OOD: fine-tuning with out-of-domain data only (clearn zero-shot setup). 
    We report recall@1 for retrieval tasks and top1 accuracy for classification tasks.
    Note that for fair comparison, we integrate baselines into our evaluation pipeline and update their improved results under the same evaluation protocol.
    }
    \label{tab:exp:core:sound_results}
    \vspace{-8pt}
\end{table*}

\PEav{} demonstrates strong zero-shot performance in both retrieval and classification tasks for sound. 
For instance, \PEav{} achieves a state-of-the-art text-to-audio retrieval score of 45.8 R@1 on AudioCaps and 35.1 R@1 on VALOR. Notably, in the zero-shot setup, where \PEav{} is trained using only out-of-domain (OOD) data, it still significantly outperforms baseline models such as CLAP—even those trained directly on in-domain downstream data like AudioCaps for downstream tasks. \PEav{} also demonstrates superior performance across different audio types: it achieves strong performance in video-to-music retrieval (our internal benchmarks) and speech-to-transcript retrieval in VCTK.
For sound classification on NSynth, GTzan, and ESC50, \PEav{} also establishes new state-of-the-art performance. 
Furthermore, \PEav{} significantly outperforms existing audio-visual baselines in audiovisual retrieval tasks. 
For example, on AudioCaps video-to-audio retrieval, \PEav{} achieves 88.3 R@1 compared to 9.1 R@1 for LanguageBind~\cite{langbind} and 51.3 R@1 for ImageBind~\cite{girdhar2023imagebind}. The similar trend is observed in VGGSound and our internal music benchmark.

These results highlight the limitations of single-anchor training when binding multiple modalities. For example in AudioCaps, text-anchored LanguageBind performs poorly on VGGSound V$\rightarrow$A retrieval (1.6 vs 48.3 R@1 by \PEav{}) when text input is absent. Also, image-anchored ImageBind underperforms on AudioCaps T$\rightarrow$A retrieval (6.6 vs 45.8 R@1 by \PEav{}) when video input is missing. 
By scaling the coverage of cross-modal pairs and caption types with an audiovisual data engine, \PEav{} closes the modality gap and generalize better.

Another new capability \PEav{} offers is transcript retrieval.
As shown in column VCTK, all the baselines fail to perform this task, leading to a zero recall rate.
With transcripts included in the pre-training stage, \PEav{} shows some capability in transcript retrieval.
After fine-tuning, \PEav{} delivers significant gains~(from 16.7 to 85.6 R@1), demonstrating the importance of including speech data in the fine-tuning stage.
Furthermore, \PEav{} outperforms baseline models in most speech-related classification tasks.
With pseudo-labeled LID and accent in the pre-training dataset, \PEav{} offers superior accuracy in language identification and accent classification even without fine-tuning.

\paragraph{Zero-Shot Video Results.}
We evaluate \PEav{} on zero-shot video classification and retrieval benchmarks by employing its video-level embedding $\mathbf{h}^v$ of the video encoder as well as the text embedding $\mathbf{}h^{t_v}$ of the text encoder. 
By default \PEav{L} is trained and evaluated under 30 fps.
We also report performance of \PEav{L} trained and evaluated with a fixed number of 16 frames with improved computational efficiency. \PEav{} at 30 fps achieved better performance.
Note that all video results adhere to the clean zero-shot (OOD) setup, with no samples from downstream datasets used.
\begin{table}[h]
        \centering
    \scriptsize
    \makebox[\linewidth][c]{
    \tablestyle{1pt}{1.15} 
    \begin{tabular}{y{51}wwbb awwwwwwwwww awwwww}
        \shline
        \multirow{2}{*}{\vspace{-2.2cm} Model}  &&&&& \multicolumn{11}{c}{\ct[c4]{\it Zero-Shot Retrieval}} %
        & \multicolumn{6}{c}{\ct[c5]{\it Zero-Shot Classification}}\\
            & \cz{V-Enc Params.}{}
            & \cz{Video Data}{}
            & \cz{Resolution} 
            & \cz{Frame/ fps\textsuperscript{*}}
            & \cz[c4]{\textit{\textbf{Avg Retrieval}}}{}
            & \cz[c4]{VTT}{T$\rightarrow$V~\cite{vtt}}
            & \cz[c4]{VTT}{V$\rightarrow$T~\cite{vtt}}
            & \cz[c4]{MSVD}{T$\rightarrow$V~\cite{msvd}}
            & \cz[c4]{MSVD}{V$\rightarrow$T~\cite{msvd}}
            & \cz[c4]{ActivityNet}{T$\rightarrow$V~\cite{anet}}
            & \cz[c4]{ActivityNet}{V$\rightarrow$T~\cite{anet}}
            & \cz[c4]{DiDeMO}{T$\rightarrow$V~\cite{didemo}}
            & \cz[c4]{DiDeMo}{V$\rightarrow$T~\cite{didemo}}
            & \cz[c4]{VATEX}{T$\rightarrow$V~\cite{vatex}}
            & \cz[c4]{VATEX}{V$\rightarrow$T~\cite{vatex}}
            & \cz[c5]{\textit{\textbf{Avg Class.}}}{}
            & \cz[c5]{Kinetics}{400~\cite{kay2017kinetics}}
            & \cz[c5]{Kinetics}{600~\cite{kay2017kinetics}}
            & \cz[c5]{Kinetics}{700~\cite{kay2017kinetics}}
            & \cz[c5]{UCF}{101~\cite{soomro2012ucf101}}
            & \cz[c5]{HMDB}{51~\cite{kuehne2011hmdb}}
            \\
        \hline     
\textit{Baselines} & & & & & \ca{} & & & & & & & & & & & \ca{} & \\         
        UMT-L~\cite{umt} & 0.3B  & 25M & 224 & 8%
        & \ca{-} & 40.7 & 37.1 & 49.0 & 74.5 & 41.9 & 39.4 & 48.6 & 49.9 & -& - & \ca{-} & - & - & - & - & - \\
        ImageBind~\cite{girdhar2023imagebind} & 0.6B & 3M & 224 & 16 &\ca{48.7} & 40.6 & 42.9 & 47.9 & 70.9 & 36.6 & 34.1 & 36.0 & 38.2 & 69.8 & 69.8 &\ca{54.4} & 55.0 & 53.5 & 42.7 & 77.1 & 43.8 \\
        LanguageBind~\cite{langbind}  & 0.3B & 10M & 224 & 16 & \ca{58.3} & 48.6 & 48.7 & 55.6 & 78.8 & 48.0 & 48.8 & 43.5 & 44.7 & 82.9 & 83.1 &\ca{63.5} & 64.3 & 63.4 & 55.4 & 81.3 & 53.0 \\
        SigLIP2-L/16~\cite{siglip2} & 0.3B & n/a & 384 & 8  
        & \ca{47.1} & 41.5 & 31.4 & 53.7 & 74.2 & 35.9 & 31.5 & 36.6 & 37.8 
        & 64.1 & 64.3 &\ca{{64.1}} & {65.3} & {62.5} & {56.8} & {86.7} & {49.3} %
        \\
        InternVL~\cite{internvl} & 5.5B & n/a & 224 & 8 
        & \ca{-} & 44.7 & 40.2 & - & -  & - & -  & - & -
        & -& - &\ca{-} & 69.1 & 68.9 & 60.6 & - & -  
        \\
        InternVideo2 ~\cite{internvideo2} & 1.0B & 102M & 224 & 8
        & \ca{62.7} & {51.9} & {50.9} & {58.1} & {83.3}  & {60.4} & {54.8} & \textbf{57.0} & \textbf{54.3}
        & 70.4 & 85.4 &\ca{70.7} & 73.1 & {72.8} & {64.9} & 88.8 & 53.9  %
        \\
        SigLIP2-g-opt~\cite{siglip2} & 1.1B & n/a & 384 & 8
        & \ca{49.4} & 43.1 & 34.2 & 55.8 & 74.6 & 38.3 & 33.4 & 39.2 & 40.1
        & 67.5 & 68.2 &\ca{68.2} & 69.8 & 67.0 & 61.8 & 90.7 & 51.8 %
        \\
        {\PE{-G}~\cite{pe}} & 1.9B & 22M & 448 & 8 
        & \ca{61.4} & {51.2} & 49.9 & 59.7 & 85.4 & {54.7} & {51.2} & 45.8 & 46.5
        & 83.6 & 85.5 &\ca{74.8} & 76.9 & 76.1 & {69.1}  & \bf{90.7} & 61.1 %
        \\
        {\PE{-L}~\cite{pe}} & 0.3B & 22M & 336 & 8
        & \ca{57.1} & 50.3 & {50.1} & {57.2} & {82.4} & {46.4} & {42.1}  & 42.4 & 41.0
        & 79.0 & 80.5 &\ca{{71.4}} & {73.4} & {72.7}  & {65.3}  & {87.1} & {58.5} %
        \\
        \hline
        \textit{16 Frames} & & & & &
        \ca{} &  &  &  &  &  &  &  &  &  & &
        \ca{} &  &  &  &  &  \\
\PEav{S} & 0.3B &  124M & 336 & 16 & \ca{65.7} & 46.7 & 49.6 & 60.1 & 86.4 & 63.4 & 64.8 & 48.7 & 49.0 & 94.2 & 93.7 &\ca{74.9} & 77.4 & 77.0 & 67.9 & 87.8 & 64.6 \\
\PEav{B} & 0.4B &  124M & 336 & 16 & \ca{65.8} & 48.6 & 50.3 & 60.8 & 87.6 & 64.0 & 64.9 & 46.2 & 47.8 & 94.3 & 93.8 &\ca{75.9} & 77.9 & 77.8 & 68.3 & 89.7 & 65.9 \\
\PEav{L} & 0.5B &  124M & 336 & 16 & \ca{66.9} & 49.0 & 50.5 & 60.5 & \textbf{88.4} & 65.4 & 66.5 & 48.9 & 50.1 & 94.9 & 94.4 &\ca{75.9} & 78.4 & 77.9 & 68.2 & 89.2 & {66.0} \\
\hline
\textit{30 fps} & & & & &
\ca{} &  &  &  &  &  &  &  &  & & &
\ca{} &  &  &  &  &  \\
\PEav{S} & 0.3B & 124M & 336 & 30\textsuperscript{*} & \ca{66.4} & 49.3 & 49.4 & 59.8 & 87.5 & 64.8 & 65.5 & 50.0 & 49.0 & 94.5 & 94.5 &\ca{76.3} & 78.7 & 78.2 & \textbf{69.1} & 89.2 & \bf{66.1} \\
\PEav{B} & 0.4B & 124M & 336 & 30\textsuperscript{*} & \ca{66.5} & 47.7 & 48.4 & 60.7 & 87.6 & 65.7 & 65.9 & 49.3 & 50.1 & 94.9 & 94.4 &\ca{76.1} & 78.5 & 78.2 & 68.9 & 89.5 & 65.7 \\
\PEav{L} & 0.5B & 124M & 336 & 30\textsuperscript{*} & \ca{\bf{67.9}} & \bf{51.9} & \bf{51.2} & \bf{60.8} & 87.6 & \bf{66.5} & \bf{67.7} & 51.6 & 51.7 & \bf{95.1} & \bf{94.8} &\ca{\bf{76.4}} & \bf{78.9} & \bf{78.3} & {69.0} & {90.4} & 65.1 \\
\shline
    \end{tabular}
}

    \caption{{\bf Zero-Shot Video Results.} Comparison of \PEav{} with recent video–language encoders. We report recall@1 for retrieval tasks and top1 accuracy for classification tasks. \PEav{} achieves state-of-the-art results on most classification and retrieval tasks with only 0.3-0.5B parameters.
    }
    \label{tab:exp:core:video_results}
\end{table}

As shown in Tab.~\ref{tab:exp:core:video_results}, \PEav{L} achieves an overall +10.8 R@1 improvement in retrieval and a +5.0 accuracy gain in classification over \PEcore{L}.
\PEav{L} even surpasses \PEcore{G}—a model with 4$\times$ more parameters—by +6.5 R@1 in retrieval and +1.6 accuracy in classification. 
This trend also holds consistently under the 16-frame sampling setting with a  lower computation cost.
We attribute \PEav{}'s superior performance to its lightweight temporal Transformers which effectively capture temporal context across longer video frames, and its broader coverage of audio-video data, particularly the inclusion of longer and semantically rich videos. 
These improvements significantly boost text-video retrieval in ActivityNet (+20.1 T$\rightarrow$V, +25.6 V$\rightarrow$T R@1)  over \PEcore{L}.

\vspace{0.5cm}
Compared to other baselines, \PEav{L} also yields a notable +5.7 accuracy improvement in zero-shot classification compared to prior SOTA model (InternVideo2~\cite{internvideo2}) and recent vision encoders such as SigLIP2~\cite{siglip2}. 
Notably, \PEav{} achieve these with significantly much fewer parameters compared to baselines such as InternVideo2~\cite{internvideo2}, VideoPrism~\cite{videoprism} and PE-G~\cite{pe}. 
\PEav{} successfully set new state-of-the-art performance in zero-shot classification and retrieval for video, sound, music and speech benchmarks.

  \flexwraptable{r}{0.52\textwidth}{-0pt}{-8pt}{
      
        \scriptsize
    \tablestyle{-1pt}{1.0} 
    \begin{tabular}{y{55}x{20}x{20}x{20}x{20}x{20}x{20}x{20}x{20}x{20}x{20}}
    \shline
    \multirow{2}{*}{\vspace{-2.0cm} Model}
    & \ct[c4]{}
    & \multicolumn{6}{c}{\ct[c4]{\it Zero-Shot Retrieval}} 
    & \multicolumn{3}{c}{\ct[c5]{\it Zero-Shot Classifi.}} \\
    & \cz[c4]{\textit{\textbf{Avg}}}{}
    & \cz[c4]{AudioCaps}{T+V$\rightarrow$A~\cite{kim2019audiocaps}}
    & \cz[c4]{VALOR}{T$\rightarrow$A+V~\cite{chen2023valor}}
    & \cz[c4]{VALOR}{T+A$\rightarrow$V~\cite{chen2023valor}}
    & \cz[c4]{VALOR}{T+V$\rightarrow$A~\cite{chen2023valor}}
    & \cz[c4]{VTT}{T+A$\rightarrow$V~\cite{vtt}}
    & \cz[c4]{DiDeMo}{T+A$\rightarrow$V~\cite{didemo}}
    & \cz[c4]{VGGSound}{A$\rightarrow$T~\cite{chen2020vggsound}}
    & \cz[c4]{VGGSound}{V$\rightarrow$T~\cite{chen2020vggsound}}
    & \cz[c4]{VGGSound}{A+V$\rightarrow$T~\cite{chen2020vggsound}} \\
    \hline
ImageBind~\cite{girdhar2023imagebind} & \ca{38.0} & 51.3 & 36.1 & 36.1 & 24.2 & 41.9 & 36.1 & 28.2 & 40.4 & 40.4 \\
LanguageBind~\cite{langbind} & \ca{37.2} & 19.7 & 46.8 & 46.8 & 6.5 & 50.9 & 44.2 & 26.0 & 45.4 & 45.4 \\
\hline
\multicolumn{1}{l}{\textit{16 Frames}} & \ca{} & \multicolumn{8}{l}{}\\
\PEav{L}\textsuperscript{$\dagger$} & \ca{65.5} & 86.1 & 70.9 & 71.6 & 71.1 & 50.7 & 61.1 & - & - & 47.3 \\
\PEav{L} & \ca{77.5} & 94.6 & \textbf{76.8} & 91.6 & 76.8 & 73.2 & 78.3 & 46.7 & 47.3 & 51.0 \\
\hline
\multicolumn{1}{l}{\textit{30 FPS}} & \ca{} & \multicolumn{8}{l}{}\\
\PEav{L}\textsuperscript{$\dagger$} & \ca{69.8} & 88.3 & 70.9 & 73.8 & 74.5 & 63.6 & 69.3 & - & - & 48.0 \\
\PEav{L} & \ca{\bf{80.2}} & \textbf{95.2} & \textbf{76.8} & \textbf{93.0} & \textbf{78.8} & \textbf{85.3} & \textbf{80.8} & \textbf{47.1} & \textbf{48.0} & \textbf{51.8} \\
    \shline
    \end{tabular}

    \caption{\textbf{Zero-shot Joint-Modal Results.} For the baselines and \PEav{} marked with\textsuperscript{$\dagger$}, joint queries are approximated via maximum over uni-modal results, i.e., T+V$\rightarrow$A=$\max($T$\rightarrow$A,V$\rightarrow$A$)$, T$\rightarrow$A+V=$\max($T$\rightarrow$A,T$\rightarrow$V$)$, and T+A$\rightarrow$V=$\max($T$\rightarrow$V,A$\rightarrow$V$)$. \PEav{} enables using native joint embeddings T+V, A+V, and T+A for retrieval and classification. 
    Using joint embeddings is beneficial when content in different modalities complement each other. 
    }
    \label{tab:exp:core:joint_results}

  }
  \vspace{-0.1cm}

\paragraph{Joint Modal Analysis.} Beyond audio-only and video-only benchmarks, \PEav{} demonstrates strong potential as a step change towards an omni-modal encoder. %
In Tab.~\ref{tab:exp:core:joint_results}, \PEav{} jointly incorporates information from multiple modalities and improves upon the best result achieved with a single modality (results marked with \textsuperscript{$\dagger$}). In all the benchmarks, \PEav{} significantly outperforms other multimodal baselines, surpassing ImageBind~\cite{girdhar2023imagebind} by $42.2\%$ and LanguageBind~\cite{langbind} by $43\%$, establishing a new state-of-the-art for audio, video, and text encoders. 

We observe that joint embeddings are helpful when the input modalities offer complementary information. 
For audio tasks such as AudioCaps, combining video and text signals (V+T$\rightarrow$A) clearly outperforms either V$\rightarrow$A or T$\rightarrow$A, yielding a $+6.9$ R@1 improvement.
Similarly, for visual tasks like DiDeMo and VTT, audio-augmented text queries (A+T$\rightarrow$V) outperform A$\rightarrow$V and T$\rightarrow$V, increasing R@1 by $21.7$ and $11.5$, respectively. Notably, when audiovisual captions are available (\eg captions in VALOR), \PEav{} also achieves stronger performance for joint audio-video retrieval compared to single-modal retrieval: 76.8 T$\rightarrow$A+V R@1 v.s. 70.9 T$\rightarrow$V and 35.1 T$\rightarrow$A R@1.

\subsection{Ablation Studies}\label{sec:exp:data_abaltions}

In Tab.~\ref{tab:ablation:caption_comparison}-\ref{tab:ablation:clip_siglip}, we ablate \PEav{B} (16-layer audio Transformer) trained for $100$K steps with 800 batch size on pre-training data. 
ACaps, GTZ, and DSUP denote AudioCaps, GTzan, and Dynamic-SUPERB, respectively. 
\vspace{-2mm}

\pagebreak

  \flexwraptable{r}{0.48\textwidth}{-0pt}{-12pt}{
        
  \scriptsize
  \tablestyle{-1.5pt}{1.0}
  \begin{tabular}
{x{32}x{20}x{28}x{23}x{23}x{23}x{23}x{22}x{22}x{22}x{22}}
\shline
\multirow{3}{*}{}
& \ct[c4]{}
& \ct[c4]{\it Sound-Ret.}
& \multicolumn{2}{c}{\ct[c4]{\it Sound-Class.}}
& \multicolumn{2}{c}{\ct[c5]{\it Speech-Class.}}
& \multicolumn{2}{c}{\ct[c6]{\it Video-Ret.}}
& \multicolumn{2}{c}{\ct[c6]{\it Video-Class.}}
\\
& \ct[c4]{}
& \ct[c4]{Acaps}
& \ct[c4]{VGG}
& \ct[c4]{GTZ}
& \ct[c5]{CV-13}
& \ct[c5]{DSUP}
& \ct[c6]{VTT}
& \ct[c6]{ANet}
& \ct[c6]{K700}
& \ct[c6]{HMDB}
\\
Caption
& \ct[c4]{\textbf{\textit{Avg}}}
& \ct[c4]{T2A}
& \ct[c4]{AV2T}
& \ct[c4]{A2T}
& \ct[c5]{accent}
& \ct[c5]{lid}
& \ct[c6]{T2V}
& \ct[c6]{T2V}
& \ct[c6]{V2T}
& \ct[c6]{V2T} \\
\hline
EnCLAP  & \ca{33.1} & 23.7 & 19.8 & 50.5 & 10.9 & 24.0 & 31.3 & 55.1 & 38.0 & 44.7 \\
CoNeTTE & \ca{35.4} & 26.8 & 29.3 & 55.6 & 12.2 & 21.5 & 31.1 & 56.8 & 38.7 & 46.4 \\
Stage-1 & \ca{38.9} & 30.3 & 39.3 & 57.2 & 15.1 & 21.5 & 36.2 & 56.8 & 44.9 & 49.1 \\
Stage-2 & \ca{\bf{41.5}} & \bf{32.2} & \bf{44.3} & \bf{59.8} & \bf{16.8} & \bf{30.0} & \bf{36.2} & \bf{57.7} & \bf{45.3} & \bf{51.1} \\
\shline
\end{tabular}

  \caption{{\bf Data Engine}.
  Compared to off-the-shelf captioners (EnCLAP~\citep{enclap} and CoNeTTE~\citep{conette}), our data engine significantly improves the data quality by taking into account video context and confidence score.
  }
  \label{tab:ablation:caption_comparison}

  }
\paragraph{Data Engine.} Tab.~\ref{tab:ablation:caption_comparison} compares captions generated by weak captioner (EnCLAP~\cite{enclap} and CoNeTTE~\cite{conette}) to the improved captions by Stage-1 and Stage-2 data engine in~\S\ref{sec:aed}. 
The proposed data engine jointly leverages video captions and confidence scores to improve the quality of audio captions over the raw EnCLAP and CoNeTTE outputs. Moreover, the improved captions from Stage-2 yield further improvements in most sound, speech, and video tasks. 

\vspace{-2mm}

\flexwraptable{r}{0.48\textwidth}{-10pt}{-8pt}{
    \tablestyle{-1.2pt}{1.0}
  \begin{tabular}
{x{20}x{20}x{20}x{22}x{22}x{20}x{20}x{24}x{22}x{22}x{22}x{22}}
\shline
\multirow{3}{*}{}
&
& \ct[c4]{}
& \ct[c4]{\it S-Ret.}
& \multicolumn{2}{c}{\ct[c4]{\it Sound-Class.}} 
& \multicolumn{2}{c}{\ct[c5]{\it Speech-Class.}} 
& \multicolumn{2}{c}{\ct[c6]{\it Video-Ret.}} 
& \multicolumn{2}{c}{\ct[c6]{\it Video-Class.}} 
\\
Real
& Syn.
& \ct[c4]{}
& \ct[c4]{Acaps}
& \ct[c4]{VGG}
& \ct[c4]{GTZ} 
& \ct[c5]{CV-13}
& \ct[c5]{DSUP}
& \ct[c6]{VTT}
& \ct[c6]{ANet}
& \ct[c6]{K700}
& \ct[c6]{HMDB}
\\
Data
& Data
& \ct[c4]{\textbf{\textit{Avg}}}
& \ct[c4]{T2A}
& \ct[c4]{AV2T}
& \ct[c4]{A2T}
& \ct[c5]{accent}
& \ct[c5]{lid}
& \ct[c6]{T2V}
& \ct[c6]{T2V}
& \ct[c6]{V2T}
& \ct[c6]{V2T} \\
\hline
0x & 1x &
\ca{43.0} & 26.1 & 44.3 & 60.7 & 18.1 & 37.5 & 32.7 & 55.4 & 57.2 & 55.4 \\

1x & 0x &
\ca{14.9} & 16.4 & 25.5 & 58.4 & 11.3 & 20.5 & 0.1 & 0.0 & 0.1 & 1.4 \\

1x & 1x &
\ca{40.5} & 27.1 & 41.3 & \bf{65.1} & 18.5 & 36.0 & 29.8 & 42.3 & 51.4 & 52.8 \\

1x & 10x &
\ca{\bf{45.4}} & \bf{32.5} & \bf{44.8} & 62.0 & \bf{23.5} & 46.0 & 32.8 & 53.8 & 54.9 & \bf{58.2} \\

1x & 20x &
\ca{43.5} & 30.6 & 44.4 & 63.0 & 16.8 & 43.0 & 33.5 & \bf{56.1} & 52.6 & 51.5 \\

1x & 30x &
\ca{44.2} & 30.8 & 43.1 & 58.9 & 23.1 & \bf{51.0} & \bf{33.8} & 55.1 & 50.3 & 51.3 \\%$\mathcal{O}(88M)$  & 32.9 & 45.7 & 68.3 & 21.8 & 33.3 & 45.5 & 62.4 & 33.2 & 17.2 & 47.5 & 30.4 & 74.7 & 33.9 & 54.1 & 57.2 & 54.2 & 43.1 & 48.9 \\
\shline
\end{tabular}

\caption{{\bf Data Mixing Ratio.} 
Mixing both types of date performs better than using only real or synthetic data. Higher synthetic ratios (till 1:10) further boost performance by improving diversity.} 
\label{tab:ablation:synthetic_data_mixing_ratio}

}

\paragraph{Real vs Synthetic Data.} Tab.~\ref{tab:ablation:synthetic_data_mixing_ratio} compares models
trained with different real–synthetic caption ratios under a fixed total number of training samples, and shows that using only real captions (row 2) underperforms using only synthetic captions (row 1), indicating that synthetic captions from our audiovisual data engine are high quality and diverse. Rows 2–5 further show that real and synthetic data are complementary: mixed training (row 4) outperforms either alone (rows 1 and 2), with a 1:10 real-to-synthetic ratio yeilding the best results.

   \vspace{-0.2cm}
  \flexwraptable{r}{0.48\textwidth}{-14pt}{-6pt}{
        
\scriptsize
\tablestyle{-1.35pt}{1.0}
  \begin{tabular}
{x{30}x{20}x{28}x{24}x{22}x{22}x{22}x{22}x{22}x{22}x{22}}
\shline
\multirow{3}{*}{}
& \ct[c4]{}
& \ct[c4]{\it Sound-Ret.}
& \multicolumn{2}{c}{\ct[c4]{\it Sound-Class.}} 
& \multicolumn{2}{c}{\ct[c5]{\it Speech-Class.}} 
& \multicolumn{2}{c}{\ct[c6]{\it Video-Ret.}} 
& \multicolumn{2}{c}{\ct[c6]{\it Video-Class.}} 
\\
& \ct[c4]{}
& \ct[c4]{Acaps}
& \ct[c4]{VGG}
& \ct[c4]{GTZ} 
& \ct[c5]{CV-13}
& \ct[c5]{DSUP}
& \ct[c6]{VTT}
& \ct[c6]{ANet}
& \ct[c6]{K700}
& \ct[c6]{HMDB}
\\
Data
& \ct[c4]{\textbf{\textit{Avg}}}
& \ct[c4]{T2A}
& \ct[c4]{AV2T}
& \ct[c4]{A2T}
& \ct[c5]{accent}
& \ct[c5]{lid}
& \ct[c6]{T2V}
& \ct[c6]{T2V}
& \ct[c6]{V2T}
& \ct[c6]{V2T} \\
\hline
$\mathcal{O}(2M)$  & \ca{38.4} & 27.0 & 40.4 & 60.2 & \textbf{20.2} & 36.0 & 32.8 & 50.4 & 36.9 & 41.6 \\
$\mathcal{O}(4M)$  & \ca{41.8} & 29.6 & 43.4 & \textbf{63.9} & 17.2 & 41.5 & 34.9 & 53.1 & 40.5 & 51.7 \\
$\mathcal{O}(8M)$  & \ca{42.0} & 31.3 & 44.7 & 61.8 & 18.9 & 39.5 & 34.5 & 55.5 & 42.8 & 48.8 \\
$\mathcal{O}(16M)$ & \ca{42.9} & 32.1 & 45.2 & 62.1 & 19.3 & 41.0 & \textbf{36.2} & 56.5 & 43.0 & 50.3 \\
$\mathcal{O}(32M)$ & \ca{42.8} & 32.8 & 45.4 & 62.0 & 18.1 & 39.0 & 35.6 & 56.5 & 43.7 & \textbf{51.9} \\
$\mathcal{O}(64M)$ & \ca{\bf{43.4}} & \textbf{33.6} & \textbf{46.2} & 63.8 & 16.0 & \textbf{43.0} & 35.6 & \textbf{57.7} & \textbf{43.9} & 50.7 \\
\shline
\end{tabular}

  \caption{{\bf Data Scaling.} 
  Performance increases and peaks at 64M, underscoring the value of diverse audio-visual data.
  }
  \vspace{-10mm}
  \label{tab:ablation:data_scaling}

  }
  
 \paragraph{Data Scaling.} Tab.~\ref{tab:ablation:data_scaling} examines the scaling behavior 
 of the proposed data engine for generating synthetic captions. We fix the data mixing ratio across datasets and data type and scale the data from 2M to 64M. 
    As shown, the model's performance increases monotonically on average as the data size grows, reaching its peak at 64M samples demonstrating the importance of the diverse audio-visual data.

  \vspace{-2mm}

  \flexwraptable{r}{0.48\textwidth}{-10pt}{-8pt}{
      \scriptsize
\tablestyle{-0.52pt}{1.10}
\begin{tabular}{x{8}x{21}x{21}x{16}
x{19}x{20}x{19}x{21}x{19}x{18}x{18}x{19}x{21}}
\shline
\multirow{3}{*}{\rotatebox[origin=c]{90}{A-layers}}
& \multirow{3}{*}{\rotatebox[origin=c]{90}{A-params}}
& \multirow{3}{*}{\rotatebox[origin=c]{90}{V-params}}
& \ct[c4]{}
& \ct[c4]{\it S-Ret.}
& \multicolumn{2}{c}{\ct[c4]{\it Sound-Class.}} 
& \multicolumn{2}{c}{\ct[c5]{\it Speech-Class.}} 
& \multicolumn{2}{c}{\ct[c6]{\it Video-Ret.}} 
& \multicolumn{2}{c}{\ct[c6]{\it Video-Class.}} 
\\
&&
& \ct[c4]{}
& \ct[c4]{Acaps}
& \ct[c4]{VGG}
& \ct[c4]{GTZ} 
& \ct[c5]{CV-13}
& \ct[c5]{DSUP}
& \ct[c6]{VTT}
& \ct[c6]{ANet}
& \ct[c6]{K700}
& \ct[c6]{HMDB}
\\
&&
& \ct[c4]{\textbf{\textit{Avg}}}
& \ct[c4]{T2A}
& \ct[c4]{AV2T}
& \ct[c4]{A2T}
& \ct[c5]{accent}
& \ct[c5]{lid}
& \ct[c6]{T2V}
& \ct[c6]{T2V}
& \ct[c6]{V2T}
& \ct[c6]{V2T} \\
\hline
8  & 0.03B & 0.34B & \ca{41.1} & 29.5 & 45.0 & 58.0 & 19.3 & 32.0 & 35.8 & 54.9 & 44.1 & 51.2 \\
12 & 0.09B & 0.35B & \ca{43.3} & 32.0 & 45.4 & 63.1 & 21.9 & 38.0 & 36.6 & 56.3 & 44.5 & 51.6 \\
16 & 0.21B & 0.38B & \ca{42.9} & 33.2 & 45.4 & 61.8 & 19.3 & 41.0 & 36.6 & 56.3 & 43.6 & 48.9 \\
20 & 0.41B & 0.42B & \ca{\bf{44.5}} & \textbf{34.4} & \textbf{46.2} & 62.8 & 21.9 & \textbf{44.0} & \textbf{37.3} & 56.7 & \textbf{44.6} & \textbf{52.4} \\
24 & 0.70B & 0.45B & \ca{43.1} & \textbf{34.4} & 45.7 & 62.6 & \textbf{22.7} & 38.0 & 35.2 & 55.7 & 43.7 & 50.1 \\
28 & 1.11B & 0.50B & \ca{42.0} & 34.3 & 44.9 & \textbf{65.0} & 16.0 & 34.0 & 35.5 & \textbf{56.7} & 43.0 & 49.0 \\
\shline
\end{tabular}

\caption{{\bf Scaling Audio Encoder}. Scaling audio encoder improves performance, with saturation beyond $\sim$20 layers likely due to limited data and training 
steps.}
\label{tab:ablation:model_scaling}

  }

\paragraph{Model Size Scaling.} Tab.~\ref{tab:ablation:model_scaling} scales Transformer layers of the audio encoder from 8 to 28 (0.03–1.11B parameters), while keeping the visual and audio-video encoders fixed. As shown, performance consistently improves with larger and deeper models up to 20 layers under the ablation setup. The scaling trend is important, as it implies strong capacity gains with depth; however, the  saturation around 20 layers is likely due to shorter training schedule. 
The 28 layers model performs the best in the full scale experiment. 

  \flexwraptable{r}{0.48\textwidth}{-10pt}{-8pt}{
      \scriptsize
\tablestyle{-0.48pt}{1.1} 
\begin{tabular}{x{7}x{7}x{7}x{7}x{7}x{7}x{7}x{7}x{18}x{18}x{20}x{18}x{20}x{20}x{17}x{18}x{18}x{22}}
\shline
\multirow{3}{*}{\rotatebox[origin=c]{90}{A-V}}
& \multirow{3}{*}{\rotatebox[origin=c]{90}{A-AT}}
& \multirow{3}{*}{\rotatebox[origin=c]{90}{A-AVT}}
& \multirow{3}{*}{\rotatebox[origin=c]{90}{V-AT}}
& \multirow{3}{*}{\rotatebox[origin=c]{90}{V-VT}}
& \multirow{3}{*}{\rotatebox[origin=c]{90}{V-AVT}}
& \multirow{3}{*}{\rotatebox[origin=c]{90}{AV-VT}}
& \multirow{3}{*}{\rotatebox[origin=c]{90}{AV-AVT}}
& \ct[c4]{}
& \ct[c4]{\it S-Ret.}
& \multicolumn{2}{c}{\ct[c4]{\it Sound-Class.}} 
& \multicolumn{2}{c}{\ct[c5]{\it Speech-Class.}} 
& \multicolumn{2}{c}{\ct[c6]{\it Video-Ret.}} 
& \multicolumn{2}{c}{\ct[c6]{\it Video-Class.}} 
\\
&&&&&&&
& \ct[c4]{}
& \ct[c4]{Acaps}
& \ct[c4]{VGG}
& \ct[c4]{GTZ} 
& \ct[c5]{CV-13}
& \ct[c5]{DSUP}
& \ct[c6]{VTT}
& \ct[c6]{ANet}
& \ct[c6]{K700}
& \ct[c6]{HMDB}
\\
&&&&&&&
& \ct[c4]{\textbf{\textit{Avg}}}
& \ct[c4]{T2A}
& \ct[c4]{AV2T}
& \ct[c4]{A2T}
& \ct[c5]{accent}
& \ct[c5]{lid}
& \ct[c6]{T2V}
& \ct[c6]{T2V}
& \ct[c6]{V2T}
& \ct[c6]{V2T} \\

\hline       
- & \checkmark & - & - & - & - & - & - &
\ca{19.0} & 31.9 & 0.5 & 60.4 & \bf{23.5} & \bf{52.5} & 0.1 & 0.0 & 0.1 & 2.2 \\
- & \checkmark & - & - & \checkmark & - & - & - &
\ca{33.9} & 32.2 & 0.3 & 62.2 & 20.6 & 33.0 & 27.2 & 55.5 & 33.0 & 40.8 \\
- & \checkmark & - & - & \checkmark & - & - & \checkmark &
\ca{32.9} & 33.0 & 0.3 & 56.6 & 18.1 & 36.0 & 26.1 & 53.0 & 31.3 & 42.1 \\
\checkmark & \checkmark & - & - & \checkmark & - & - & \checkmark & 
\ca{31.9} & 31.9 & 0.4 & 53.3 & 21.9 & 38.5 & 27.3 & 47.0 & 30.1 & 36.4 \\
\checkmark & \checkmark & - & \checkmark & \checkmark & - & \checkmark & \checkmark & 
\ca{42.6} & 31.4 & 45.1 & 61.0 & 18.1 & 43.5 & \bf{34.5} & 56.7 & \bf{43.8} & \bf{49.7} \\
\checkmark & \checkmark & \checkmark & \checkmark & \checkmark & \checkmark & \checkmark & \checkmark & 
\ca{\bf{43.2}} & \bf{32.9} & \bf{45.5} & \bf{62.4} & 17.2 & 47.5 & 33.9 & \bf{57.2} & 43.1 & 48.9 \\
\shline
\end{tabular}

\caption{\textbf{Scaling Coverage of Contrastive Objective.} A: Audio, V: Video, AT: Audio caption, VT: Video caption. Expanding the coverage of contrastive objectives to more modality pairs strengthens cross-modal alignment and improves zero-shot performance.
Performance peaks when the objective includes all eight pairs.
} 
\label{tab:ablation:clip_siglip}

  }

\paragraph{Scaling Contrastive Objective.} 
Tab.~\ref{tab:ablation:clip_siglip} shows how scaling coverage and types of contrastive loss pairs impact model performance.
As can be seen, expanding the contrastive objective to cover a greater number of possible alignments between modalities and caption types leads to improved results. For example, the contrastive objective, when applied to all eight modality-caption pairs, outperforms the audio-text only contrastive training, which is limited to only audio to audio caption pairs.
Interestingly, we also find that adding cross-modality pairs for example \emph{video to audio caption} and \emph{audio-visual to video caption} in row 5 leads to improvements in text to video retrieval and zero-shot classification, showcasing the value of richer cross-modal alignment strengthens the shared embedding space. These findings highlight the importance of scaling the coverage of cross-modal alignments for learning aligned audio, visual, and text representations. \PEav{} achieves its peak performance when the contrastive objective considers eight possible cross-modal pairings.

\subsection{Qualitative Results}
\label{app:qualitative_results}
\begin{figure*}[h]
  \centering
  \begin{subfigure}[b]{0.95\textwidth}
    \centering
    \begin{minipage}{0.9\linewidth}
    {\footnotesize
        \textbf{Query: } in the ocean a man on a surfboard rides a wave \\
    }
    \end{minipage}

    \begin{minipage}{0.9\linewidth}
    {\footnotesize
        \href{https://youtu.be/VZ5XnXlZQsM?t=9}{\textbf{Top~1 video} }: (\textbf{ground truth})\\
        \includegraphics[width=\linewidth]{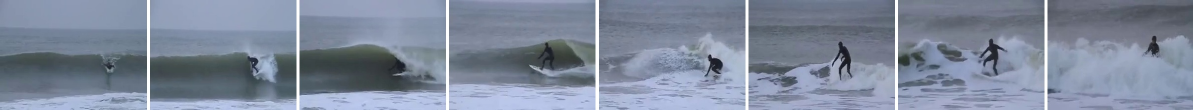}
        
        \href{https://youtu.be/xrt27dZ7DOA?t=39}{\textbf{Top~2 video} }:\\
        \includegraphics[width=\linewidth]{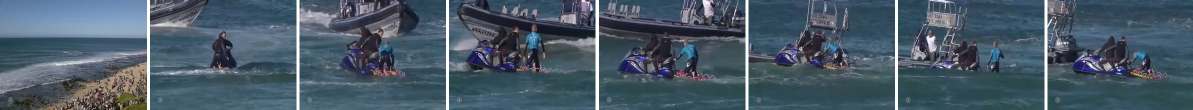}

        \href{https://youtu.be/ukYr_C1Tnzk?t=261}{\textbf{Top~3 video} }:\\
        \includegraphics[width=\linewidth]{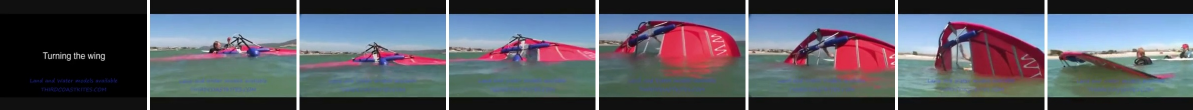}

    }
    \end{minipage}
  \end{subfigure}
  \vspace{0.5em}
  \caption{Video-caption $\rightarrow$ Video retrieval results from \PEav{}. Ground truth (if present) is bolded.}
  \label{fig:peav_t_v_retrieval_demo}
\end{figure*}

\begin{figure*}[h]
  \centering
  \begin{subfigure}[b]{0.95\textwidth}
    \centering
    \centering
    \begin{minipage}{0.9\linewidth}
    {\footnotesize
         \href{https://www.youtube.com/watch?v=wov1DA-Jtjc&t=32s}{{\textbf{Query Video}}} : \\[0.2em]
            \includegraphics[width=\linewidth]{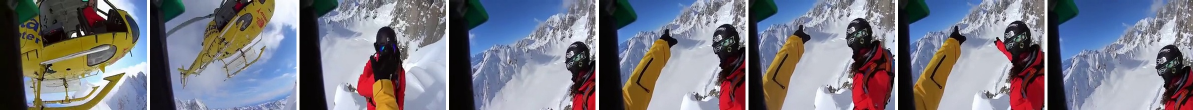} \\

        \textbf{Top 1:} a helicopter moving in air and red and yellow dress man hand touching speaking in snow land wearing helmet displaying on screen (\textbf{ground truth}) \\[0.2em]
        \textbf{Top 2:} a man rides a lift to the top of a mountain \\[0.2em]
        \textbf{Top 3:} flight is shaken and the pilots trying to land the flight while they opened the air
    }
    \end{minipage}
  \end{subfigure}
  \vspace{0.5em}
  \caption{Video $\rightarrow$ Video-caption retrieval results from \PEav{}. Ground truth (if present) is bolded.}
  \label{fig:peav_v_t_retrieval_demo}
\end{figure*}

Fig.~\ref{fig:peav_t_v_retrieval_demo} and \ref{fig:peav_v_t_retrieval_demo} demonstrate qualitative video$\rightarrow$text and text$\rightarrow$video retrieval results by \PEav{}.
In Fig.~\ref{fig:peav_t_v_retrieval_demo}, the ground truth video is successfully retrieved, while the top 2 and 3 retrieved videos show similar scenarios as well~(water sports).
Fig.~\ref{fig:peav_v_t_retrieval_demo} shows a similar phenomenon, but retrieved in the opposite direction.
These examples showcase \PEav{}'s natural capabilities for capturing information from unimodal data and aligning content across modalities.

\begin{figure*}[h!]
  \centering
  \begin{subfigure}[b]{0.95\textwidth}
    \centering

    \begin{minipage}{0.9\linewidth}
    {\footnotesize
        \textbf{Text Query:} In the room, a man pressed the alarm with his index finger, and the alarm rang. \\
        \textbf{Audio Query:} \\
        \includegraphics[width=0.98\linewidth]{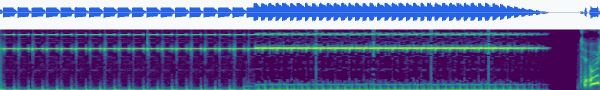}
        \vspace{0.1em}
    }
    \end{minipage}

    \begin{minipage}{0.9\linewidth}
    {\footnotesize
    \begin{tabular}{@{}p{0.98\linewidth}@{}}
    \toprule
      \multicolumn{1}{c}{\textbf{Text $\rightarrow$ \href{https://youtu.be/B8HCQ3rEXM4?t=26}{Video}}} \\[0.3em]
      \includegraphics[width=\linewidth]{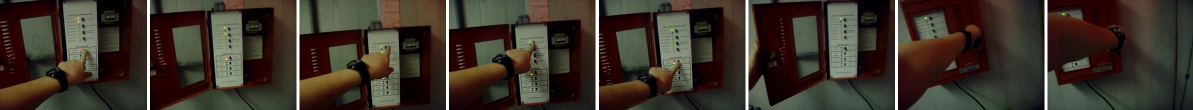}\\
      (\textbf{Caption:} A person presses the button of the instrument watch on the wall, and the instrument drips.)
      \\[0.6em]
    \midrule
      \multicolumn{1}{c}{\textbf{Audio $\rightarrow$ \href{https://youtu.be/ZS4Chf9yh8s?t=20}{Video}}} \\[0.3em]
      \includegraphics[width=\linewidth]{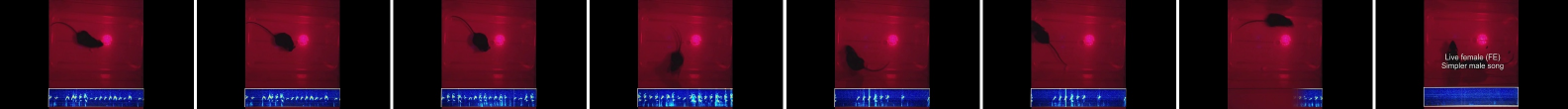}\\
      (\textbf{Caption:} In the enclosed space, a mouse whirled in the dripping sound, the picture turned into two rats.)
      \\[0.6em]
    \midrule
      \multicolumn{1}{c}{\textbf{Text $+$ Audio $\rightarrow$  \href{https://youtu.be/omVDsmGevlI?t=100}{Video}} (\textbf{ground truth})}\\[0.3em]
      \includegraphics[width=\linewidth]{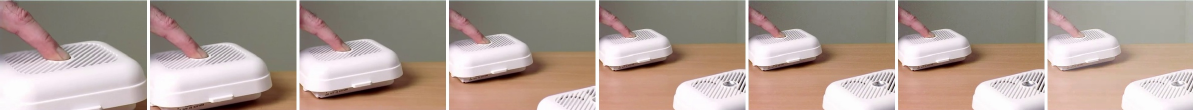}\\
      \bottomrule
    \end{tabular}
    }
    \end{minipage}
  \end{subfigure}
  \vspace{0.5em}
  \caption{T $+$ A $\rightarrow$ Video retrieval results from \PEav{}. Ground truth (if present) is bolded.
  }
  \label{fig:peav_t+a_v_retrieval_demo}
\end{figure*}

\begin{figure*}[h!]
  \centering
  \begin{subfigure}[b]{0.95\textwidth}
    \centering
    \includegraphics[width=0.882\linewidth]{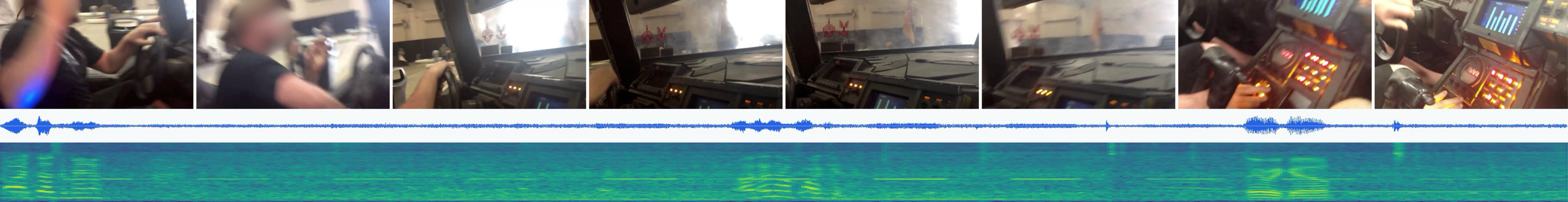}
    \vspace{0.1em}

    \begin{minipage}{0.9\linewidth}
    {\footnotesize
    \begin{tabular}{@{}p{0.98\linewidth}@{}}
    \multicolumn{1}{c}{\href{https://youtu.be/XPyCWnbii9Y?t=180}{\textbf{Video} }} \\[0.3em]
    \toprule
      \multicolumn{1}{c}{\textbf{Audio} $\rightarrow$ \textbf{Text}} \\[0.3em]
      \textbf{Top 1:} A man in a life-saving suit stood beside the manhole cover, directing the rumbling engineering vehicle to reverse, and then angling the vehicle's gear to the sewer. \\[0.2em]
      \textbf{Top 2:} In the field, a command officer in a fluorescent green work suit was waving a flag to direct a farm machine vehicle, with the sound and beeping of vehicles and the voice of men. \\[0.2em]
      \textbf{Top 3:} Outside, a man sits in a car talking while driving slowly as the car dribbles. (\textbf{ground truth})
      \\[0.6em]
    \midrule
      \multicolumn{1}{c}{\textbf{Video} $\rightarrow$ \textbf{Text}} \\[0.3em]
      \textbf{Top 1:} A man fiddled with the steering wheel in the driver's seat, making a rustling sound, and the roar of machine operation from time to time in the distance. \\[0.2em]
      \textbf{Top 2:} The man was sitting in the driver's seat talking, the picture shaking, saw the co-pilot and the windows open inside and behind the car. \\[0.2em]
      \textbf{Top 3:} A man is introducing virtual technology to his buddies at the creaking edge of the wheel.
      \\[0.6em]
    \midrule
      \multicolumn{1}{c}{\textbf{Audio} $+$ \textbf{Video} $\rightarrow$ \textbf{Text}} \\[0.3em]
      \textbf{Top 1:} {Outside, a man sits in a car talking while driving slowly as the car dribbles. (\textbf{ground truth})} \\[0.2em]
      \textbf{Top 2:} A man fiddled with the steering wheel in the driver's seat, making a rustling sound, and the roar of machine operation from time to time in the distance. \\[0.2em]
      \textbf{Top 3:} The man was sitting in the driver's seat talking, the picture shaking, saw the co-pilot and the windows open inside and behind the car. \\
      \bottomrule
    \end{tabular}
    }
    \end{minipage}
  \end{subfigure}
  \vspace{0.5em}
  \caption{A/V/AV $\rightarrow$ AV-caption retrieval results from \PEav{}. Ground truth (if present) is bolded.}
\vspace{-10pt}  
  \label{fig:peav_a-v-av_avt_retrieval_demo}
\end{figure*}

Next, the following examples demonstrate \PEav{}'s novel capability to extract and relate multiple modalities.
Fig.~\ref{fig:peav_t+a_v_retrieval_demo} showcases joint audio+text$\rightarrow$video retrieval results.
The additional audio context helps break the ties of the video and retrieve the corresponding video correctly compared with using either text or audio as the query.
Moreover, in Fig.~\ref{fig:peav_a-v-av_avt_retrieval_demo}, retrieval based solely on video or audio omits key information.
E.g., ``audio $\rightarrow$ text'' is unsuccessful because the visual cue of ``car'' is challenging to extract from the audio.
By leveraging joint multimodal retrieval, \PEav{} incorporates both audio and video context, enabling it to correctly identify the top-1 result.

\begin{figure*}[h!]
  \centering
  \begin{subfigure}[b]{0.95\textwidth}
    \centering
    \centering
    \includegraphics[width=0.95\linewidth]{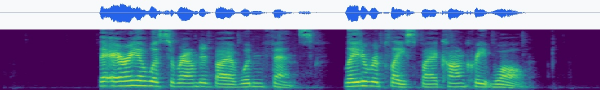}
    \vspace{0.2em}

    \begin{minipage}{\linewidth}
    {\footnotesize
    \begin{tabular}{@{}p{0.9\linewidth}c@{}}
    \toprule
      \multicolumn{1}{c}{\textbf{Caption}} & \textbf{Score} \\[0.3em]
      \textbf{Correct:} A middle-aged female voice, spoken at a normal pace, with a normal pitch and quality. & \textbf{0.571} \\[0.2em]
      \textbf{Similar 1:} A \textcolor{red}{young} female voice, spoken at a normal pace, with a normal pitch and quality. & 0.250 \\[0.2em]
      \textbf{Similar 2:} A middle-aged \textcolor{red}{male} voice, spoken at a normal pace, with a normal pitch and quality. & $-$0.207 \\[0.2em]
      \textbf{Similar 3:} A middle-aged female voice, spoken at a \textcolor{red}{fast} pace, with a normal pitch and quality. & 0.109 \\[0.2em]
      \textbf{Wrong 1:} A \textcolor{red}{young} female voice, spoken at a \textcolor{red}{fast} pace, with a \textcolor{red}{high} pitch and \textcolor{red}{low} recording quality. & $-$0.538 \\[0.2em]
      \textbf{Wrong 2:} A middle-aged \textcolor{red}{male} voice, spoken at a \textcolor{red}{slow} pace, with a \textcolor{red}{low} pitch and normal quality. & $-$0.582 \\[0.2em]
      \textbf{Wrong 3:} A \textcolor{red}{young male} voice, spoken at a \textcolor{red}{fast} pace, with a normal pitch and quality. & $-$0.739 \\[0.6em]
    \midrule
      \multicolumn{1}{c}{\textbf{Transcript}} & \textbf{Score} \\[0.3em]
      \textbf{Correct:} The person says: ``The area was surrounded by a wooden fence, later replaced by a concrete wall.'' & \textbf{0.399} \\[0.2em]
      \textbf{Similar Pronunciation 1:} \\
      The person says: ``The \textcolor{red}{era} was surrounded by a wooden \textcolor{red}{sense}, later \textcolor{red}{replayed} by a \textcolor{red}{convict call}.'' & 0.227 \\[0.2em]
      \textbf{Similar Pronunciation 2:} \\
      The person says: ``The \textcolor{red}{aria} was \textcolor{red}{confounded} by a \textcolor{red}{warden} fence, later replaced by a \textcolor{red}{con fleet} wall.'' & 0.268 \\[0.2em]
      \textbf{Similar Pronunciation 3:} \\
      The person says: ``The \textcolor{red}{airy} was \textcolor{red}{surrendered} by a wooden \textcolor{red}{lens}, later \textcolor{red}{rephrased} by a concrete \textcolor{red}{mall}.'' & 0.172 \\[0.2em]
      \textbf{Similar Meaning 1:} \\
      The person says: ``The yard was enclosed by a timber fence, later swapped for a stone wall.'' & $-$0.331 \\[0.2em]
      \textbf{Similar Meaning 2:} \\
      The person says: ``The field was bordered by a wooden fence, which was later rebuilt in concrete.'' & $-$0.233 \\[0.2em]
      \textbf{Similar Meaning 3:} \\
      The person says: ``A wooden fence once circled the property, but it was replaced by a solid wall.'' & $-$0.478 \\[0.2em]
      \textbf{Wrong 1:} \\
      The person says: ``Man in red tshirt and baseball cap viewed from above he is has a pile of posters.'' & $-$0.956 \\[0.2em]
      \textbf{Wrong 2:} \\
      The person says: ``Hash trees allow efficient and secure verification of the contents of large data structures.'' & $-$2.469 \\[0.2em]
      \textbf{Wrong 3:} \\
      The person says: ``Armand immigrated to the United States from France and sold hats as an occupation.'' & $-$1.981 \\[0.6em]
    \midrule
      \multicolumn{1}{c}{\textbf{Caption + Transcript}} & \textbf{Score} \\[0.3em]
      \textbf{Correct:} A middle-aged female voice, spoken at a normal pace, with a normal pitch and quality. The person says: ``The area was surrounded by a wooden fence, later replaced by a concrete wall.'' & \textbf{0.984} \\[0.2em]
      \textbf{Wrong Caption $+$ Correct Transcript 1:} \\
      A young female voice, spoken at a fast pace, with a high pitch and low recording quality. The person says: ``The area was surrounded by a wooden fence, later replaced by a concrete wall.'' & 0.170 \\[0.2em]
      \textbf{Wrong Caption $+$ Correct Transcript 2:} \\
      A middle-aged male voice, spoken at a slow pace, with a low pitch and normal quality. The person says: ``The area was surrounded by a wooden fence, later replaced by a concrete wall.'' & 0.205 \\[0.2em]
      \textbf{Wrong Caption $+$ Correct Transcript 3:} \\
      A young male voice, spoken at a fast pace, with a normal pitch and quality. The person says: ``The area was surrounded by a wooden fence, later replaced by a concrete wall.'' & 0.018 \\[0.6em]
      \textbf{Correct Caption $+$ Wrong Transcript 1:} \\
      A middle-aged female voice, spoken at a normal pace, with a normal pitch and quality. The person says: ``Man in red tshirt and baseball cap viewed from above he is has a pile of posters.'' & $-$0.536 \\[0.2em]
      \textbf{Correct Caption $+$ Wrong Transcript 2:} \\
      A middle-aged female voice, spoken at a normal pace, with a normal pitch and quality. The person says: ``Hash trees allow efficient and secure verification of the contents of large data structures.'' & $-$1.737 \\[0.2em]
      \textbf{Correct Caption $+$ Wrong Transcript 3:} \\
      A middle-aged female voice, spoken at a normal pace, with a normal pitch and quality. The person says: ``Armand immigrated to the United States from France and sold hats as an occupation.'' & $-$1.425 \\[0.6em]
    \bottomrule
    \end{tabular}
    }
    \end{minipage}
  \end{subfigure}
  \vspace{0.5em}
  \caption{Speech $\rightarrow$ Audio-caption and transcript retrieval results from \PEav{}. The scores indicate the embedding similarity scores between the [CLS-A] and [CLS-AT].
  }
  \label{fig:peav_transcript_retrieval_demo_2}
\end{figure*}

Furthermore, Fig.~\ref{fig:peav_transcript_retrieval_demo_2} demonstrates the speech $\rightarrow$ audio caption/transcript retrieval capabilities.
First, when the audio caption is perturbed~(similar and wrong examples), the retrieval score decreases, indicating the success of identifying the correct speaker, speaking style, and recording environment.
In the second section, we replace some words in the correct transcript with similar-sounding words and find slightly lower scores.
In contrast, rewriting the transcript with different words while preserving meaning significantly decreases scores, implying that \PEav{} captures pronunciation more than meaning in speech.
Moreover, completely irrelevant transcripts lead to even worse scores.
The final section shows the case in which both the caption and the transcript are present in the retrieved text.
The highest score is achieved when both the caption and the transcript are correct, indicating that providing more textual information helps retrieve the desired speech signal.
Finally, we replace the captions and transcripts with incorrect ones and find that incorrect transcripts decrease scores the most.
The results reveal transcripts have a higher impact on the score than audio descriptions, offering more accurate retrieval between speech and text when the transcript is presented.
Overall, the results strongly suggest the usefulness of \PEav{} for speech-related tasks.

\subsection{Implementation Details}
\paragraph{Architecture.} Tab.~\ref{tab:exp:model_config} summarize \PEav{} model configurations. We utilize a pre-trained \PEcore{L}~\cite{pe} as the base frame encoder to capture spatial context and stack 4 lightweight temporal Transformer layers as the video encoder to capture temporal context across frames. 
For the audio, we encode raw audio using a pre-trained DAC-VAE~\cite{moviegen} followed by a Transformer audio encoder, which contains 28 layers and 1.11B parameters in \PEav{L}. 
We scale the hidden dimension proportionally to the number of layers with a factor of 64, and adjust the number of heads with a factor of 0.5. 
The audio-video transformer comprises of 6 layers with the same scaling principle for its hidden size and number of heads. For the text encoder, to support transcript data which requires long context length, we use pre-trained ModernBERT with 28 layers with 512 context length.

  \flexwraptable{r}{0.48\textwidth}{-10pt}{-8pt}{
       \scriptsize
       \centering
       {
           \tablestyle{-0pt}{1.25} 
\begin{tabular}{x{15}x{65}x{25}x{25}x{20}x{20}x{20}x{20}}
    \shline
        \ct{Scale} &
        \ct{Tower} &
        \ct[c1]{Params} & \ct[c2]{Width} & \ct[c3]{Depth} & \ct[c4]{MLP} & \ct[c5]{Heads} & \ct[c6]{Dim} \\
        \hline
       \addpadding
        \multirow{5}{*}{S} & Audio & 0.09B & 768 & 12 & 2048 & 6 & 1024  \\
        & Video (Spatial, \PEcore{L}) & 0.32B & 1024 & 24 & 4096 & 16 & 1024 \\
        & Video (Temporal) & 0.03B & 768 & 4 & 2048 & 6 & 1024 \\
        & Audio-Video & 0.05B & 768 & 6 & 2048 & 6 & 1024 \\
        & Text   & 0.39B & 1024 & 28 & 5248 & 16 & 1024 \\
       \hline
       \multirow{5}{*}{B} & Audio & 0.21B & 1024 & 16 & 2752 & 8 & 1024 \\
        & Video (Spatial, \PEcore{L}) & 0.32B & 1024 & 24 & 4096 & 16 & 1024 \\
        & Video (Temporal) & 0.06B & 1024 & 4 & 2752 & 8 & 1024 \\
        & Audio-Video & 0.08B & 1024 & 6 & 2752 & 8 & 1024 \\
        & Text   & 0.39B & 1024 & 28 & 5248 & 16 & 1024 \\
       \hline
       \addpadding
        \multirow{5}{*}{L} & Audio & 1.11B & 1792 & 28 & 4800 & 14 & 1024\\
        & Video (Spatial, \PEcore{L}) & 0.32B & 1024 & 24 & 4096 & 16 & 1024\\
        & Video (Temporal) & 0.18B & 1792 & 4 & 4800 & 14 & 1024\\
        & Audio-Video & 0.25B & 1792 & 6 & 4800 & 14 & 1024\\
        & Text   & 0.39B & 1024 & 28 & 5248 & 16 & 1024\\
                           \hline
    \shline
\end{tabular}

       }
       \caption{{{\bf Model Configurations.} Total Parameters: \PEav{S}: 0.9B; \PEav{B}: 1.1B; \PEav{L}: 2.2B.}
       }
       \label{tab:exp:model_config}
  }

\paragraph{Training.} We pre-train \PEav{} for 250K steps using batch size of 3024 and a learning rate of 10\textsuperscript{-4}. We leverage pre-trained \PEcore{L}~\cite{pe} and ModernBERT~\cite{modernbert} as the frame and text enocoder respectively, and randomly initialize the rest of modules (video encoder, audio encoder, and audio-video fusion encoder).
The 92M pre-training data composition is in 
Tab.\ref{tab:training_data:stats} and the synthetic captions are generated using the audiovisual data engine in~\S\ref{sec:aed}.
We pre-train \PEav{L} on 216 GPUs for around 9 days. For fine-tuning, we train with the same learning rate for 50K steps with  32M data composition described in Tab.~\ref{tab:training_data:stats} under the fine-tuning data. For the out-of-domain setup we exclude 8M data from the public datasets and internal datasets used in the evaluation in Tab.~\ref{tab:exp:core:sound_results}-\ref{tab:exp:core:video_results}.

\section{Downstream Application of \PEaframe{}: Sound Event Detection (SED)}\label{sec:sed}

We evaluate \PEaframe{} on the task of polyphonic sound event detection \ac{SED}, which is defined as the detection of sound events from multiple classes, where sound events can occur simultaneously~\cite{mesaros2016metrics}.
Traditional (closed-vocabulary) \ac{SED} typically targets a fixed set of classes, predicting a binary label for each class per time frame. 
In contrast, open-vocabulary SED aims to detect the temporal boundaries of arbitrary sound events conditioned on a free-form textual description~\cite{wu2025flam, hai2025flexsed}.
Our model is designed to address both closed- and open-vocabulary \ac{SED}, supporting free-form textual queries for arbitrary sound events.
For closed-vocabulary evaluation, it is prompted with each class from the predefined ontology, and detection proceeds as in the traditional setting. 
For open-vocabulary evaluation, we assume access to a free-form textual description of the sound events present in the audio, and the model predicts the precise onset and offset boundaries of every instance of the specified events.

\paragraph{Test sets and metrics.} 
To evaluate performance, we employ both open-vocabulary (Internal Bench, ASFX-SED~\cite{Wu2025_ASFX_SED}) and closed-vocabulary test sets (AudioSet-Strong~\cite{hershey2021benefit}, DESED~\cite{turpault2019sound}—a.k.a. ``Youtube'' subset in DCASE19~\cite{Turpault2019_DESED_public_eval}, and UrbanSED~\cite{salamon2017scaper}), where Internal Bench denotes an internal benchmark.
We assess model performance using two threshold-independent metrics: the intersection-based \ac{PSDS}~\cite{bilen2020framework} and the segment-based \ac{AUROC}.
For open-vocabulary \ac{SED} datasets, \ac{AUROC} is computed only over the true positive and false positive rates of events that actually occur in each audio clip.
For closed-vocabulary test sets, predictions are generated for all classes included in the respective datasets.
We apply a median filter of 9 to the raw predictions and use the \texttt{sed\_scores\_eval} package~\cite{ebbers2022threshold} with parameters $\rho_\mathrm{DTC} = 0.7$, $\rho_\mathrm{GTC} = 0.7$, $\alpha_\mathrm{ST} = 1$, $\alpha_\mathrm{CT} = 0$, and $e_\text{max} = 100$, corresponding to the standard PSDS1 configuration~\cite{bilen2020framework}.
However, consistent with \cite{li2024self, schmid2025effective}, we omit the variance penalty ($\alpha_\mathrm{ST} = 0$) for AudioSet-Strong, as PSDS was originally designed for datasets with fewer and less imbalanced classes~\cite{bilen2020framework}.
Following~\cite{hai2025flexsed}, we refer to PSDS1 computed across all classes as PSDS1\textsubscript{A}, which emphasizes accurate temporal alignment but still penalizes false positives, and adopt its variant, PSDS1\textsubscript{T}, which focuses solely on target sounds.

\flexwraptable{r}{0.48\textwidth}{-0pt}{-14pt}{
      \centering
{
  \centering
    \begin{subfigure}{\linewidth}
        \includegraphics[width=\linewidth]{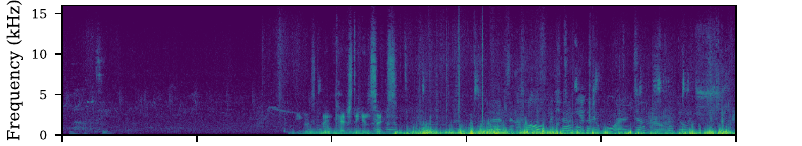}
        \caption{Input spectrogram}
        \vspace{0.8em} 
    \end{subfigure}
    \begin{subfigure}{\linewidth}
        \includegraphics[width=\linewidth]{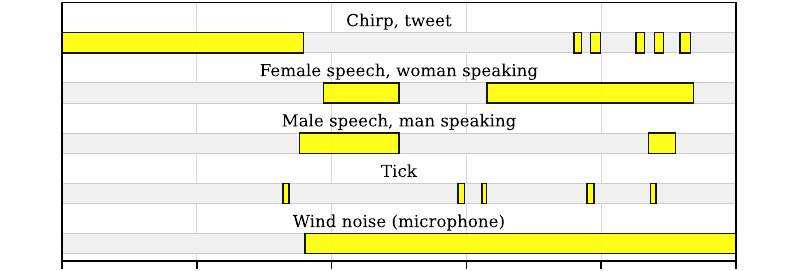}
        \caption{Annotated ``ground-truth'' labels}
            \vspace{0.8em}
    \end{subfigure}
    \begin{subfigure}{\linewidth}
        \includegraphics[width=\linewidth]{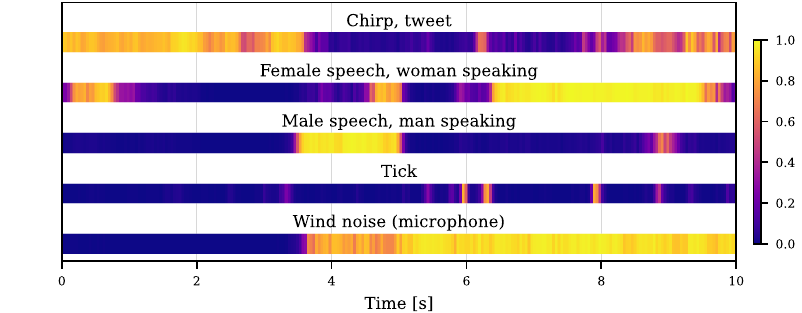}
        \caption{Predicted scores}
    \end{subfigure}
    \caption{\textbf{Sound event detection example using \PEaframe.} 
    The model successfully detects all sound events, accurately distinguishing between male and female speech, and capturing short transient events such as “Tick” enabled by its high temporal resolution (25 Hz). 
    }
    \label{fig:pe_a_frame_example}
}

  }

\paragraph{Baselines.} 
We compare our model with PretrainedSED \cite{schmid2025effective}, FLAM~\cite{wu2025flam}, and FlexSED~\cite{hai2025flexsed}.
PretrainedSED. We use the best-performing checkpoint based on the BEATs transformer~\cite{chen2023beats} as it includes a final task-specific layer that outputs probabilities for the Audioset-Strong classes.
Its training pipeline employs a balanced sampler, extensive data augmentation, and ensemble knowledge distillation, making it highly optimized for the Audioset-Strong test set.
In contrast, FLAM and FlexSED accept free-form textual descriptions and are therefore also suitable for open-vocabulary \ac{SED}.

\paragraph{Results.} 
Table~\ref{tab:sed_results} presents the results across the different \ac{SED} test sets, grouped into open-vocabulary and closed-vocabulary categories.
Overall, \PEaframe{} demonstrates strong performance, achieving the highest scores on all test sets in PSDS1\textsubscript{T}, indicating superior capability in accurately detecting temporal boundaries of target sounds. 
In particular, \PEaframe{} attains the best performance on DESED across all metrics, highlighting its robustness in real-world acoustic environments, as DESED comprises real recordings with fine-grained human annotations~\cite{turpault2019sound}. 
As expected, PretrainedSED performs well on AudioSet-Strong, as it is optimized for that specific ontology; however, its closed-vocabulary nature prevents its application to the other test sets. 
Finally, FLAM performs best on UrbanSED, which is a synthetic and relatively unrealistic dataset. 
We hypothesize that this is because, unlike our model, their system was also  trained on UrbanSED, giving it an inherent advantage on this benchmark.
Moreover, FLAM operates with a coarser frame rate of 3.2 Hz, which limits its ability to perform fine-grained temporal detection and may lead to lower performance on intersection-based metrics.

\begin{table*}[t]
    \centering
    \scriptsize
    \tablestyle{0pt}{1.0}
    \begin{tabular}{>{\raggedright \arraybackslash}p{50pt}x{12}x{20}|x{44}|x{44}|x{33}x{33}x{33}|x{33}x{33}x{33}|x{33}x{33}x{33}}
        \shline
        \multicolumn{1}{l}{} & \multicolumn{1}{c}{\multirow{3}{*}[+1em]{\cz{Freeform}{}}} & \multicolumn{1}{c|}{\multirow{3}{*}[+1em]{\cz{Rate [Hz]}{}}}
        & \multicolumn{2}{c|}{\ct[c4]{\textbf{Open-vocabulary SED}}}
        & \multicolumn{9}{c}{\ct[c5]{\textbf{Closed-vocabulary SED}}} \\
        \multicolumn{1}{l}{} & \multicolumn{1}{l}{} & \multicolumn{1}{l|}{} &
        \multicolumn{1}{c|}{\ct[c4]{ {\it Internal Bench}}} &
        \multicolumn{1}{c|}{\ct[c4]{ {\it ASFX-SED}}} &
        \multicolumn{3}{c|}{\ct[c5]{ {\it AudioSet-Strong} (407 classes)}}  &
        \multicolumn{3}{c|}{\ct[c5]{{\it DESED} (10 classes)}} &
        \multicolumn{3}{c}{\ct[c5]{{\it UrbanSED} (10 classes)}} \\
        & & \multicolumn{1}{l|}{} & \ct[c4]{AUROC} & \ct[c4]{AUROC} &
        \ct[c5]{PSDS1\textsubscript{A}} & \ct[c5]{PSDS1\textsubscript{T}} & \ct[c5]{AUROC} &
        \ct[c5]{PSDS1\textsubscript{A}} & \ct[c5]{PSDS1\textsubscript{T}} & \ct[c5]{AUROC} &
        \ct[c5]{PSDS1\textsubscript{A}} & \ct[c5]{PSDS1\textsubscript{T}} & \ct[c5]{AUROC} \vspace{6pt}  \\ \hline

PretrainedSED~\cite{schmid2025effective} & \xmark & 25 & - & - & \bf 0.47 & 0.52 & \bf 0.98 & - & - & - & - & - \\
FLAM~\cite{wu2025flam} & \cmark & 3.2 & - & 0.81 & 0.35 & - & 0.95 & 0.09 & - & 0.92 & \bf 0.30 & - & \bf 0.94 \\
FlexSED~\cite{hai2025flexsed} & \cmark & 25 & 0.62 & 0.74 & 0.45 & 0.58 & 0.96 & 0.16 & 0.27 & 0.93 & 0.05 & 0.11 & 0.71 \\
\bf \PEaframe{{}} & \cmark & 25 & \bf 0.91 & \bf 0.83 & 0.43 & \bf 0.61 & 0.96 & \bf 0.34 & \bf 0.58 & \bf 0.97 & 0.12 & \bf 0.22 & 0.89 \\
        \shline
    \end{tabular}
    \caption{\textbf{Sound Event Detection Results.} Performance of \PEaframe{} on open-vocabulary (Internal Bench, ASFX-SED) and closed-vocabulary (AudioSet-Strong, DESED, UrbanSED) \ac{SED} test sets. \PEaframe{} achieves the best PSDS1\textsubscript{T} across all benchmarks, indicating superior temporal localization. PretrainedSED is strong but limited to AudioSet-Strong due to its closed-vocabulary design, while FLAM mainly excels on UrbanSED, likely because it is trained on this synthetic dataset and operates at a coarser 3.2 Hz frame rate.}
    \label{tab:sed_results}
\end{table*}

\begin{table*}[h!t!]
    \centering
    \scriptsize
    \tablestyle{0pt}{1.0}
    \begin{tabular}{>{\raggedright \arraybackslash}p{40pt}|x{44}|x{44}|x{33}x{33}x{33}|x{33}x{33}x{33}|x{33}x{33}x{33}}
        \shline
        \multicolumn{1}{l|}{}
        & \multicolumn{2}{c|}{\ct[c4]{\textbf{Open-vocabulary SED}}}
        & \multicolumn{9}{c}{\ct[c5]{\textbf{Closed-vocabulary SED}}} \\
        \multicolumn{1}{l|}{} &
        \multicolumn{1}{c|}{\ct[c4]{ {\it Internal Bench}}} &
        \multicolumn{1}{c|}{\ct[c4]{ {\it ASFX-SED}}} &
        \multicolumn{3}{c|}{\ct[c5]{ {\it AudioSet-Strong} (407 classes)}}  &
        \multicolumn{3}{c|}{\ct[c5]{{\it DESED} (10 classes)}} &
        \multicolumn{3}{c}{\ct[c5]{{\it UrbanSED} (10 classes)}} \\
        \multicolumn{1}{l|}{} & \ct[c4]{AUROC} & \ct[c4]{AUROC} &
        \ct[c5]{PSDS1\textsubscript{A}} & \ct[c5]{PSDS1\textsubscript{T}} & \ct[c5]{AUROC} &
        \ct[c5]{PSDS1\textsubscript{A}} & \ct[c5]{PSDS1\textsubscript{T}} & \ct[c5]{AUROC} &
        \ct[c5]{PSDS1\textsubscript{A}} & \ct[c5]{PSDS1\textsubscript{T}} & \ct[c5]{AUROC} \vspace{2pt} \\ \hline
    
\bf \PEaframe{{L}} & 0.91 & 0.83 & 0.43 & 0.61 & 0.96 & 0.34  & 0.58 & 0.97 & 0.12 & 0.22 & 0.89 \\
\bf \PEaframe{{B}} & 0.92 & 0.83 & 0.42 & 0.60 & 0.96 & 0.39  & 0.56 & 0.98 & 0.12 & 0.21 & 0.89 \\
\bf \PEaframe{{S}} & 0.91 & 0.83 & 0.39 & 0.59 & 0.96 & 0.32  & 0.54 & 0.96 & 0.09 & 0.19 & 0.88 \\
\noalign{\vskip 2pt\hrule height 0.4pt\vskip 2pt}
\makecell[l]{\bf \PEaframe{B}\\ \tiny{$\;$(from scratch)}} & 0.89 & 0.76 & 0.22 & 0.55 & 0.89 & 0.10  & 0.52 & 0.89 & 0.01 & 0.08 & 0.82 \\
\noalign{\vskip 2pt}
        \shline
    \end{tabular}
    \caption{Ablation results for model sizes small (S), base (B), large (L), and a base model trained from scratch. Larger models give only slight gains, while training from scratch leads to a substantial drop, highlighting the importance of large-scale pretraining.}
    \label{tab:sed_model_size}
\end{table*}

\subsection{Ablations studies for \PEaframe}

Figure~\ref{fig:sed_ablation} presents an ablation study on the sampling probability $p_{\text{local}}$, which controls the ratio between the local-activity and the global-activity objective (see \S\ref{sec:pe_a_frame}), conducted with the base model trained for 10$\,$k steps.
We observe that higher values of $p_{\text{local}}$ lead to improved PSDS\textsubscript{T} scores, emphasizing that the local-activity loss benefits the detection of target sounds under the intersection-based evaluation metric. 
However, this comes at the cost of reduced PSDS\textsubscript{A}, which penalizes false positives that may arise because the model contrasts fewer non-target sound events and instead focuses more narrowly on local alignment. Furthermore, we observe a monotonic increase in \ac{AUROC} on the internal benchmark and an increase up to $p_{\text{local}} = 0.8$ on the ASFX-SED benchmark, followed by a drop thereafter. 
A value of $p_{\text{local}} = 0.7$ provides a favorable trade-off between these metrics, and we therefore adopt it as the default setting in all subsequent experiments.

\flexwrapfigure{r}{0.48\textwidth}{-10pt}{-8pt}{
      \centering
{
  \centering
  \begin{tikzpicture}[scale=0.79, every node/.append style={scale=0.79}] 
    \definecolor{cbBlue}{RGB}{56,108,176}
        \definecolor{cbGreen}{RGB}{240,173,78}
        \definecolor{cbPurple}{RGB}{77,175,74}
        \definecolor{cbOrange}{RGB}{152,78,163}
        \begin{axis}[
    width=7cm, height=7cm,
    xlabel={$p_{\text{local}}$},
    xmin=0, xmax=1,
    xtick={0,0.2,0.4,0.6,0.8,1.0},
    ylabel={PSDS1\textsubscript{A} / PSDS1\textsubscript{T}},
    ymin=0.0, ymax=0.65,
    ytick distance=0.1,
    axis y line*=left,
    axis x line*=bottom,
    ymajorgrids,
    clip=false,
]

    \addplot+[thick,mark=square*,cbBlue,solid,mark options={fill=cbBlue,draw=cbBlue}] coordinates {
(0.0, 0.4012339122263864)
(0.1, 0.41674571583738457)
(0.2, 0.42731515999157027)
(0.3, 0.42814139408617335)
(0.4, 0.4290787145912568)
(0.5, 0.4273283049316403)
(0.6, 0.40845637244272864)
(0.7, 0.4018937630125366)
(0.8, 0.3809284833330986)
(0.9, 0.3202510215631761)
(1.0, 0.20627868524528004)
};
    \addplot+[thick,mark=triangle*,cbBlue,densely dotted,mark options={fill=cbBlue,draw=cbBlue,solid}] coordinates {
(0.0, 0.5061272379814107)
(0.1, 0.5450202653882316)
(0.2, 0.5606844697782594)
(0.3, 0.5753701294932282)
(0.4, 0.5894737762193532)
(0.5, 0.591447417954849)
(0.6, 0.6053524919887312)
(0.7, 0.6073020723539396)
(0.8, 0.6096514219989106)
(0.9, 0.6159245863553747)
(1.0, 0.6180278182449027)
};

    \end{axis}

    \begin{axis}[
        width=7cm, height=7cm,
        xmin=0, xmax=1,
        xtick={0,0.2,0.4,0.6,0.8,1.0},
        ylabel={AUROC},
        ymin=0.78, ymax=0.94,
        ytick distance=0.02,
        axis y line*=right,
        axis x line=none,
        clip=false,
        legend to name=groupedLegend,
        legend columns=4,      %
        column sep=0.8em,
        legend style={draw=none, fill=none, font=\small},
    ]

    \addplot+[thick,mark=o,cbGreen,solid,mark options={fill=cbGreen,draw=cbGreen}] coordinates {
(0.0, 0.8308)
(0.1, 0.8704)
(0.2, 0.8795)
(0.3, 0.8791)
(0.4, 0.8958)
(0.5, 0.8961)
(0.6, 0.9094)
(0.7, 0.9153)
(0.8, 0.9202)
(0.9, 0.9232)
(1.0, 0.9266)
};
    \addplot+[thick,mark=o,cbPurple,solid,mark options={fill=cbPurple,draw=orange,cbPurple}] coordinates {
(0.0, 0.7951)
(0.1, 0.8013)
(0.2, 0.8122)
(0.3, 0.8158)
(0.4, 0.8222)
(0.5, 0.8258)
(0.6, 0.826)
(0.7, 0.8307)
(0.8, 0.8341)
(0.9, 0.8226)
(1.0, 0.7959)
};

    \end{axis}

\begin{axis}[
    hide axis,
    xmin=0, xmax=1, ymin=0, ymax=1,
    legend to name=groupedLegendFix,
    legend columns=3,            %
    column sep=0.8em,
    legend cell align=left,
    legend style={draw=none, fill=none, font=\small},
    ]

    \addlegendimage{empty legend}
    \addlegendentry{{\textcolor{cbBlue}{Audioset-Strong:}}}
    \addlegendimage{thick,cbBlue,solid,mark=square*,mark options={fill=cbBlue,draw=cbBlue}}
    \addlegendentry{\textcolor{black}{PSDS1\textsubscript{A}}}
    \addlegendimage{thick,cbBlue,densely dotted,mark=triangle*,mark options={fill=cbBlue,draw=cbBlue,solid}}
    \addlegendentry{\textcolor{black}{PSDS1\textsubscript{T}}}

    \addlegendimage{empty legend}
    \addlegendentry{{\textcolor{cbGreen}{Internal Bench:}}}
    \addlegendimage{thick,cbGreen,solid,mark=o,mark options={fill=cbGreen,draw=cbGreen}}
    \addlegendentry{\textcolor{black}{AUROC}}
    \addlegendimage{empty legend}
    \addlegendentry{}

    \addlegendimage{empty legend}
    \addlegendentry{{\textcolor{cbPurple}{ASFX-SED:}}}
    \addlegendimage{thick,cbPurple,solid,mark=o,mark options={fill=cbPurple,draw=cbPurple}}
    \addlegendentry{\textcolor{black}{AUROC}}
    \addlegendimage{empty legend}
    \addlegendentry{}

\end{axis}

    \node[
    anchor=south,
    draw=gray!40,
    fill=white,
    rounded corners=2pt,
    inner sep=4pt
    ] at ($(current bounding box.north)+(0,06pt)$)
    {\pgfplotslegendfromname{groupedLegendFix}};

  \end{tikzpicture}
  \caption{{\bf Effect of local-activity sampling}. $p_{\text{local}}$ trades off local vs.\ global activity, with $p_{\text{local}}=0.7$ giving the best balance between PSDS\textsubscript{T}, PSDS\textsubscript{A}, and AUROC.}

  \label{fig:sed_ablation}
}

  }

An optimal value of $p_{\text{local}}$ ultimately depends on the intended application of the SED system. 
If the goal is to precisely detect sound event boundaries for a given set of target events, it is recommended to use a higher $p_{\text{local}}$ value. 
Conversely, if the application prioritizes minimizing false positives, such as in continuous environmental monitoring or safety-critical detection scenarios (e.g., false alarms in smart home or surveillance systems), a lower $p_{\text{local}}$ may be more favorable.

Table~\ref{tab:sed_model_size} reports ablation results for different model sizes (large, base, and small), and the base model trained from scratch without pretrained weights. 
The results show that larger models yield only a slight improvement in performance metrics. However, there is a notable drop when the model is trained from scratch, highlighting the importance of large-scale pretraining for achieving strong performance.

\section{Summary}
We have presented \PEav{}, a family of audio-video-text encoders trained with contrastive objectives that jointly align information across all modalities at scale. 
Our audiovisual data engine expands data scale and diversity, and produces high-quality synthetic captions that outperform weak audio captioners while being complementary to real captions.
Using this large-scale and diverse data, we scaled the contrastive objective to ten cross-modal pairs, yielding unified audio-video-text representations for perception. 
Our resulting unified embeddings achieved state-of-the-art zero-shot performance on a broad suite of benchmarks. %
Our audio encoder shows broad coverage across sound, speech, and music domains. 
Given the strong performance of \PEav{}, we hope that the community builds upon it as a foundation for future work in omni-modal  perception and generation.

\section{Acknowlegement}
We would like to thank Yi-Chiao Wu, Andros Tjandra, Bowen Shi, Dangna Li, Peng-Jen Chen,  Robin San Roman, Helin Wang, Carleigh Wood, Andrew Westbury, Vanessa Stark, George Orlin, Anushka Sagar, Vivian Lee, Josh Terry, Helen Klein, Mallika Malhotra, Ty Toledano, Cynthia Gao, Ana Laraia, Mitesh Kumar Singh, John Hoffmann, Andrea Madotto, Muhammad Maaz, Shuming Hu, Daniel Bolya, Vincent Cho, Jianwei Yang, Rafael Valle, Manohar Paluri, Parth Malani, Natacha Supper, Amit Gala, Kyle Bendsen, for their inspiring discussions and timely support throughout this work.

\clearpage
\appendix
\begin{center}
  \vspace*{2mm}
  {\huge\sffamily\bfseries Appendix \par}
  \vspace{2mm}
\end{center}

\section{Overview}
The Appendix is organized as follows. We first discuss the related work in \S\ref{sec:related_work}. Then we provide the details of building PE-AV's audiovisual data engine and the stage-1 and stage-2 prompts used for generating synthetic captions in \S\ref{sec:app:data_engine}. 
Next, in \S\ref{app:impl_details}, we provide additional implementation details of PE-AV including the full hyperparameter setup, training recipe (\S\ref{app:impl_arch_training}), and an efficient implementation (\S\ref{appx:efficient_implementation}) to expand sigmoid contrastive loss for audio-video-text training. We also provide more details for our evaluation protocol to ensure reproducibility in \S\ref{appx:zeroshot_settings}.
Finally, we present additional experiments in \S\ref{app:additional_restuls}.

\section{Related Work}\label{sec:related_work}

Learning visual, acoustic, and textual representations has become central to building multimodal foundation models for perception. By aligning images, video, audio, and language in a shared embedding space, contrastive vision–language and audio–language encoders enable strong zero-shot performance across diverse benchmarks: zero-shot audio retrieval on AudioCaps~\cite{laion_clap, ms_clap, internvideo2}, zero-shot image classification on ImageNet~\cite{imagenet}, and zero-shot text-to-video retrieval~\cite{clip, openclip, laion} on MSR-VTT~\cite{vtt}. Furthermore, these encoders now serve as critical perception front-ends for multi-modal large language models (MLLMs)~\cite{qwen-vl, kosmos-2, llava, paligemma, mm1, cambrian, qwen3_omni}.

\paragraph{Vision-Language Representation Learning.}
Vision–language contrastive pretraining was established by early works such as Virtex~\cite{desai2021virtex}, ICMLM~\cite{sariyildiz2020icmlm}, and ConVIRT~\cite{pmlr-v182-zhang22a}, and later scaled up by CLIP~\cite{clip,openclip} and ALIGN~\cite{align} on significantly larger datasets and models.

Subsequently, a series of open-weight contrastive models~\cite{EVA-CLIP, siglip, li2023clipav2, dfn, metaclip, laion} have been developed to enhance CLIP’s performance and robustness. Notably, SigLIP~\cite{siglip} replaces softmax with a sigmoid objective, and FLIP~\cite{flip} employs masking to accelerate training.
Additionally, researchers have explored incorporating auxiliary objectives, such self-supervised losses~\cite{mu2021slip, silc, dinotext} and captioning losses~\cite{coca, cappa, locca}.
In the data part, a series of works~\cite{datacomp, laion, dfn, metaclip} have studied large-scale sourcing and filtering of web data. These efforts aim to boost model performance by scaling high-quality data through efficient data curation strategies.
To further improve alignment and reduce noise in web-crawled data, several works~\cite{rewrite, veclip, Nguyen2023recap, altogether} explore re-captioning training images using MLLMs or VLMs. This strategy seeks to enhance text quality, thereby strengthening the robustness of the learned representations.

Recently, Perception Encoder (PE)~\cite{pe} modernizes CLIP-style training and, with the Perception Language Model (PLM)~\cite{plm} as a video data engine, scales image–video–language pretraining. Building upon PE and PLM, in this work, we further extend PE to build \PEav{}, an audio-video-text encoder by incorporating the audio modality through model and 
data scaling with an audio-video data engine.

\paragraph{Audio/Speech Representation Learning.}
Self-supervised learning (SSL) has become a dominant approach for audio representation learning, leveraging large amounts of unlabeled data. Notable models for speech representation include wav2vec 2.0~\cite{w2v}, HuBERT~\cite{hsu2021hubert}, and WavLM~\cite{chen2022wavlm}. SSAST~\cite{gong2022ssast}, Audio-MAE~\cite{huang2022masked}, data2vec~\cite{baevski2022data2vec}, and BEATs~\cite{chen2022beats} were developed for general audio. Moreover, MERT~\cite{li2023mert} and MuQ~\cite{zhu2025muq} have advanced music-domain audio representations. Recent advancements aim to learn audio representations at lower cost~\cite{dinosr,chen2024eat} or across multiple domains within a single model~\cite{usad}. However, these methods are limited to single-modality learning and do not use cross-modal information. 

Recently, there has also been growing interest in leveraging paired audio–text data to better align audio and text modalities. Inspired by the success of CLIP in vision–language learning, CLAP~\cite{ms_clap} introduced a contrastive language–audio pre-training objective; subsequent work such as LAION-CLAP~\cite{laion_clap}, M2D-CLAP~\cite{m2d}, FLAP~\cite{flap}, and AF-CLAP~\cite{af2} scaled this paradigm to more data, added SSL objectives, and incorporated synthetic captions, yielding stronger and more transferable audio encoders (including for LLMs).

\paragraph{Toward Unified Audio–Video–Text Encoders.}
To move beyond audio-only or audio–text alignment, several works exploit the video modality to learn audiovisual representations. CAV-MAE~\cite{cav} and MAViL~\cite{mavil} use video as complementary supervision and show strong results on classification and cross-modal retrieval. More recent “hub-style’’ approaches such as ImageBind~\cite{girdhar2023imagebind}, LanguageBind~\cite{langbind}, and InternVid 2~\cite{internvideo2} connect multiple modalities via a single anchor (image or language), but still suffer from scale mismatches between modality-pair datasets, which can hurt less-represented modalities, especially non-speech audio. In contrast, \PEav{} focuses on large-scale, language-guided audiovisual representation learning by utilizing a robust audio-video data engine. This enables broader coverage of contrastive objectives, facilitating the learning of more robust audiovisual and text representations.

\paragraph{Sound Event Detection.}
Traditional \ac{SED} systems operate under a closed-vocabulary setting, targeting a predefined and limited set of sound classes, where each class is assigned a binary label at every time frame \cite{mesaros2021sound}. 
The performance of \ac{SED} models has improved considerably on small-scale datasets centered on domestic environments~\cite{turpault2019sound,nam22_interspeech,shao2024fine}. More recently, \ac{SSL} and large-scale pretraining of audio spectrogram transformers have dramatically advanced \ac{SED} capabilities, enabling the detection of diverse and complex acoustic scenes across hundreds of sound categories~\cite{li2024self,schmid2025effective}. 
These developments mark a significant shift from traditional, closed-vocabulary \ac{SED} to flexible, open-vocabulary paradigms~\cite{wu2025flam, hai2025flexsed}, which aim to identify the temporal boundaries of any sound event described by natural language.

\section{Audio-Video Data Engine}\label{sec:app:data_engine}
In the following, we provide details of the prompts and examples for the stage-1 and stage-2 pipelines used to generate audio, video, and audiovisual captions using the audio-video data engine.

\subsection{Stage-1 Prompts}
\label{sec:appx_video_caption}
The stage-1 prompt used in the data engine is as follows. We leverage CoNeTTe~\cite{conette} and ENCLAP~\cite{enclap}, as well as an internal video captioner.

\promptbox{LLM Prompts for Visual Captions\label{appx_stage1_video_caption_prompts}}{
Create primarily visual captions that focus on what can be seen in the video. Video captions are your reliable source -- \textbf{ALWAYS} create a caption from them, even if audio doesn't match. Handle repetitive video descriptions by summarizing while preserving all unique details. Audio can optionally enhance but should never drive the caption. Remember:
(1) NEVER output an empty caption
(2) ALWAYS create a caption from video content
(3) If audio doesn't match, use only video details
\subsection*{Video Caption Principles}
\begin{itemize}
    \item \textbf{Primary Source:} ALWAYS use video captions
    \item \textbf{Detail Preservation:} Keep all distinct visual elements
    \item \textbf{Redundancy:} Clean up repetitive descriptions
    \item \textbf{Flow:} Create natural, coherent sentences
\end{itemize}
\subsection*{Required Output}
\begin{itemize}
    \item Video summary (clean, non-repetitive)
    \item Audio context (if used)
    \item Visual-focused caption between \texttt{<BOS>} and \texttt{<EOS>}
    \item Explanation of choices
\end{itemize}
\subsection*{Example 1}
\textbf{Input:} INPUT FOR AUDIO-VISUAL:
\begin{itemize}
    \item VIDEO CAPTIONS:
    \begin{itemize}
        \item A professional surfer in a black wetsuit performs an aerial maneuver on a bright red surfboard against massive white waves.
        \item An experienced surfer wearing dark gear rides along the crest of a towering ocean wave on their red board.
        \item A skilled surfer executes a 360-degree turn while surfing on crystal clear blue waters.
    \end{itemize}
    \item AUDIO CAPTIONS (WITH CONFIDENCE LABELS):
    \begin{itemize}
        \item The thunderous crash of ocean waves fills the air. [confidence: HIGH\_CONF]
        \item Cat and dog meowing. [confidence: LOW\_CONF]
    \end{itemize}
\end{itemize}
\textbf{Main Goal:} Create a primarily visual caption that focuses on what can be seen in the video.

\textbf{Video Caption Handling:}
\begin{itemize}
    \item Video captions are your primary and reliable source -- ALWAYS use them
    \item Preserve all distinct visual details (colors, actions, numbers, descriptions)
    \item If video captions are repetitive, summarize while keeping all unique details
    \item Combine multiple video perspectives into natural-flowing sentences
\end{itemize}
\textbf{Audio Caption Handling (Optional):}
\begin{itemize}
    \item Audio is strictly optional -- visual details should stand alone
    \item Only consider HIGH\_CONF audio that directly matches video content
    \item When using audio, add it at the end of the caption without disrupting visual flow
\end{itemize}
\textbf{Output:}
\begin{itemize}
    \item Video summary: A professional surfer in a black wetsuit performs aerial maneuvers and a 360-degree turn on a bright red surfboard, riding along the crest of towering white waves in crystal clear blue waters.
    \item Audio summary: Wave sounds [HIGH\_CONF] align with visible wave activity.
    \item Merged caption: \texttt{<BOS> A professional surfer in a black wetsuit executes impressive aerial maneuvers and a 360-degree turn on their bright red surfboard, riding along the crest of towering white waves in crystal clear blue waters. <EOS>}
    \item Explanation: Focused on rich visual details (wetsuit color, specific moves, board color, wave description). Though wave sounds matched, kept focus on visual elements.
\end{itemize}
}

\promptbox{LLM Prompts for Rewriting Audio Captions\label{appx_stage1_audio_caption_prompts}}{
Your task is to generate an \textbf{audio-focused caption} from model-generated video and audio captions for the same audio-visual input. All video captions are equally likely to be correct.
Audio captions are generated using only audio, and different objects can produce similar sounds (e.g., machine low hum can be confused with crickets, lawn mower cutting grass may sound similar to engine whirring). Sometimes the video and audio may not correspond to each other as the recorded object may be far away. Consider if the described sounds are plausible given the video context. Each audio caption has a confidence label: \texttt{LOW\_CONF}, \texttt{MED\_CONF}, or \texttt{HIGH\_CONF}: \textbf{LOW\_CONF}: This caption is likely incorrect; when only LOW\_CONF captions are present, use details that align with video. \textbf{MED\_CONF}: At least one sound in the caption is correct; others may be incorrect. Only use details aligned with the video. \textbf{HIGH\_CONF}: Generally reliable caption. Include all information with minor video-based adjustments. Prioritize over video caption if clear conflict.
Your task is to merge the video and audio captions into a single caption, focusing on details determinable from audio. Summarize redundant captions before merging. If captions conflict, favor HIGH\_CONF audio after verifying plausibility. When only LOW\_CONF captions exist, try using common elements with video to create an audio-focused summary with general details.
\subsection*{Steps}
\begin{enumerate}
    \item Summarize the video captions.
    \item Identify the audio captions relevant to the video scene and summarize them (dropping irrelevant ones). Verify HIGH\_CONF details from given video context.
    \item Merge the summarized captions into a single caption, focusing on plausible audio-based details, ignoring non-audio details like color.
\end{enumerate}
Keep your answer succinct. Provide: (1) Summarized video caption.(2) Summarized audio caption. Include all plausible HIGH\_CONF details. (3) A final merged caption enclosed between \texttt{<BOS>} and \texttt{<EOS>}.
\subsection*{Example 1}
\textbf{Input:} INPUT FOR AUDIO-VISUAL:
\begin{itemize}
    \item VIDEO CAPTIONS:
    \begin{itemize}
        \item A person surfing in the clear blue ocean in front of the large white waves.
        \item A tan man is surfing in the middle of the vast ocean and it starts to wave in the ocean.
        \item A surfer on a bright red surfboard rides in a towering wave, then jumps up and starts surfing.
    \end{itemize}
    \item AUDIO CAPTIONS (WITH CONFIDENCE LABELS):
    \begin{itemize}
        \item The ocean waves can be heard. [confidence: HIGH\_CONF]
        \item Men talking. [confidence: HIGH\_CONF]
    \end{itemize}
\end{itemize}
\textbf{YOUR TASK:}
\begin{itemize}
    \item Write an audio-focused summary caption.
    \item Summarize redundant or incomplete captions, focusing on details determinable from audio alone, such as gender, species, or car model, while ignoring vision-specific details like colors.
    \item If audio captions conflict, prioritize HIGH\_CONF captions and include details, even if they seem to conflict with video. Make adjustments only when video clearly suggests a more plausible sound source, as the audio source may be off-camera or far away.
    \item Ignore LOW\_CONF.
    \item Include MED\_CONF details that seem plausible given the video.
    \item If no captions remain, create a short summary from the video. Output empty merged caption like \texttt{<BOS><EOS>} when caption cannot be created.
\end{itemize}
\textbf{Output:}
\begin{itemize}
    \item Video summary: A tan man is surfing in the clear blue ocean with a bright red surfboard in front of large white waves.
    \item Audio summary: Sounds of waves and men speaking are heard.
    \item Merged caption: \texttt{<BOS> A man is surfing in the ocean, riding large waves, accompanied by the sounds of waves and men speaking. <EOS>}
    \item Explanation: Combining 'ocean waves' from first audio caption and 'men talking' from second caption to summarize using both HIGH\_CONF. Added large waves and man details using video caption in the merged caption.
\end{itemize}
}

\promptbox{LLM Prompts for Rewriting Audio-Video Captions\label{appx_stage1_audio_video_caption_prompts}}{
You are tasked with generating \textbf{comprehensive audio-visual captions} that effectively combine information from both modalities. You will receive model-generated video captions and audio captions, each audio caption having a confidence label: \texttt{LOW\_CONF}, \texttt{MED\_CONF}, or \texttt{HIGH\_CONF}.
\subsection*{Key Guidelines}
\begin{itemize}
    \item \textbf{Video captions are generally reliable} -- ALWAYS use video information.
    \item \textbf{Audio captions} are based on only audio, and different objects can produce similar sounds (e.g., lawn mower and car engine). For audio captions: (1) \texttt{HIGH\_CONF}: Include if there's any plausible connection to the video context.
        (2) \texttt{MED\_CONF}: Use only when clearly complementing video information.
        (3) \texttt{LOW\_CONF}: Ignore these captions.
    \item When \texttt{HIGH\_CONF} audio seems unrelated to video:
        (1) Include both video and audio information.
        (2) Add note that they might not correspond.
    \item Look for creative ways to interpret audio captions in the video context using generic terms.
    \item Be generous in finding plausible connections between modalities by using broader categories. For example:     (1) If audio mentions specific vehicles (car/truck) and video shows any vehicle -- use generic terms like \emph{vehicle engine/sounds}.
    (2) If audio describes \emph{water sounds} and video shows any liquid -- include it.
    (3) If audio mentions \emph{speaking/talking} and video shows people -- connect them.
\end{itemize}

\subsection*{For each input, provide:}
\begin{itemize}
    \item \textbf{Video summary:} Key visual elements and actions.
    \item \textbf{Audio summary:} Relevant sounds and speech from \texttt{HIGH\_CONF} captions.
    \item \textbf{Merged caption:} Natural combination of both modalities between \texttt{<BOS>} and \texttt{<EOS>}.
    \item \textbf{Brief explanation} of your integration choices.
\end{itemize}
Remember: (1) NEVER output an empty caption. (2) ALWAYS include video information. (3) Be generous in finding plausible audio-video connections. (4) When using unrelated \texttt{HIGH\_CONF} audio, note the potential mismatch.

\subsection*{Example 1}
\textbf{INPUT FOR AUDIO-VISUAL CAPTIONING:}
\begin{itemize}
    \item VIDEO CAPTIONS:
    \begin{itemize}
        \item A drain on a tiled floor shows signs of clogging with debris.
        \item The metallic drain cover is partially lifted, revealing dirty water inside.
        \item The surrounding tiles appear stained and poorly maintained.
    \end{itemize}
    \item AUDIO CAPTIONS (WITH CONFIDENCE LABELS):
    \begin{itemize}
        \item Water is splashing with a scraping sound. [confidence: HIGH\_CONF]
        \item People are talking in a foreign language. [confidence: HIGH\_CONF]
    \end{itemize}
\end{itemize}
\textbf{YOUR TASK:}
\begin{itemize}
    \item Always create a comprehensive caption using the video information as primary source.
    \item For audio captions:
    \begin{itemize}
        \item Use \texttt{HIGH\_CONF} audio captions that seem plausible given the video context.
        \item Include \texttt{MED\_CONF} details only when they clearly complement video information.
        \item Ignore \texttt{LOW\_CONF} audio captions.
    \end{itemize}
    \item When video and \texttt{HIGH\_CONF} audio seem unrelated:
    \begin{itemize}
        \item Include both video and audio information.
        \item Note that they might not correspond to each other.
    \end{itemize}
    \item Never output an empty caption -- always use at least the video information.
    \item Be generous in finding plausible connections between audio and video.
\end{itemize}
\textbf{Output:}
\begin{itemize}
    \item Video summary: A clogged metallic drain on a tiled floor shows debris and dirty water with a partially lifted cover.
    \item Audio summary: Sounds of water splashing and scraping are heard.
    \item Merged caption: <BOS> A clogged drain on a tiled floor contains dirty water, with splashing and scraping sounds as the water moves through the debris. <EOS>
    \item Explanation: Combined visual details of the drain with relevant \texttt{HIGH\_CONF} audio of water and scraping sounds, ignoring unrelated audio about people talking.
\end{itemize}

}

\subsection{PLM-AV Training Details}
\label{app:plm_av_training}

  \flexwraptable{r}{0.48\textwidth}{-10pt}{-8pt}{
    \centering
  \scriptsize
  \begin{tabular}{@{}l lll}
    \toprule
    \textbf{Dataset} & \textbf{Ground Truth} & \textbf{PLM-AV (NPVP)} & \textbf{PLM-AV (caption)} \\
    \midrule
    Audiocaps & 0.52  & 0.54 & 0.46 \\
    Clotho-V2 & 0.57  & 0.34 & 0.54 \\
    \bottomrule
  \end{tabular}
  \caption{Performance comparison with ground-truth capptions using CLAP scores from LAION-CLAP model on AudioCaps and Clotho-V2 datasets. We find that PLM-AV produces high quality tags and captions even on the out-of-domain datasets.
  }
  \label{tab:plm_av_performance}
 
    \vspace{-0.1cm}
  }

In the warm-up phase we only train a MLP layer on $4$M synthetic captions to project the audio-visual / audio embeddings to the same dimensions as LLM embeddings. In the mid-training phase, we fine-tune the entire model on a $30$M synthetic captions previously described. In the final stage, we fine-tune on a curated mix of data focused on audio-visual QA, captioning, instrument recognition, sound tagging, and paralinguistic attributes. Our key goal is to have a model that can produce an audio caption in natural language, or produce a list of sound events in Noun-Verb format. We additionally focus on improving the understanding of the acoustic environment. In Tab.~\ref{tab:plm_av_performance}, we show the performance of the PLM-AV models when measured on held-out Audiocaps~\cite{kim2019audiocaps} and Clotho-V2~\cite{drossos2020clotho} datasets. We use CLAP score from LAION \cite{laion_clap} model to measure the quality of audio captions. We find that even in out-of-domain settings the model produces CLAP scores in same ball-park as ground-truth captions.

\vspace{-0.25cm}
\subsection{Stage-2 Prompts}
\promptbox{Stage-2: PLM Prompts for Fine-Grained Video Caption}{
Create a fine-grained caption of a video using the provided metadata (if applicable), video caption, and frame captions.

\textbf{Task}: Extract key information from the captions and combine it into an alt text format using single phrase or set of phrases that includes all relevant details.

\textbf{Steps to Follow}:
\begin{itemize}
\item Review the metadata if (title and description) for general context, you can rely it for entity names but do not rely on it as the primary source of information for your caption. 
\item Blend title / description with video caption and frame captions for the main storyline
\item Extract the most relevant and concise information.
\item Combine extracted information into a alt text format using short phrase or set of phrases with approximately 120 tokens, considering special characters like comma as part of the token count.
\item Prioritize including all key information over sentence structure or grammar.
\item Minimize the use of special characters and focus of key information.
\end{itemize}

\textbf{What to Avoid:}
\begin{itemize}
\item Avoid adding or inferring information not present in the original metadata and captions.
\item Avoid using complex sentence structures or prioritizing sentence flow.
\end{itemize}

Create a concise caption with the full details of the video based on the metadata, video caption, and frame captions.}\label{app:audio-video-caption:appx_stage2_plm_prompts}

\vspace{-0.05cm}
\promptbox{Stage-2: Final LLM Summarization Prompts for Stage-2 Video Captions}{
\textbf{Task:} \\
You are provided with two types of captions for the same video:
\begin{enumerate}
    \item \textbf{Video-level captions:} A high-level summary of the entire video.
    \item \textbf{Fine-grained captions:} Detailed descriptions of specific events within the video.
\end{enumerate}
\textbf{Goal:} \\
Write a single, concise, and coherent summary that:
\begin{itemize}
    \item Clearly captures the main events of the video.
    \item Preserves important actions, objects, and contextual information from both caption types.
    \item Avoids unnecessary repetition of frame-by-frame details.
    \item Resolves any inconsistencies by prioritizing the video-level caption, unless the fine-grained caption adds essential information.
    \item Is written as a single sentence, not exceeding 72 words.
\end{itemize}
\textbf{Input Format:}
\begin{itemize}
    \item Video-level captions: \texttt{<stage1\_vcap>}
    \item Fine-grained captions: \texttt{<plm\_vcap>}
\end{itemize}
\textbf{Output:} \\
A single-sentence summary describing the video.

}\label{app:audio-video-caption:stage2_llm_summarize}

\section{Implementation Details}\label{app:impl_details}

\subsection{Architecture and Training Setups}
\label{app:impl_arch_training}
\paragraph{Model Architecture.} 
We provide the detailed parameters of \PEav{} in Tab.~\ref{tab:app_model_arch}.
For the frame encoder, we utilize the pre-trained \PEcore{L}~\cite{pe} ($\sim$320M parameters) as the base frame encoder to capture spatial context, and employ a video encoder on top of it to encode temporal context.
By default video frames are sampled under 30 frames-per-second (fps) from up to 30 seconds videos. Each frame is transformed into 336$\times$336 resolution and then encoded into one \textit{CLS} token via \PEcore{L}.
To capture temporal context across frames, we stack 4 additional shallow Transformer layers (30M$\sim$180M parameters) as the video encoder on top of the frame encoder outputs. 
Note that we choose to freeze the frame encoder in \PEav{} to ensure comparable image-only performance as in \PEcore{L}.

For the audio modality, we pre-train a DAC-VAE with in-house audio data. We use it to encode raw audio into 25×128-dimensional vectors for a 1-second audio clip (and 750×128 for a 30-second audio), which serve as input to \PEav{}. For the random-projection quantization module in BEST-RQ, we first project DAC-VAE features to a 16-dimensional latent space and quantize these features with four codebooks, each consisting of 16384 codewords.
The base audio encoder, \PEav{B}, is composed of 16 Transformer layers with around 0.21B parameters, while the large variant, \PEav{L}, contains 28 layers and 1.11B parameters. The small model \PEav{S} contains 12 layers and 0.09B parameters. We scale the hidden dimension proportionally to the number of layers with a factor of 64, and adjust the number of heads with a factor of 0.5.

To integrate audio and video features, we interpolate the video and audio token sequences to the same sequence length for alignment, then concatenate them along the channel dimension. This combined representation is fed into the audio-visual fusion module, which comprises 6 Transformer layers designed to incorporate both audio and video context within the video. The hidden dimension of the fusion tower also scales with the number of layers in the audio encoder, using a factor of 64.
For the text encoder, to extend our support to transcript data which requires long context length, instead of using paired text encoder in \PEcore{L}, we use pre-trained ModernBERT with 28 layers to accommodate input texts up to 512 tokens. Based on early experiments, we use the $22$nd-layer output, and we keep the text tower unfrozen during training.

We employ customized Transformer configurations as detailed in Tab.~\ref{tab:app_model_arch}. For pooling, we add an attention pooling block in the last-layer of video, audio, and audio-video Transformer. Regarding positional embedding,we use 2D RoPE~\cite{rope} for relative positional embeddings. We additionally add a 2D learnable absolute positional embeddings (abs) the same size as the model's input resolution for the frame encoder. Finally, for simplicity, we use an input mean and standard deviation of $(0.5, 0.5, 0.5)$.

\begin{table*}[h!]
\centering
\scriptsize
\tablestyle{0pt}{1.1} 
\begin{tabular}{x{20}x{60}x{25}x{20}x{20}x{20}x{20}x{20}x{40}x{40}x{40}x{40}x{40}}
    \shline
        \ct{Scale} & \ct{Tower} & \ct[c1]{Params} & \ct[c2]{Width} & \ct[c3]{Depth} & \ct[c4]{MLP} & \ct[c5]{Heads} & \ct[c6]{CLIP Dim} & \ct[c7]{Pooling} & \ct[c8]{Positional Embedding} & \ct[c9]{Resolution \& Context Len}  & \ct[c10]{Patch Size}\\
        \hline
       \addpadding
        \multirow{5}{*}{S} & Audio & 0.09B & 768 & 12 & 2048 & 6 & 1024 & Attn Pool & RoPE & - & -  \\
        & Frame (\PEcore{L}) & 0.32B & 1024 & 24 & 4096 & 16 & - & Attn Pool & RoPE+Abs & 336 & 14  \\
        & Vision (Temporal) & 0.03B & 768 & 4 & 2048 & 6 & 1024 & Attn Pool & RoPE & - & - \\
        & Audio-Video & 0.05B & 768 & 6 & 2048 & 6 & 1024 & Attn Pool & RoPE & - & - \\
        & Text   & 0.39B & 1024 & 28 & 5248 & 16 & 1024 & First Token & RoPE & 512 & - \\
       \hline
       \multirow{5}{*}{B} & Audio & 0.21B & 1024 & 16 & 2752 & 8 & 1024 & Attn Pool & RoPE & - & -  \\
        & Frame (\PEcore{L}) & 0.32B & 1024 & 24 & 4096 & 16 & - & Attn Pool & RoPE+Abs & 336 & 14  \\
        & Vision (Temporal) & 0.06B & 1024 & 4 & 2752 & 8 & 1024 & Attn Pool & RoPE & - & - \\
        & Audio-Video & 0.08B & 1024 & 6 & 2752 & 8 & 1024 & Attn Pool & RoPE & - & - \\
        & Text   & 0.39B & 1024 & 28 & 5248 & 16 & 1024 & First Token & RoPE & 512 & - \\
       \hline
       \addpadding
        \multirow{5}{*}{L} & Audio & 1.11B & 1792 & 28 & 4800 & 14 & 1024 & Attn Pool & RoPE & - & - \\
        & Frame (\PEcore{L}) & 0.32B & 1024 & 24 & 4096 & 16 & - & Attn Pool & RoPE+Abs & 336 & 14  \\
        & Vision (Temporal) & 0.18B & 1792 & 4 & 4800 & 14 & 1024 & Attn Pool & RoPE & - & - \\
        & Audio-Video & 0.25B & 1792 & 6 & 4800 & 14 & 1024 & Attn Pool & RoPE & - & - \\
        & Text   & 0.39B & 1024 & 28 & 5248 & 16 & 1024 & First Token & RoPE & 512 & - \\
                           \hline
    \shline
\end{tabular}
\captionsetup{justification=centering}
\caption{\PEav{} Model Configurations.
} 
\label{tab:app_model_arch}
\end{table*}

\subsection{Efficient Scaling of Contrastive Pairs}
\label{appx:efficient_implementation}

\begin{figure}[h]
\centering
\begin{subfigure}{0.48\linewidth}
    \centering
    \includegraphics[width=\linewidth]{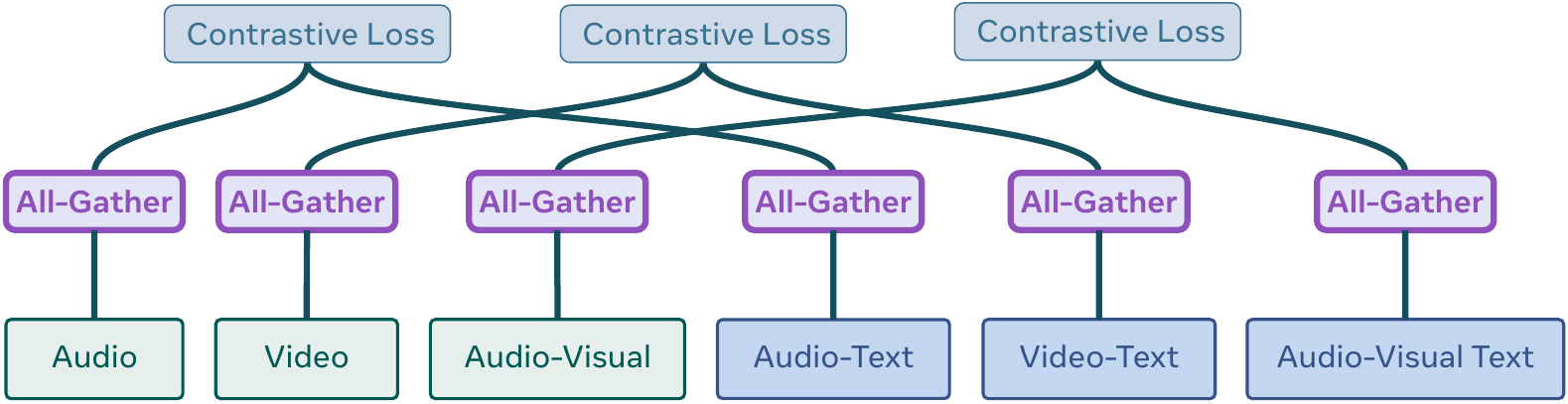}
    \caption{Default implementation with all-gather for each loss computation}
\end{subfigure}
\hfill
\begin{subfigure}{0.48\linewidth}
    \centering
    \includegraphics[width=\linewidth]{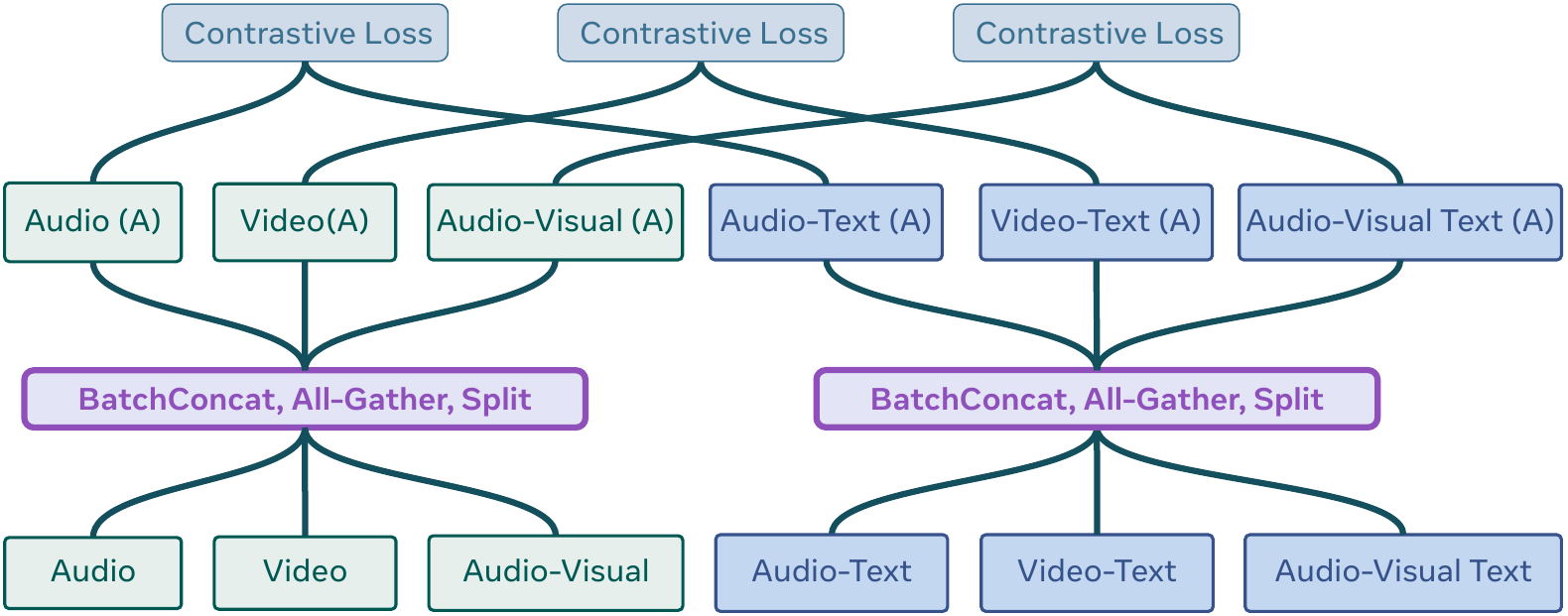}
    \caption{Efficient implementation with fewer all-gather}
\end{subfigure}
\caption{\textbf{Efficient implementation for SigLIP scaling.}
\emph{Left:} Naïve computation involves \textbf{two} \texttt{all\_gather} calls per loss pair which makes it hard to scale as the number of loss terms increase.
\emph{Right:} Our approach concatenates the first terms across all pairs along the batch axis (and likewise for the second terms), performs a \textbf{single} \texttt{all\_gather} independent of the number of pairs, then splits by batch sizes before computing losses. This reduces collectives and improves throughput; in our setup (4 pairs, 8 nodes) we observe \(\sim\)40–50\% speedup.
}
\label{fig:efficient_impl}
\end{figure}

Fig.~\ref{fig:efficient_impl} sketches our efficient implementation for contrastive loss scaling. The default strategy performs two \texttt{all\_gather} operations for every loss pair, so with \(P\) pairs the step performs \(2P\) collectives. As node count grows, the all gather operations dominate runtime.

In  our efficient implementation, we \emph{reduce the number of all gather calls to two} irrespective of the number of loss pairs involved. We stack the first and second arguments of all \(P\) pairs along the batch dimension. We then issue a single \texttt{all\_gather} over the stacked tensors to collect all modalities, and then split the result using recorded batch sizes before evaluating each loss.

Batch-wise concat/split is far cheaper than multiple cross-node collectives, yielding fewer synchronizations and better bandwidth utilization; in practice (4 pairs, 8 nodes) this reduces step time by approximately 40–50\%.

\paragraph{Training Recipe}
\label{sec:appx_joint_train}
As discussed in \S{2} in the Main text, 
the training of \PEav{} involves two stages: 1) audio-video pre-training; 2) video and speech fine-tuning. These two stages work together to develop a robust and effective \PEav{} model.
We first provide the complete training recipes for 1) audio-video pre-training in Tab.~\ref{tab:app:pt-detail} and 2) video and speech fine-tuning in Tab.~\ref{tab:app:ft-detail}.

\label{sec:appx_image_pre_train}

\begin{table*}[!ht]
\begin{minipage}[t]{0.32\textwidth}
\vspace{0pt}
\centering
\scriptsize
{
\tablestyle{5pt}{1.1}
\begin{tabular}{l | c }
config  & values \\
\hline
\addpadding
optimizer & AdamW \\ 
$\beta_1, \beta_2$ & (0.9, 0.999)\\
weight decay & 0.0 \\
learning rate & 1e-4\\
batch size & 3024 \\
warm-up steps & 500 \\
training steps & 250K \\
data quantity & 92M \\
samples seen & 750M \\
\end{tabular} \\
}
\captionsetup{justification=centering}
\caption{\bf Detailed Pre-training Setup.}
\label{tab:app:pt-detail}
\end{minipage}
\hfill
\begin{minipage}[t]{0.32\textwidth}
\vspace{0pt}
\centering
\scriptsize
{
\tablestyle{5pt}{1.1}
\begin{tabular}{l | c }
config  & values \\
\hline
\addpadding
optimizer & AdamW \\ 
$\beta_1, \beta_2$ & (0.9, 0.999)\\
weight decay & 0.0 \\
learning rate &  1e-4 \\
batch size & 1344\\
warm-up steps & 500 \\
training steps & 50K \\
data quantity &  32M(stage2) + 92M(stage1) \\
samples seen & 67M \\
\end{tabular} \\
}
\captionsetup{justification=centering}
\caption{\bf Detailed Fine-tuning Setup.}
\label{tab:app:ft-detail}
\end{minipage}
\hfill
\begin{minipage}[t]{0.30\textwidth}
\vspace{0pt}
\centering
\scriptsize
{
\tablestyle{5pt}{1.1}
\begin{tabular}{l | c }
config  & values \\
\hline
\addpadding
optimizer & AdamW \\ 
$\beta_1, \beta_2$ & (0.9, 0.999)\\
weight decay & 0.0 \\
learning rate &  1e-4 \\
batch size & 800\\
warm-up steps & 500 \\
training steps & 100K \\
data quantity & 92M \\
samples seen & 80M \\
\end{tabular} \\
}
\captionsetup{justification=centering}
\caption{\bf Detailed Ablation Setup.}
\label{tab:app:ablation-detail}
\end{minipage}
\end{table*}

\subsection{Zero-Shot Classification and Retrieval}
\label{appx:zeroshot_settings}

\paragraph{Zero-Shot Evaluation on Images and Videos.} We use CLIPBench\footnote{\url{https://github.com/LAION-AI/CLIP\_benchmark}} for zero-shot classification and retrieval benchmarking. The benchmark datasets and splits are obtained from the original dataset websites or HuggingFace. We extend zero-shot classification and retrieval in CLIPBench to include additional audio and video datasets such as AudioCaps, MSR-VTT, and Kinetics. 
We release our model checkpoints, evaluation code, and scripts for reproducibility.

\paragraph{Prompt Design.} For zero-shot video-text retrieval, we rely solely on the original captions without any additional prompts. 
In contrast, for zero-shot classification, we utilize task-specific prompts graciously provided by the InternVL~\cite{internvl} authors. 
All additional dataset-specific prompts are released for reproducibility. 
For example, we employ specific prompts for zero-shot video classification on Kinetics datasets (e.g., K400, K600, K700).

\label{text:Zero-shot Video Classification Prompts}
\promptbox{Zero-Shot Video Classification Prompts - Kinetics}{
a photo of \{c\}. a photo of a person \{c\}. a photo of a person using \{c\}. a photo of a person doing \{c\}. a photo of a person during \{c\}. a photo of a person performing \{c\}. a photo of a person practicing \{c\}. a video of \{c\}. a video of a person \{c\}. a video of a person using \{c\}. a video of a person doing \{c\}. a video of a person during \{c\}. a video of a person performing \{c\}. a video of a person practicing \{c\}. a example of \{c\}. a example of a person \{c\}. a example of a person using \{c\}. a example of a person doing \{c\}. a example of a person during \{c\}. a example of a person performing \{c\}. a example of a person practicing \{c\}. a demonstration of \{c\}. a demonstration of a person \{c\}. a demonstration of a person using \{c\}. a demonstration of a person doing \{c\}. a demonstration of a person during \{c\}. a demonstration of a person performing \{c\}. a demonstration of a person practicing \{c\}.
}

\paragraph{Evaluation Method.}
\label{appx:zeroshot_eval_method}
For all the zero-shot evaluation, we follow \cite{internvl} in using \textit{retrieval reweighting} (DSL) to apply normalization over the softmax score distribution to the similarities used for retrieval:
\begin{equation}
{\texttt{sims = sims * softmax(sims, dim=0)}}
\end{equation}
This slightly improves retrieval for most models, so we do it for all models we evaluate for fairness.
Notably, we were able to reproduce the reported numbers for most papers with these techniques, but for cases where we could not, we default to the reported number. 

For all retrieval tasks, we use dual-softmax \cite{dsl} for both our models and other baselines. Empirically, we find that sharpening the final logits by a factor of 10 improves downstream performance. This adjustment aligns with the intuition that the model was trained with a scaled logit space to classify paired samples effectively. 
We also ignore the bias term, as it does not affect relative rankings and softmax is invariant to additive shifts.

\section{Additional Experimental Results}\label{app:additional_restuls}
In this section, we provide the complete benchmark results for the data ablation in Tab.~\ref{tab:ablation:caption_comparison_full} (data engine), Tab.~\ref{tab:ablation:synthetic_data_mixing_ratio} (real vs synthetic data), and Tab.~\ref{tab:ablation:data_scaling_full} (data scaling). The complete model scaling results are shown in Tab.~\ref{tab:ablation:model_scaling_full} and for contrastive loss in Tab.~\ref{tab:ablation:clip_siglip_full}.

Additionally, we include additional ablation studies on (1) choice of the text encoder; (2) impact of video frame rate; and (3) BEST-RQ loss in audio encoder training.
Moreover, we provide extensive retrieval results of Tab.~{2}-{3}, and joint-modal results of Tab.~{4} in the main paper.

\begin{table*}[h!]
\centering
\scriptsize
\tablestyle{0pt}{1.00} 
\begin{tabular}
{x{45}x{22}x{22}x{22}x{22}x{22}x{22}x{22}x{22}x{22}x{22}x{22}x{22}x{22}x{22}x{22}x{22}x{22}x{22}}
\shline
\multirow{3}{*}{Caption Type}
& \multicolumn{4}{c}{\ct[c4]{\it Sound-Retrieval}} 
& \multicolumn{4}{c}{\ct[c4]{\it Sound-Classification}} 
& \multicolumn{4}{c}{\ct[c5]{\it Speech-Classification}} 
& \multicolumn{3}{c}{\ct[c6]{\it Video-Retrieval}} 
& \multicolumn{3}{c}{\ct[c6]{\it Video-Classification}} 
\\
& \multicolumn{2}{c}{\ct[c4]{AudioCaps}}
& \ct[c4]{VALOR} 
& \ct[c4]{Internal} 
& \multicolumn{2}{c}{\ct[c4]{VGGSound}}
& \ct[c4]{GTzan} 
& \ct[c4]{Cremad} 
& \ct[c5]{CV-13}
& \multicolumn{3}{c}{\ct[c5]{D-SUPERB}}
& \ct[c6]{VTT}
& \ct[c6]{MSVD}
& \ct[c6]{ANet}
& \ct[c6]{K400}
& \ct[c6]{K700}
& \ct[c6]{HMDB}
\\
& \ct[c4]{T$\rightarrow$A} & \ct[c4]{A$\rightarrow$V}
& \ct[c4]{T$\rightarrow$AV}
& \ct[c4]{A$\rightarrow$V}
& \ct[c4]{A$\rightarrow$T} & \ct[c4]{AV$\rightarrow$T}
& \ct[c4]{A$\rightarrow$T}
& \ct[c4]{A$\rightarrow$T}
& \ct[c5]{accent}
& \ct[c5]{lid} & \ct[c5]{emo} & \ct[c5]{vocal}
& \ct[c6]{T$\rightarrow$V}
& \ct[c6]{T$\rightarrow$V}
& \ct[c6]{T$\rightarrow$V}
& \ct[c6]{V$\rightarrow$T}
& \ct[c6]{V$\rightarrow$T}
& \ct[c6]{V$\rightarrow$T} \\
\hline
EnCLAP & 23.7 & 30.8 & 56.8 & 24.0 & 20.3 & 19.8 & 50.5 & 35.9 & 10.9 & 24.0 & \bf{40.0} & 57.1 & 31.3 & 47.1 & 55.1 & 49.1 & 38.0 & 44.7   \\
CoNeTTE &
26.8 & \bf{36.1} & 59.6 & 24.4 & 25.2 & 29.3 & 55.6 & 28.8  & 12.2 & 21.5 & 30.8 & 67.4 & 31.1 & 49.2 & 56.8 & 49.5 & 38.7 & 46.4\\
Stage-1 & 30.3 & 31.6 & 62.1 & 22.6 & 28.4 & 39.3 & 57.2 & \bf{38.4} & 15.1 & 21.5 & 32.1 & 61.3 & 36.2 & 53.7 & 56.8 & \bf{56.6} & 44.9 & 49.1 \\
Stage-2 & \bf{32.2} & 32.0 & \bf{64.6} & \bf{25.2} & \bf{29.7} & \bf{44.3} & \bf{59.8} & 32.8 & \bf{16.8} & \bf{30.0} & 34.2 & \bf{73.6} & \bf{36.2} & \bf{53.9} & \bf{57.7} & 55.8 & \bf{45.3} & \bf{51.1} \\
\shline
\end{tabular}
\captionsetup{justification=centering}
\caption{{\bf Data Engine}. 
Compared to off-the-shelf captioners (EnCLAP~\citep{enclap} and CoNeTTE~\citep{conette}), the proposed data engine significantly improves the data quality by taking into account video context and confidence score.
}
\label{tab:ablation:caption_comparison_full}
\end{table*}

\begin{table*}[h!]
\centering
\scriptsize
\tablestyle{0pt}{1.00} 
\begin{tabular}{x{30}x{30}x{22}x{22}x{22}x{22}x{22}x{22}x{22}x{22}x{22}x{22}x{22}x{22}x{22}x{22}x{22}x{22}x{22}x{22}}
\shline
\multirow{3}{*}{Real Data}
& \multirow{3}{*}{Syn. Data}
& \multicolumn{4}{c}{\ct[c4]{\it Sound-Retrieval}} 
& \multicolumn{4}{c}{\ct[c4]{\it Sound-Classification}} 
& \multicolumn{4}{c}{\ct[c5]{\it Speech-Classification}} 
& \multicolumn{3}{c}{\ct[c6]{\it Video-Retrieval}} 
& \multicolumn{3}{c}{\ct[c6]{\it Video-Classification}} 
\\
&
& \multicolumn{2}{c}{\ct[c4]{AudioCaps}}
& \ct[c4]{VALOR} 
& \ct[c4]{Internal} 
& \multicolumn{2}{c}{\ct[c4]{VGGSound}}
& \ct[c4]{GTzan} 
& \ct[c4]{Cremad} 
& \ct[c5]{CV-13}
& \multicolumn{3}{c}{\ct[c5]{D-SUPERB}}
& \ct[c6]{VTT}
& \ct[c6]{MSVD}
& \ct[c6]{ANet}
& \ct[c6]{K400}
& \ct[c6]{K700}
& \ct[c6]{HMDB}
\\
&
& \ct[c4]{T$\rightarrow$A} & \ct[c4]{A$\rightarrow$V}
& \ct[c4]{T$\rightarrow$AV}
& \ct[c4]{A$\rightarrow$V}
& \ct[c4]{A$\rightarrow$T} & \ct[c4]{AV$\rightarrow$T}
& \ct[c4]{A$\rightarrow$T}
& \ct[c4]{A$\rightarrow$T}
& \ct[c5]{accent}
& \ct[c5]{lid} & \ct[c5]{emo} & \ct[c5]{vocal}
& \ct[c6]{T$\rightarrow$V}
& \ct[c6]{T$\rightarrow$V}
& \ct[c6]{T$\rightarrow$V}
& \ct[c6]{V$\rightarrow$T}
& \ct[c6]{V$\rightarrow$T}
& \ct[c6]{V$\rightarrow$T} \\
\hline
0x & 1x & 26.1 & 39.1 & 58.4 & 31.0 & 30.2 & 44.3 & 60.7 & 28.2 & 18.1 & 37.5 & 15.0 & 69.7 & 32.7 & 47.1 & 55.4 & 68.4 & 57.2 & 55.4 \\
1x & 0x & 16.4 & 28.1 & 0.0 & 1.7 & 18.3 & 25.5 & 58.4 & 31.1 & 11.3 & 20.5 & 27.9 & 54.6 & 0.1 & 0.3 & 0.0 & 0.3 & 0.1 & 1.4  \\
1x & 1x & 27.1 & 34.9 & 53.7 & 14.4 & 27.3 & 41.3 & \bf{65.1} & 28.8 & 18.5 & 36.0 & 25.4 & 64.9 & 29.8 & 46.7 & 42.3 & 63.7 & 51.4 & 52.8 \\
1x & 10x & 32.5 & \bf{44.3} & \bf{63.2} & 25.2 & \bf{31.9} & \bf{44.8} & 62.0 & 30.7 & \bf{23.5} & \bf{46.0} & \bf{37.5} & 74.2 & 32.8 & \bf{47.8} & 53.8 & \bf{66.4} & \bf{54.9} & \bf{58.2} \\
1x & 20x & 30.6 & 42.9 & 63.7 & 26.5 & 31.2 & 44.4 & 63.0 & \bf{31.3} & 16.8 & 43.0 & 23.8 & \bf{77.1} & 33.5 & 47.2 & \bf{56.1} & 63.6 & 52.6 & 51.5 \\
1x & 30x & \bf{30.8} & 43.6 & 60.5 & \bf{29.0} & 30.4 & 43.1 & 58.9 & 30.2 & 23.1 & 51.0 & 30.4 & 75.6 & \bf{33.8} & 46.9 & 55.1 & 61.7 & 50.3 & 51.3 \\
\shline
\end{tabular}
\captionsetup{justification=centering}
\caption{{\bf Comparing different mixing ratios of real and synthetic caption data.} Mixing both data types outperforms using only real or synthetic data. Higher synthetic ratios (till 1:10) further boost performance by improving diversity.} 
\label{tab:ablation:synthetic_data_mixing_ratio_full}
\end{table*}

\begin{table*}[h!]
\centering
\scriptsize
\tablestyle{0pt}{1.05} 
\begin{tabular}{x{45}x{22}x{22}x{22}x{22}x{22}x{22}x{22}x{22}x{22}x{22}x{22}x{22}x{22}x{22}x{22}x{22}x{22}x{22}}
\shline
\multirow{3}{*}{Data Scale}
& \multicolumn{4}{c}{\ct[c4]{\it Sound-Retrieval}} 
& \multicolumn{4}{c}{\ct[c4]{\it Sound-Classification}} 
& \multicolumn{4}{c}{\ct[c5]{\it Speech-Classification}} 
& \multicolumn{3}{c}{\ct[c6]{\it Video-Retrieval}} 
& \multicolumn{3}{c}{\ct[c6]{\it Video-Classification}} 
\\
& \multicolumn{2}{c}{\ct[c4]{AudioCaps}}
& \ct[c4]{VALOR} 
& \ct[c4]{Internal} 
& \multicolumn{2}{c}{\ct[c4]{VGGSound}}
& \ct[c4]{GTzan} 
& \ct[c4]{Cremad} 
& \ct[c5]{CV-13}
& \multicolumn{3}{c}{\ct[c5]{D-SUPERB}}
& \ct[c6]{VTT}
& \ct[c6]{MSVD}
& \ct[c6]{ANet}
& \ct[c6]{K400}
& \ct[c6]{K700}
& \ct[c6]{HMDB}
\\
& \ct[c4]{T$\rightarrow$A} & \ct[c4]{A$\rightarrow$V}
& \ct[c4]{T$\rightarrow$AV}
& \ct[c4]{A$\rightarrow$V}
& \ct[c4]{A$\rightarrow$T} & \ct[c4]{AV$\rightarrow$T}
& \ct[c4]{A$\rightarrow$T}
& \ct[c4]{A$\rightarrow$T}
& \ct[c5]{accent}
& \ct[c5]{lid} & \ct[c5]{emo} & \ct[c5]{vocal}
& \ct[c6]{T$\rightarrow$V}
& \ct[c6]{T$\rightarrow$V}
& \ct[c6]{T$\rightarrow$V}
& \ct[c6]{V$\rightarrow$T}
& \ct[c6]{V$\rightarrow$T}
& \ct[c6]{V$\rightarrow$T} \\
\hline
$\mathcal{O}(2M)$ & 27.0 & 38.1 & 57.8 & 18.1 & 27.4 & 40.4 & 60.2 & 30.6 & \bf{20.2} & 36.0 & 32.9 & 66.3 & 32.8 & 47.6 & 50.4 & 48.0 & 36.9 & 41.6  \\
$\mathcal{O}(4M)$  & 29.6 & 46.1 & 64.6 & 22.1 & 30.1 & 43.4 & 63.9 & 28.6 & 17.2 & 41.5 & 25.8 & 66.4 & 34.9 & 51.6 & 53.1 & 51.5 & 40.5 & 51.7 \\
$\mathcal{O}(8M)$  & 31.3 & 44.8 & 65.2 & 23.9 & 32.0 & 44.7 & 61.8 & 29.9 & 18.9 & 39.5 & \bf{39.2} & 73.9 & 34.5 & 52.8 & 55.5 & 54.0 & 42.8 & 48.8 \\
$\mathcal{O}(16M)$  & 32.1 & 48.7 & 65.9 & 24.1 & 33.1 & 45.2 & 62.1 & 26.1 & 19.3 & 41.0 & 35.4 & 74.4 & \bf{36.2} & \bf{54.0} & 56.5 & 54.2 & 43.0 & 50.3 \\
$\mathcal{O}(32M)$  & 32.8 & \bf{50.6} & \bf{67.5} & 23.7 & 32.9 & 45.4 & 62.0 & 27.9 & 18.1 & 39.0 & 30.4 & \bf{76.5} & 35.6 & 53.7 & 56.5 & \bf{55.8} & 43.7 & \bf{51.9} \\
$\mathcal{O}(64M)$  & \bf{33.6} & 47.0 & 67.0 & \bf{26.2} & \bf{34.3} & \bf{46.2} & \bf{63.8} & \bf{33.3} & 16.0 & \bf{43.0} & 24.2 & 71.8 & 35.6 & 53.7 & \bf{57.7} & 55.1 & \bf{43.9} & 50.7 \\
\shline
\end{tabular}
\captionsetup{justification=centering}
\caption{{\bf Comparing performance as synthetic-caption data scale increases.} Performance increases with synthetic-caption data scale (peaking at 64M), underscoring the value of diverse set of audio-visual-text data.
}
\label{tab:ablation:data_scaling_full}
\vspace{-5pt}
\end{table*}

\begin{table*}[h!]
\centering
\scriptsize
\tablestyle{0pt}{1.05}
\begin{tabular}{x{12}x{18}x{18}x{22}x{22}x{22}x{22}x{22}x{22}x{22}x{22}x{22}x{22}x{22}x{22}x{22}x{22}x{22}x{22}x{22}x{22}}
\shline
\multirow{3}{*}{\rotatebox[origin=c]{90}{A-layers}}
& \multirow{3}{*}{\rotatebox[origin=c]{90}{A-params}}
& \multirow{3}{*}{\rotatebox[origin=c]{90}{V-params}}
& \multicolumn{4}{c}{\ct[c4]{\it Sound-Retrieval}} 
& \multicolumn{4}{c}{\ct[c4]{\it Sound-Classification}} 
& \multicolumn{4}{c}{\ct[c5]{\it Speech-Classification}} 
& \multicolumn{3}{c}{\ct[c6]{\it Video-Retrieval}} 
& \multicolumn{3}{c}{\ct[c6]{\it Video-Classification}} 
\\
&&
& \multicolumn{2}{c}{\ct[c4]{AudioCaps}}
& \ct[c4]{VALOR} 
& \ct[c4]{Internal} 
& \multicolumn{2}{c}{\ct[c4]{VGGSound}}
& \ct[c4]{GTzan} 
& \ct[c4]{Cremad} 
& \ct[c5]{CV-13}
& \multicolumn{3}{c}{\ct[c5]{D-SUPERB}}
& \ct[c6]{VTT}
& \ct[c6]{MSVD}
& \ct[c6]{ANet}
& \ct[c6]{K400}
& \ct[c6]{K700}
& \ct[c6]{HMDB}
\\
&&
& \ct[c4]{T$\rightarrow$A} & \ct[c4]{A$\rightarrow$V}
& \ct[c4]{T$\rightarrow$AV}
& \ct[c4]{A$\rightarrow$V}
& \ct[c4]{A$\rightarrow$T} & \ct[c4]{AV$\rightarrow$T}
& \ct[c4]{A$\rightarrow$T}
& \ct[c4]{A$\rightarrow$T}
& \ct[c5]{accent}
& \ct[c5]{lid} & \ct[c5]{emo} & \ct[c5]{vocal}
& \ct[c6]{T$\rightarrow$V}
& \ct[c6]{T$\rightarrow$V}
& \ct[c6]{T$\rightarrow$V}
& \ct[c6]{V$\rightarrow$T}
& \ct[c6]{V$\rightarrow$T}
& \ct[c6]{V$\rightarrow$T} \\
\hline
8 & 0.03B & 0.34B & 29.5 & 38.3 & 67.1 & 16.6 & 28.7 & 45.0 & 58.0 & 28.4 & 19.3 & 32.0 & 26.7 & 70.7 & 35.8 & \bf{54.1} & 54.9 & 55.3 & 44.1 & 51.2 \\
12 & 0.09B & 0.35B & 32.0 & 46.0 & 67.8 & 23.1 & 31.4 & 45.4 & 63.1 & 27.9 & 21.9 & 38.0 & 35.4 & 72.2 & 36.6 & 54.0 & 56.3 & 55.7 & 44.5 & 51.6 \\
16 & 0.21B & 0.38B & 33.2 & 48.4 & \bf{68.1} & 25.6 & 33.3 & 45.4 & 61.8 & 32.0 & 19.3 & 41.0 & 32.9 & 73.5 & 36.6 & 53.7 & 56.3 & 55.2 & 43.6 & 48.9 \\
20 & 0.41B & 0.42B & \bf{34.4} & \bf{57.2} & 67.9 & \bf{27.3} & 33.6 & \bf{46.2} & 62.8 & \bf{35.9} & 21.9 & \bf{44.0} & 31.7 & 74.0 & \bf{37.3} & 53.2 & \bf{56.7} & \bf{56.0} & \bf{44.6} & \bf{52.4} \\
24 & 0.70B & 0.45B & \bf{34.4} & 53.4 & 66.9 & 24.4 & 33.2 & 45.7 & 62.6 & 30.1 & \bf{22.7} & 38.0 & \bf{35.8} & 76.9 & 35.2 & 53.3 & 55.7 & 54.4 & 43.7 & 50.1 \\
28 & 1.11B & 0.50B & 34.3 & 56.7 & 66.6 & 24.7 & \bf{34.2} & 44.9 & \bf{65.0} & 32.7 & 16.0 & 34.0 & 32.5 & \bf{78.1} & 35.5 & 53.3 & \bf{56.7} & 53.3 & 43.0 & 49.0 \\
\shline
\end{tabular}
\captionsetup{justification=centering}
\caption{{\bf Scaling the audio encoder}. Scaling from 0.03B to 1.11B parameters shows consistent performance gains with depth. 
The observed saturation around 20 layers is likely due to limited training steps and data in the ablation setup.
}
\label{tab:ablation:model_scaling_full}
\end{table*}

\begin{table*}[h!]
\centering
\scriptsize
\tablestyle{-.0pt}{1.05} 
\begin{tabular}{x{20}x{8}x{8}x{8}x{8}x{8}x{8}x{8}x{8}x{20}x{20}x{20}x{20}x{20}x{20}x{20}x{20}x{20}x{20}x{20}x{20}x{20}x{20}x{20}x{20}x{20}x{20}}
\shline
\multirow{3}{*}{Loss}
& \multirow{3}{*}{\rotatebox[origin=c]{90}{A-V}}
& \multirow{3}{*}{\rotatebox[origin=c]{90}{A-AT}}
& \multirow{3}{*}{\rotatebox[origin=c]{90}{A-AVT}}
& \multirow{3}{*}{\rotatebox[origin=c]{90}{V-AT}}
& \multirow{3}{*}{\rotatebox[origin=c]{90}{V-VT}}
& \multirow{3}{*}{\rotatebox[origin=c]{90}{V-AVT}}
& \multirow{3}{*}{\rotatebox[origin=c]{90}{AV-VT}}
& \multirow{3}{*}{\rotatebox[origin=c]{90}{AV-AVT}}
& \multicolumn{4}{c}{\ct[c4]{\it Sound-Retrieval}} 
& \multicolumn{4}{c}{\ct[c4]{\it Sound-Classification}} 
& \multicolumn{4}{c}{\ct[c5]{\it Speech-Classification}} 
& \multicolumn{3}{c}{\ct[c6]{\it Video-Retrieval}} 
& \multicolumn{3}{c}{\ct[c6]{\it Video-Classification}} 
\\
&&&&&&&&
& \multicolumn{2}{c}{\ct[c4]{AudioCaps}}
& \ct[c4]{VALOR} 
& \ct[c4]{Internal} 
& \multicolumn{2}{c}{\ct[c4]{VGGSound}}
& \ct[c4]{GTzan} 
& \ct[c4]{Cremad} 
& \ct[c5]{CV-13}
& \multicolumn{3}{c}{\ct[c5]{D-SUPERB}}
& \ct[c6]{VTT}
& \ct[c6]{MSVD}
& \ct[c6]{ANet}
& \ct[c6]{K400}
& \ct[c6]{K700}
& \ct[c6]{HMDB}
\\
&&&&&&&&
& \ct[c4]{T$\rightarrow$A} & \ct[c4]{A$\rightarrow$V}
& \ct[c4]{T$\rightarrow$AV}
& \ct[c4]{A$\rightarrow$V}
& \ct[c4]{A$\rightarrow$T} & \ct[c4]{AV$\rightarrow$T}
& \ct[c4]{A$\rightarrow$T}
& \ct[c4]{A$\rightarrow$T}
& \ct[c5]{accent}
& \ct[c5]{lid} & \ct[c5]{emo} & \ct[c5]{vocal}
& \ct[c6]{T$\rightarrow$V}
& \ct[c6]{T$\rightarrow$V}
& \ct[c6]{T$\rightarrow$V}
& \ct[c6]{V$\rightarrow$T}
& \ct[c6]{V$\rightarrow$T}
& \ct[c6]{V$\rightarrow$T} \\
\hline       
SigLIP & - & \checkmark & - & - & - & - & - & - &
31.9 & 0.1 & 0.0 & 0.0 & 32.4 & 0.5 & 60.4 & 30.8 & \bf{23.5} & \bf{52.5} & 30.4 & 76.1 & 0.1 & 0.2 & 0.0 & 0.3 & 0.1 & 2.2 \\
SigLIP & - & \checkmark & - & - & \checkmark & - & - & - &
32.2 & 0.1 & 0.1 & 0.0 & 31.2 & 0.3 & 62.2 & 27.3 & 20.6 & 33.0 & \bf{32.5} & 71.4 & 27.2 & 44.5 & 55.5 & 42.2 & 33.0 & 40.8 \\
SigLIP & - & \checkmark & - & - & \checkmark & - & - & \checkmark &
33.0 & 0.0 & 47.1 & 0.1 & 31.9 & 0.3 & 56.6 & 30.6 & 18.1 & 36.0 & 32.1 & 71.8 & 26.1 & 42.5 & 53.0 & 41.8 & 31.3 & 42.1 \\
SigLIP & \checkmark & \checkmark & - & - & \checkmark & - & - & \checkmark & 
31.9 & 45.6 & 47.5 & 24.7 & 30.6 & 0.4 & 53.3 & 25.2 & 21.9 & 38.5 & 32.1 & 70.0 & 27.3 & 44.2 & 47.0 & 39.7 & 30.1 & 36.4 \\
SigLIP & \checkmark & \checkmark & - & \checkmark & \checkmark & - & \checkmark & \checkmark & 
31.4 & \bf{49.5} & 66.3 & \bf{25.1} & 32.5 & 45.1 & 61.0 & 26.2 & 18.1 & 43.5 & \bf{32.5} & \bf{76.5} & \bf{34.5} & 53.9 & 56.7 & \bf{55.8} & \bf{43.8} & \bf{49.7} \\
SigLIP & \checkmark & \checkmark & \checkmark & \checkmark & \checkmark & \checkmark & \checkmark & \checkmark & 
\bf{32.9} & 45.7 & \bf{68.3} & 21.8 & \bf{33.3} & \bf{45.5} & \bf{62.4} & \bf{33.2} & 17.2 & 47.5 & 30.4 & 74.7 & 33.9 & \bf{54.1} & \bf{57.2} & 54.2 & 43.1 & 48.9 \\
\shline
\end{tabular}
\captionsetup{justification=centering}
\caption{\textbf{Scaling the SigLIP objective.} A: Audio, V: Video, AT: Audio text caption, VT: Video text caption. Expanding the contrastive objective to cover more modality pairs strengthens cross-modal alignment and improves zero-shot retrieval and classification. Audio–text–only training lags behind, while adding cross-modality pairs (e.g., V$\rightarrow$AT, AV$\rightarrow$VT) yields further gains. Performance peaks when the objective includes all eight pairs (bottom row).
} 
\label{tab:ablation:clip_siglip_full}
\end{table*}

\begin{table*}[h!]
\centering
\scriptsize
\tablestyle{0pt}{1.05} 
\begin{tabular}{x{45}x{22}x{22}x{22}x{22}x{22}x{22}x{22}x{22}x{22}x{22}x{22}x{22}x{22}x{22}x{22}x{22}x{22}x{22}}
\shline
\multirow{3}{*}{Text Encoder}
& \multicolumn{4}{c}{\ct[c4]{\it Sound-Retrieval}} 
& \multicolumn{4}{c}{\ct[c4]{\it Sound-Classification}} 
& \multicolumn{4}{c}{\ct[c5]{\it Speech-Classification}} 
& \multicolumn{3}{c}{\ct[c6]{\it Video-Retrieval}} 
& \multicolumn{3}{c}{\ct[c6]{\it Video-Classification}} 
\\
& \multicolumn{2}{c}{\ct[c4]{AudioCaps}}
& \ct[c4]{VALOR} 
& \ct[c4]{Internal} 
& \multicolumn{2}{c}{\ct[c4]{VGGSound}}
& \ct[c4]{GTzan} 
& \ct[c4]{Cremad} 
& \ct[c5]{CV-13}
& \multicolumn{3}{c}{\ct[c5]{D-SUPERB}}
& \ct[c6]{VTT}
& \ct[c6]{MSVD}
& \ct[c6]{ANet}
& \ct[c6]{K400}
& \ct[c6]{K700}
& \ct[c6]{HMDB}
\\
& \ct[c4]{T$\rightarrow$A} & \ct[c4]{A$\rightarrow$V}
& \ct[c4]{T$\rightarrow$AV}
& \ct[c4]{A$\rightarrow$V}
& \ct[c4]{A$\rightarrow$T} & \ct[c4]{AV$\rightarrow$T}
& \ct[c4]{A$\rightarrow$T}
& \ct[c4]{A$\rightarrow$T}
& \ct[c5]{accent}
& \ct[c5]{lid} & \ct[c5]{emo} & \ct[c5]{vocal}
& \ct[c6]{T$\rightarrow$V}
& \ct[c6]{T$\rightarrow$V}
& \ct[c6]{T$\rightarrow$V}
& \ct[c6]{V$\rightarrow$T}
& \ct[c6]{V$\rightarrow$T}
& \ct[c6]{V$\rightarrow$T} \\
\hline
PE-Text & 30.5 & \bf{49.1} & \bf{66.6} & 23.1 & 33.3 & \bf{47.3} & 62.1 & 25.9 & \bf{19.3} & 46.5 & 23.3 & 59.7 & \bf{45.3} & \bf{60.4} & 52.2 & \bf{66.7} & \bf{57.6} & \bf{55.3} \\ 
ModernBERT & \bf{34.1} & 49.0 & 66.5 & \bf{25.3} & \bf{34.1} & 46.1 & \bf{63.0} & \bf{29.5} & 18.1 & \bf{47.0} & \bf{37.9} & \bf{74.2} & 36.2 & 53.3 & \bf{57.4} & 54.6 & 44.0 & 50.2 \\
\shline
\end{tabular}
\captionsetup{justification=centering}
\caption{{\bf Comparison of text encoder choices for \PEav{} using the \PEcore{L} visual encoder.} ModernBERT outperforms the original \PEcore{L} text encoder on audio-focused tasks due to its longer context (512 tokens vs. 32) and its support of general text domain, while \PEcore{L} performs better on video-centric tasks. As noted in main results, ModernBERT catches up with and surpasses the \PEcore{L} text encoder after fine-tuning, making it the preferred choice for \PEav{}.}
\label{tab:ablation:modernbert_petext}
\end{table*}

\begin{table*}[h!]
\centering
\scriptsize
\tablestyle{0pt}{1.05} 
\begin{tabular}{x{30}x{30}x{22}x{22}x{22}x{22}x{22}x{22}x{22}x{22}x{22}x{22}x{22}x{22}x{22}x{22}x{22}x{22}x{22}x{22}}
\shline
\multirow{3}{*}{PT frames}
& \multirow{3}{*}{FT frames}
& \multicolumn{4}{c}{\ct[c4]{\it Sound-Retrieval}} 
& \multicolumn{4}{c}{\ct[c4]{\it Sound-Classification}} 
& \multicolumn{4}{c}{\ct[c5]{\it Speech-Classification}} 
& \multicolumn{3}{c}{\ct[c6]{\it Video-Retrieval}} 
& \multicolumn{3}{c}{\ct[c6]{\it Video-Classification}} 
\\
&
& \multicolumn{2}{c}{\ct[c4]{AudioCaps}}
& \ct[c4]{VALOR} 
& \ct[c4]{Internal} 
& \multicolumn{2}{c}{\ct[c4]{VGGSound}}
& \ct[c4]{GTzan} 
& \ct[c4]{Cremad} 
& \ct[c5]{CV-13}
& \multicolumn{3}{c}{\ct[c5]{D-SUPERB}}
& \ct[c6]{VTT}
& \ct[c6]{MSVD}
& \ct[c6]{ANet}
& \ct[c6]{K400}
& \ct[c6]{K700}
& \ct[c6]{HMDB}
\\
&
& \ct[c4]{T$\rightarrow$A} & \ct[c4]{A$\rightarrow$V}
& \ct[c4]{T$\rightarrow$AV}
& \ct[c4]{A$\rightarrow$V}
& \ct[c4]{A$\rightarrow$T} & \ct[c4]{AV$\rightarrow$T}
& \ct[c4]{A$\rightarrow$T}
& \ct[c4]{A$\rightarrow$T}
& \ct[c5]{accent}
& \ct[c5]{lid} & \ct[c5]{emo} & \ct[c5]{vocal}
& \ct[c6]{T$\rightarrow$V}
& \ct[c6]{T$\rightarrow$V}
& \ct[c6]{T$\rightarrow$V}
& \ct[c6]{V$\rightarrow$T}
& \ct[c6]{V$\rightarrow$T}
& \ct[c6]{V$\rightarrow$T} \\
\hline
16 & 16 & 43.1 & 87.1 & 80.4 & 25.5 & 45.7 & 51.2 & 73.8 & 34.2 & 24.4 & \bf{65.0} & 45.4 & 85.0 & 46.8 & 58.6 & 63.4 & 75.8 & 66.2 & 62.2 \\
16 & All & 42.0 & 87.5 & 79.3 & 43.2 & 45.9 & 51.3 & \bf{74.2} & 37.5 & 21.0 & 59.5 & 38.3 & 85.3 & 46.4 & 58.9 & 63.9 & 75.9 & 66.5 & 62.4 \\
All & 16 & 44.7 & 85.9 & \bf{83.7} & 23.7 & 46.7 & 51.8 & 72.3 & 42.0 & 23.1 & 62.0 & \bf{47.9} & 85.4 & 49.0 & 60.5 & 65.4 & 78.4 & 68.2 & \bf{66.0} \\
All & All & \bf{45.8} & \bf{89.0} & \bf{83.7} & \bf{49.0} & \bf{47.1} & \bf{52.4} & 72.2 & \bf{43.3} & \bf{25.6} & 64.5 & 43.8 & \bf{86.1} & \bf{51.9} & \bf{60.8} & \bf{66.5} & \bf{78.9} & \bf{69.0} & 65.1 \\
\shline
\end{tabular}
\captionsetup{justification=centering}
\caption{{\bf Impact of video frame rate.} ``All'': Encode all frames at 30 FPS. ``16'': Uniformly sample and encode 16 frames per video. Both configurations yield similar performance overall. 
However, a notable exception arises in the internal video-music retrieval task, which involves videos with wide variations in duration. In this case, the 30 FPS models capture duration information more effectively and achieve better performance.
} 
\label{tab:ablation:fps}
\end{table*}

\begin{table*}[h!]
\centering
\scriptsize
\tablestyle{0pt}{1.05} 
\begin{tabular}{x{30}x{22}x{22}x{22}x{22}x{22}x{22}x{22}x{22}x{22}x{22}x{22}x{22}x{22}x{22}x{22}x{22}x{22}x{22}}
\shline
\multirow{3}{*}{SSL Loss}
& \multicolumn{4}{c}{\ct[c4]{\it Sound-Retrieval}} 
& \multicolumn{4}{c}{\ct[c4]{\it Sound-Classification}} 
& \multicolumn{4}{c}{\ct[c5]{\it Speech-Classification}} 
& \multicolumn{3}{c}{\ct[c6]{\it Video-Retrieval}} 
& \multicolumn{3}{c}{\ct[c6]{\it Video-Classification}} 
\\
& \multicolumn{2}{c}{\ct[c4]{AudioCaps}}
& \ct[c4]{VALOR} 
& \ct[c4]{Internal} 
& \multicolumn{2}{c}{\ct[c4]{VGGSound}}
& \ct[c4]{GTzan} 
& \ct[c4]{Cremad} 
& \ct[c5]{CV-13}
& \multicolumn{3}{c}{\ct[c5]{D-SUPERB}}
& \ct[c6]{VTT}
& \ct[c6]{MSVD}
& \ct[c6]{ANet}
& \ct[c6]{K400}
& \ct[c6]{K700}
& \ct[c6]{HMDB}
\\
& \ct[c4]{T$\rightarrow$A} & \ct[c4]{A$\rightarrow$V}
& \ct[c4]{T$\rightarrow$AV}
& \ct[c4]{A$\rightarrow$V}
& \ct[c4]{A$\rightarrow$T} & \ct[c4]{AV$\rightarrow$T}
& \ct[c4]{A$\rightarrow$T}
& \ct[c4]{A$\rightarrow$T}
& \ct[c5]{accent}
& \ct[c5]{lid} & \ct[c5]{emo} & \ct[c5]{vocal}
& \ct[c6]{T$\rightarrow$V}
& \ct[c6]{T$\rightarrow$V}
& \ct[c6]{T$\rightarrow$V}
& \ct[c6]{V$\rightarrow$T}
& \ct[c6]{V$\rightarrow$T}
& \ct[c6]{V$\rightarrow$T} \\
\hline
None  & 30.7 & 44.1 & 65.7 & \bf{25.7} & 31.4 & 44.6 & 61.2 & 30.3 & 17.7 & 35.5 & \bf{33.3} & \bf{74.9} & 34.6 & 53.6 & 55.6 & 54.2 & 43.1 & 49.3 \\
NCE & 31.9 & 46.9 & 67.7 & 25.3 & 31.6 & 45.0 & \bf{63.1} & 26.3 & 14.7 & 36.0 & 32.9 & 72.2 & 35.6 & \bf{53.9} & \bf{56.7} & 54.0 & 43.0 & 48.9 \\
BEST-RQ & \bf{33.2} & \bf{48.4} & \bf{68.1} & 25.6 & \bf{33.3} & \bf{45.4} & 61.8 & \bf{32.0} & \bf{19.3} & \bf{41.0} & 32.9 & 73.5 & \bf{36.6} & 53.7 & 56.3 & \bf{55.2} & \bf{43.6} & \bf{50.9} \\
    \shline
\end{tabular}
\captionsetup{justification=centering}
\caption{\textbf{Audio encoder losses: NCE vs. BEST-RQ.} NCE follows wav2vec~2.0 contrastive objective using DAC-VAE features as negatives but skips quantization. BEST-RQ delivers the strongest results, outperforming NCE and no-SSL baselines. Speech and sound tasks benefit the most from the finer-grained representations encouraged by BEST-RQ.
} 
\label{tab:ablation:bestrq_nce}
\end{table*}

\paragraph{Text encoder}. In Tab.~\ref{tab:ablation:modernbert_petext}, we present an ablation of different choices of text encoders for \PEav{} with \PEcore{L} as the default visual encoder. We compare the performance of ModernBERT~\cite{modernbert} with the original paired CLIP text encoder from \PEcore{L}~\cite{pe}.
After audio-video-text pre-training, we observe that the original \PEcore{L} text encoder performs better on video-centric tasks such as VTT~\cite{vtt} and Kinetics~\cite{kay2017kinetics}. %
However, its performance lags on out-of-visual-domain concepts, such as audio events, in audio-focused tasks like text-to-audio retrieval in AudioCaps~\cite{kim2019audiocaps}, and LID, emotion, vocal classification in Dynamic-SUPERB~\cite{dsuperb}, where ModernBERT excels. 
Additionally, the \PEcore{L} text encoder supports a shorter context length (32 tokens) compared to ModernBERT (512 tokens). 
Therefore, we choose the pre-trained ModernBERT as our default text encoder. After the later fine-tuning phase, \PEav{} catches up and outperforms the original PE models as shown in 
Tab.~{3} in the main paper.

\paragraph{Video Frame Rate}. Tab.~\ref{tab:ablation:fps} compares different video frame rates—30 FPS versus sampling a fixed set of 16 frames—during pre-training and fine-tuning.
Because this ablation addresses a critical design choice and training various models at both frame rates is computationally intensive, we performed this ablation with our largest model \PEav{L}.
We train for the full pre-training duration of 250K steps followed by 50K steps of fine-tuning, ensuring that conclusions are drawn from the strongest configuration.

For most tasks, models trained with these different frame rate configurations exhibit similar performance, while 30 FPS sampling provides a modest advantage, especially during in the pre-training phase. However, models operating at higher frame rates achieve better results on downstream tasks that require fine-grained temporal understanding, such as ActivityNet and audio-video retrieval on the internal dataset with a wide duration variation. 
Notably, models trained with 30 FPS sampling inherently encode duration information, enabling them to retrieve audios or videos of similar length as the query. 
This also highlights a key limitation of existing audio-video retrieval benchmarks, which do not evaluate robustness to variation in duration. 
With this, for all other models, we adopt 30 FPS sampling during pre-training and later fine-tune with the same setup or with fixed 16-frame inputs.

\paragraph{BEST-RQ vs NCE Loss}. 
In Tab.~\ref{tab:ablation:bestrq_nce}, we compare different SSL losses for the audio encoder.
The NCE loss is similar to the contrastive loss in wav2vec 2.0, except that we skip the quantization step and use the DAC-VAE features as the negative samples directly.
BEST-RQ offers the best overall results, significantly outperforming NCE loss and no SSL loss conditions.
Video tasks retain performance while some even show slight improvement when the BEST-RQ loss is present.
The results demonstrate the necessity of including BEST-RQ loss to enhance the performance of sound and speech tasks without compromising video capabilities, corroborating the hypothesis that encouraging fine-grained representations benefits speech-related tasks.

\begin{table*}[h!]
    \centering
    \scriptsize
    \makebox[\linewidth][c]{
    \tablestyle{-1.5pt}{1.2} 
    \begin{tabular}{y{48} x{16}x{16}x{16}x{16}x{16}x{16}x{16}x{16}x{16}x{16} x{16}x{16} x{16}x{16}x{16}x{16}x{16}x{16} x{16} x{16}x{16} x{16}x{16} x{16}x{16} x{16}x{16} x{16}x{16} x{16}x{16}x{16}x{16}}
        \shline
        \multirow{2}{*}{\vspace{+10pt} Model} %
        & \cz[c3]{AudioCaps}{A$\rightarrow$T}
        & \cz[c3]{AudioCaps}{T$\rightarrow$A}
        & \cz[c3]{AudioCaps}{V$\rightarrow$T}
        & \cz[c3]{AudioCaps}{T$\rightarrow$V}
        & \cz[c3]{AudioCaps}{A$\rightarrow$V}
        & \cz[c3]{AudioCaps}{V$\rightarrow$A}
        & \cz[c3]{AudioCaps}{A+V$\rightarrow$T}
        & \cz[c3]{AudioCaps}{T$\rightarrow$A+V}
        & \cz[c3]{AudioCaps}{A+VT$\rightarrow$V}
        & \cz[c3]{AudioCaps}{V+AT$\rightarrow$A}
        & \cz[c4]{Clotho}{T$\rightarrow$A}
        & \cz[c4]{Clotho}{A$\rightarrow$T}
        & \cz[c4]{Valor}{A$\rightarrow$T}
        & \cz[c4]{Valor}{T$\rightarrow$A}
        & \cz[c4]{Valor}{V$\rightarrow$T}
        & \cz[c4]{Valor}{T$\rightarrow$V}
        & \cz[c4]{Valor}{A+V$\rightarrow$T}
        & \cz[c4]{Valor}{T$\rightarrow$A+V}
        & \cz[c4]{VCTK}{A$\rightarrow$T}
        & \cz[c4]{VGGSound}{V$\rightarrow$A}
        & \cz[c4]{VGGSound}{A$\rightarrow$V}
        & \cz[c4]{Internal}{V$\rightarrow$A}
        & \cz[c4]{Internal}{A$\rightarrow$V}
        & \cz[c5]{MSR-VTT}{T$\rightarrow$V}
        & \cz[c5]{MSR-VTT}{V$\rightarrow$T}
        & \cz[c5]{MSVD}{T$\rightarrow$V}
        & \cz[c5]{MSVD}{V$\rightarrow$T}
        & \cz[c5]{ActivityNet}{T$\rightarrow$V}
        & \cz[c5]{ActivityNet}{V$\rightarrow$T}
        & \cz[c5]{DiDeMo}{T$\rightarrow$V}
        & \cz[c5]{DiDeMo}{V$\rightarrow$T} 
        & \cz[c5]{VATEX}{T$\rightarrow$V}
        & \cz[c5]{VATEX}{V$\rightarrow$T} 
        \\
        \hline
        \textit{Baselines} &  \\ 
         
         AFlamingo2 \cite{af2} & 45.7 & 29.8 & - & - & - & - & - & - & - & - & 20.4 & 16.9 & 7.4 & 7.3 & - & - & - & - & 0.3 & - & - & - & - & - & - & - & - & - & - & - & - & - & - \\
        ImageBind~\cite{girdhar2023imagebind} & 9.6 & 6.6 & 11.3 & 7.6 & 51.6 & 51.3 & - & - & - & - & 5.5 & 3.9 & 4.9 & 5.4 & 35.8 & 36.1 & - & - & 0.4 & 10.5 & 10.8 & 2.8 & 2.8 & 40.6 & 42.9 & 47.9 & 70.9 & 36.6 & 34.1 & 36.0 & 38.2 & 69.8 & 69.8 \\
        CLAP-Fusion~\cite{laion_clap} & 43.3 & 35.4 & - & - & - & - & - & - & - & - & 20.2 & 17.7 & 5.4 & 5.5 & - & - & - & - & 0.3 & - & - & - & - & - & - & - & - & - & - & - & - & - & -\\
        CLAP~\cite{laion_clap} & 43.7 & 31.6 & - & - & - & - & - & - & - & - & 21.0 & 16.6 & 6.5 & 5.8 & - & - & - & - & 0.2 & - & - & - & - & - & - & - & - & - & - & - & - & - & -\\
        LanguageBind~\cite{langbind} & 27.1 & 19.7 & 14.2 & 10.6 & 10.7 & 9.1 & - & - & - & - & 17.1 & 13.3 & 5.6 & 6.5 & 46.9 & 46.8 & - & - & 0.2 & 1.8 & 1.6 & 1.3 & 1.4 & 48.6 & 48.7 & 55.6 & 78.8 & 48.0 & 48.8 & 43.5 & 44.7 & 82.9 & 83.1 \\
        M2D-CLAP~\cite{m2d} & 27.4 & 27.4 & - & - & - & - & - & - & - & - & 11.4 & 10.5 & 5.9 & 6.3 & - & - & - & - & 0.1 & - & - & - & - & - & - & - & - & - & - & - & - & - & -\\
        MS-CLAP\textsuperscript{23}~\cite{ms_clap} & 32.4 & 23.4 & - & - & - & - & - & - & - & - & 23.4 & 17.8 & 8.0 & 5.9 & - & - & - & - & 0.3 & - & - & - & - & - & - & - & - & - & - & - & - & - & -\\
        \hline
        \textit{16 Frames} & \\
        \PEav{S} & 59.4 & 41.2 & 26.2 & 18.6 & 75.4 & 75.4 & 56.1 & 40.9 & 71.9 & 89.7 & 33.6 & 24.0 & 30.2 & 29.8 & 70.3 & 70.1 & 75.9 & 76.3 & \textbf{96.1} & 33.3 & 34.1 & 17.7 & 17.9 & 46.7 & 49.6 & 60.1 & 86.4 & 63.4 & 64.8 & 48.7 & 49.0 & 94.2 & 93.7 \\
        \PEav{B} & 59.7 & 43.1 & 26.9 & 19.8 & 81.6 & 80.6 & 57.1 & 41.1 & 78.1 & 91.6 & 33.4 & 23.4 & 31.2 & 31.9 & 70.7 & 70.0 & 76.6 & 76.0 & 94.8 & 38.4 & 39.0 & 20.4 & 20.4 & 48.6 & 50.3 & 60.8 & 87.6 & 64.0 & 64.9 & 46.2 & 47.8 & 94.3 & 93.8 \\
        \PEav{L} & 62.0 & 44.7 & 26.9 & 19.5 & 85.9 & 86.1 & 58.0 & 41.6 & 83.1 & 94.6 & 32.6 & 22.8 & 35.2 & 35.0 & 70.8 & 70.9 & 76.0 & 76.8 & 85.6 & 44.6 & 45.2 & 23.7 & 23.9 & 49.0 & 50.5 & 60.5 & \textbf{88.4} & 65.4 & 66.5 & 48.9 & 50.1 & 94.9 & 94.4 \\
        \hline
        \textit{30 FPS} & \\
        \PEav{S}-OOD & 55.2 & 40.2 & 25.4 & 17.7 & 76.6 & 75.3 & 52.6 & 36.8 & 73.0 & 89.1 & 32.3 & 23.4 & 28.9 & 28.0 & 70.5 & 69.8 & 76.0 & 76.1 & 73.1 & 27.7 & 28.5 & 41.4 & 40.8 & 50.9 & 51.0 & 60.8 & 87.6 & 66.7 & 65.9 & 51.4 & 51.8 & 95.1 & 94.6 \\
        \PEav{B}-OOD & 56.8 & 40.4 & 24.6 & 17.7 & 81.8 & 82.2 & 54.0 & 38.0 & 77.3 & 92.6 & 32.7 & 24.3 & 30.5 & 30.7 & 70.0 & 70.0 & 76.5 & 75.6 & 57.3 & 31.0 & 31.8 & 46.7 & 45.1 & 50.5 & 49.7 & \textbf{61.2} & 86.3 & 67.3 & 67.7 & 51.8 & \bf{52.4} & 94.0 & 94.0 \\
        \PEav{L}-OOD & 60.0 & 43.4 & 26.2 & 18.2 & 87.5 & 86.1 & 53.3 & 38.7 & 82.9 & 93.3 & 33.3 & 23.7 & 34.7 & 34.2 & 71.4 & 70.2 & 75.8 & 76.0 & 50.7 & 35.6 & 36.5 & \textbf{52.2} & \textbf{50.3} & 50.4 & \textbf{51.5} & 61.0 & 87.5 & \textbf{67.6} & \textbf{68.0} & \bf{51.9} & 51.5 & 95.1 & 94.4 \\
        \PEav{S} & 58.2 & 41.8 & 27.2 & 18.8 & 77.7 & 77.4 & 56.5 & 40.1 & 73.6 & 90.3 & 33.2 & 23.9 & 30.1 & 29.3 & 71.6 & 70.9 & 76.6 & 76.4 & 94.9 & 35.4 & 35.4 & 41.0 & 40.5 & 49.3 & 49.4 & 59.8 & 87.5 & 64.8 & 65.5 & 50.0 & 49.0 & 94.5 & 94.5 \\
        \PEav{B} & 60.0 & 42.7 & 28.3 & 19.6 & 83.5 & 83.7 & 56.1 & 41.0 & 79.9 & 93.2 & 33.7 & 23.8 & 31.0 & 30.8 & \textbf{72.1} & \textbf{71.2} & \textbf{76.9} & \textbf{76.9} & 94.9 & 40.7 & 40.7 & 45.9 & 44.6 & 47.7 & 48.4 & 60.7 & 87.6 & 65.7 & 65.9 & 49.3 & 50.1 & 94.9 & 94.4 \\
        \PEav{L} (PT) & 48.5 & 33.7 & 22.6 & 14.7 & 83.4 & 83.3 & 47.9 & 33.2 & - & - & 26.3 & 17.5 & 24.2 & 24.0 & 56.1 & 57.1 & 62.6 & 63.3 & 16.7 & 32.6 & 33.9 & 50.6 & 47.8 & 35.5 & 36.6 & 49.8 & 79.6 & 62.0 & 64.0 & 44.8 & 46.3 & 87.1 & 87.2 \\
        \PEav{L} & \textbf{63.3} & \textbf{45.8} & \textbf{29.1} & \textbf{20.8} & \textbf{89.0} & \textbf{88.3} & \textbf{58.2} & \textbf{42.6} & \textbf{84.0} & \textbf{95.2} & 32.7 & 23.0 & \textbf{36.4} & \textbf{35.1} & 71.6 & 70.9 & \textbf{76.9} & 76.8 & 85.6 & \textbf{47.8} & \textbf{48.3} & 49.0 & 46.5 & \textbf{51.9} & 51.2 & 60.8 & 87.6 & 66.5 & 67.7 & 51.6 & 51.7 &\bf{95.1} & \bf{94.8} \\
    \shline
    \end{tabular}
    }
    \caption{{\bf Full Zero-Shot Retrieval Results}. Per-dataset Recall@1 for all audio–text, video–text, and audio–video retrieval directions corresponding to the main audio and video tables. \PEav{} consistently outperforms baseline models  across most benchmarks and retrieval directions.}
    \label{tab:app:full_retrieval_results}
\end{table*}

\begin{table*}[h!]
    \centering
    \scriptsize
    \tablestyle{-0pt}{1.2} 
    \begin{tabular}{y{60}x{24}x{24}x{24}x{24}x{24}x{24}x{24}x{24}x{24}x{24}x{24}x{24}x{24}x{24}x{24}}
    \shline
    \multirow{2}{*}{\vspace{-2cm} Model}
    & \multicolumn{12}{c}{\ct[c3]{\it Zero-Shot Retrieval}} 
    & \multicolumn{3}{c}{\ct[c5]{\it Zero-Shot Classification}} 
    \\
    & \cz[c3]{AudioCaps}{T$\rightarrow$A+V}
    & \cz[c3]{AudioCaps}{T+A$\rightarrow$V}
    & \cz[c3]{AudioCaps}{T+V$\rightarrow$A}
    & \cz[c3]{VALOR}{T$\rightarrow$A+V}
    & \cz[c3]{VALOR}{T+A$\rightarrow$V}
    & \cz[c3]{VALOR}{T+V$\rightarrow$A}
    & \cz[c3]{VTT}{T$\rightarrow$A+V}
    & \cz[c3]{VTT}{T+A$\rightarrow$V}
    & \cz[c3]{VTT}{T+V$\rightarrow$A}    
    & \cz[c3]{DiDeMo}{T$\rightarrow$A+V}
    & \cz[c3]{DiDeMo}{T+A$\rightarrow$V}
    & \cz[c3]{DiDeMo}{T+V$\rightarrow$A} 
    & \cz[c5]{VGGSound}{A$\rightarrow$T}
    & \cz[c5]{VGGSound}{V$\rightarrow$T}
    & \cz[c5]{VGGSound}{A+V$\rightarrow$T}       
    \\
    \hline
    ImageBind\textsuperscript{$\dagger$}~\cite{girdhar2023imagebind} & 7.6 & 51.6 & 51.3 & 36.1 & 36.1 & 24.2 & 41.9 & 41.9 & 23.9 & 36.1 & 36.1 & 18.3 & 28.2 & 40.4 & 40.4 \\
    LanguageBind\textsuperscript{$\dagger$}~\cite{langbind} & 19.7 & 10.7 & 19.7 & 46.8 & 46.8 & 6.5 & 50.9 & 50.9 & 3.7 & 44.2 & 44.2 & 5.5 & 26.0 & 45.4 & 45.4 \\ \hline
    \PEav{S}-OOD\textsuperscript{$\dagger$} & 40.2 & 76.6 & 75.3 & 69.8 & 69.8 & 57.7 & 52.0 & 64.0 & 65.3 & {53.7} & 55.2 & 55.5 & 40.0 & 46.3 & 46.3 \\
    \PEav{S}-OOD & 36.8 & 73.0 & 89.1 & 76.1 & 88.8 & 65.2 & 48.3 & 86.7 & 63.9 & 48.0 & 76.2 & 51.4 & 40.0 & 46.3 & 52.5 \\
    \PEav{B}-OOD\textsuperscript{$\dagger$} & 40.4 & 81.8 & 82.2 & 70.0 & 70.0 & 63.9 & 51.4 & 64.4 & 66.0 & 52.0 & 60.6 & 62.0 & 41.4 & 47.0 & 47.0 \\
    \PEav{B}-OOD & 38.0 & 77.3 & 92.6 & 75.6 & 90.5 & 70.9 & 50.1 & 87.8 & 65.8 & 48.0 & 81.0 & 57.8 & 41.4 & 47.0 & 52.1 \\
    \PEav{L}-OOD\textsuperscript{$\dagger$} & 43.4 & 87.5 & 86.1 & 70.2 & 72.2 & 73.1 & 51.5 & 70.0 & {71.6} & 52.7 & 66.8 & 67.7 & 43.9 & 46.7 & 46.7 \\
    \PEav{L}-OOD & 38.7 & 82.9 & 93.3 & 76.0 & 92.0 & 77.2 & 49.9 & {89.0} & 69.8 & 48.8 & {82.1} & 62.1 & 43.9 & 46.7 & 52.0 \\
    \PEav{S}\textsuperscript{$\dagger$} & 41.8 & 77.7 & 77.4 & 70.9 & 70.9 & 60.1 & 50.1 & 59.6 & 61.1 & 49.8 & 53.2 & 56.9 & 43.0 & 47.3 & 47.3 \\
    \PEav{S} & 40.1 & 73.6 & 90.3 & 76.4 & 89.8 & 67.6 & 48.3 & 85.9 & 58.6 & 40.0 & 75.5 & 44.5 & 43.0 & 47.3 & 52.2\\
    \PEav{B}\textsuperscript{$\dagger$} & 42.7 & 83.5 & 83.7 & 71.2 & 71.2 & 65.3 & 50.1 & 62.3 & 63.8 & 49.6 & 61.6 & 63.4 & 44.5 & 47.8 & 47.8 \\
    \PEav{B} & 41.0 & 79.9 & 93.2 & {76.9} & 91.1 & 72.0 & 48.1 & 86.3 & 60.2 & 39.8 & 80.7 & 51.9 & 44.5 & 47.8 & {52.7}\\
    \PEav{L}\textsuperscript{$\dagger$} & {45.8} & {89.0} & 88.3 & 70.9 & 73.8 & 74.5 & {52.9} & 63.6 & 67.4 & 51.4 & 69.3 & {69.5} & {47.1} & {48.0} & 48.0 \\
    \PEav{L} & 42.6 & 84.0 & {95.2} & 76.8 & {93.0} & {78.8} & 49.0 & 85.3 & 65.1 & 43.0 & 80.8 & 61.6 & {47.1} & {48.0} & 51.8 \\

    \shline
    \end{tabular}
    \caption{\textbf{Zero-shot joint-modal retrieval and classification.} For baselines and \PEav{} variants marked with \textsuperscript{$\dagger$},  joint queries are approximated via the max over unimodal results: T+V$\rightarrow$A = $\max($T$\rightarrow$A, V$\rightarrow$A$)$, T$\rightarrow$A+V = $\max($T$\rightarrow$A, T$\rightarrow$V$)$, and T+A$\rightarrow$V = $\max($T$\rightarrow$V, A$\rightarrow$V$)$. All other \PEav{} variants use native joint embeddings for T+V, A+V, and T+A. For all \PEav{} models, we observe that joint embeddings are helpful when the input modalities are complimentary to each others. Specifically (i) for audio-only captions (\textit{AudioCaps}) V+T$\rightarrow$A significantly outperforms V$\rightarrow$A and T$\rightarrow$A; (ii) for visual captions (\textit{DiDeMo} \& \textit{MSR-VTT}) A+T$\rightarrow$V improves over A$\rightarrow$V and T$\rightarrow$V; (iii) for audio-visual captions (\textit{VALOR}) both V+T$\rightarrow$A and A+T$\rightarrow$V help; (iv) \textit{VGGSound}: A+V$\rightarrow$T exceeds A$\rightarrow$T and V$\rightarrow$T.}
    \label{tab:app:exp:core:joint_results}
\end{table*}

{
    \small
    \bibliographystyle{ieeenat_fullname}
    \bibliography{reference}
}

\end{document}